\documentclass[a4paper,12pt,final]{article}
\usepackage[sfdefault]{biolinum}  
\usepackage{amsfonts,amsmath,amsthm,amssymb}
\usepackage[margin=1.25in]{geometry}
\usepackage[onehalfspacing]{setspace}
\usepackage{soul}
\usepackage[multiple]{footmisc}
\usepackage{natbib}
\usepackage{tikz}
\usepackage[colorlinks,linkcolor=links,citecolor=cites,urlcolor=MyDarkBlue]{hyperref}
\usepackage[nameinlink]{cleveref}
\usepackage{bbm}
\usepackage[justification=centering]{caption}
\usepackage[justification=centering]{subfig}
\usepackage[section]{placeins}
\usepackage{parskip}
\usepackage{nicefrac}

\definecolor{MyDarkBlue}{rgb}{0,0.08,0.45}
\definecolor{cites}{HTML}{324b13}
\definecolor{links}{HTML}{1a663b}
\definecolor{MyLightMagenta}{cmyk}{0.1,0.8,0,0.1}

\newtheorem{lemma}{Lemma} 
\newtheorem{proposition}{Proposition}

\newtheorem{corollary}{Corollary}

\newtheorem{remark}{Remark}
\newtheorem{observation}{Observation}

\usetikzlibrary{arrows}
\usetikzlibrary{trees,calc,snakes,fit,shapes,positioning}
\usetikzlibrary{decorations.pathmorphing}
\usetikzlibrary{decorations.markings}
\usetikzlibrary{decorations.shapes}
\usetikzlibrary{patterns}
\usepackage{pgfplots}
 \usepgfplotslibrary{fillbetween}
 \usepackage[multiple]{footmisc}
 
 \tikzset{ every node/.style={inner sep=0pt,minimum size=1mm},
  nsnode/.style={draw,circle,black},
    nsnode2/.style={draw,circle,blue},
  nnnode/.style={draw,circle,black,fill=black},
  asnode/.style={draw,circle,myblue,fill=myblue},
  bsnode/.style={draw,circle,mygreen,fill=mygreen},
  csnode/.style={draw,circle,red,fill=red, minimum size=2mm},
  every fit/.style={inner sep=-1.5pt,text width=1cm}  }
  \tikzset{nero/.style={decorate,draw=black}}
\tikzset{bianco/.style={decorate,draw=bg}}

 \tikzset{ every node/.style={inner sep=0pt,minimum size=1mm},
  nsnode/.style={draw,circle,black},
    wsnode/.style={draw,circle,white},
  nnnode/.style={draw,circle,black,fill=black},
  asnode/.style={draw,circle,myblue,fill=myblue},
  bsnode/.style={draw,circle,mygreen,fill=mygreen},
  csnode/.style={draw,circle,red,fill=red, minimum size=2mm},
  every fit/.style={inner sep=-1.5pt,text width=1cm}  }
\tikzset{
  treenode/.style = {align=center, inner sep=0pt, text centered,
    font=\sffamily},
  arn_n/.style = {treenode, circle, black, font=\sffamily\bfseries, draw=black,
    fill=white, text width=1.5em},
  arn_r/.style = {treenode, circle, newblue, draw=newblue, font=\sffamily\bfseries,
    text width=1.5em},
  arn_x/.style = {treenode, circle, orange,draw=orange,font=\sffamily\bfseries,
   text width=2em}
}

\newcommand{\type}{\ensuremath{\theta}}
\newcommand{\Types}{\ensuremath{\Theta}}
\newcommand{\priorf}{\ensuremath{F_1}}
\newcommand{\priorpdf}{\ensuremath{f_1}}
\newcommand{\Posteriors}{\ensuremath{\Delta(\Types)}}
\newcommand{\typeb}{\ensuremath{\type^\prime}}
\newcommand{\typec}{\ensuremath{\tilde{\type}}}
\newcommand{\transfer}{\ensuremath{x}}
\newcommand{\virtual}{\ensuremath{\hat{\type}}}
\newcommand{\vh}{\ensuremath{\val_\high}}
\newcommand{\vl}{\ensuremath{\val_\low}}
\newcommand{\dv}{\ensuremath{\Delta\val}}

\newcommand{\delay}{\ensuremath{\overline{\mu}}}
\newcommand{\posteriorf}{\ensuremath{F_2}}

\newcommand{\maxq}{\ensuremath{Q}}
\newcommand{\price}{\ensuremath{P}}
\newcommand{\mean}{\ensuremath{m}}
\newcommand{\posteriormean}{\ensuremath{\mu_{\posteriorf}}}
\newcommand{\posteriorpmean}{\ensuremath{p_{\posteriorf}}}
\newcommand{\measurablet}{\ensuremath{\widetilde{\Types}}}
\newcommand{\measurablem}{\ensuremath{\widetilde{U}}}

\newcommand{\high}{\ensuremath{H}}
\newcommand{\low}{\ensuremath{L}}

\newcommand{\bsplit}{\ensuremath{\tau}}
\newcommand{\Revenue}{\ensuremath{\Pi}}
\newcommand{\mpc}{\ensuremath{G}}
\newcommand{\gap}{\ensuremath{G}}

\newcommand{\val}{\ensuremath{v}}
\newcommand{\valb}{\ensuremath{\val^\prime}}

\newcommand{\commission}{\ensuremath{\gamma}}
\newcommand{\cost}{\ensuremath{c}}
\newcommand{\costb}{\ensuremath{\tilde{c}}}

\newcommand{\qual}{\ensuremath{q}}
\newcommand{\qualb}{\ensuremath{\qual^\prime}}
\newcommand{\qualone}{\ensuremath{\qual_1}}
\newcommand{\qualoneb}{\ensuremath{\qualb_1}}
\newcommand{\pricetwo}{\ensuremath{\qual_2^*}}

\newcommand{\consumer}{\text{consumer}}

\newcommand{\profits}{\ensuremath{\Pi}}
\newcommand{\welfare}{\ensuremath{W}}

\newcommand{\upstream}{\text{upstream}}
\newcommand{\downstream}{\text{downstream}}
\newcommand{\Upstream}{\text{Upstream}}
\newcommand{\Downstream}{\text{Downstream}}

\newcommand{\ufirm}{\text{U-firm}}
\newcommand{\dfirm}{\text{D-firm}}
\newcommand{\menu}{\ensuremath{\mathcal{M}}}
\newcommand{\upscale}{premium}
\newcommand{\Upscale}{Premium}
\newcommand{\mass}{mass}
\newcommand{\Mass}{Mass}
\newcommand{\premium}{premium}

\newcommand{\rec}{\ensuremath{\transfer_2}}

\newcommand{\transferone}{\ensuremath{\transfer_1}}
\newcommand{\allocations}{\ensuremath{A}}
\newcommand{\mechanism}{\ensuremath{\mathbb{M}}}

\newcommand{\mint}{\ensuremath{\underline{\type}}}
\newcommand{\maxt}{\ensuremath{\overline{\type}}}

\title{Purchase history and product personalization\thanks{We thank the Editor, David Myatt, and three anonymous referees for feedback that has greatly improved this paper. We also thank Nageeb Ali, Anthony Dukes, Ran Eilat, Shota Ichihashi, Nikhil Vellodi, and seminar participants in the Symposium on Communication and Persuasion, EARIE (Virtual, 2020), AMETS, Columbia University, Texas A\&M, University of Essex, University College London, University of Montreal, Durham University, IO Day 2021 (New York, 2020), CEPR 2022 Workshop on Contracts, Incentives, and Information (Turin, 2022), TOI Workshop 2022 (Santiago, 2022),  Workshop in Antitrust Economics (Rochester, 2023), and the 24th Economics and Computation Conference (London, 2023) for helpful comments. We owe special thanks to  Maher Said for his insightful discussion at IO Day 2021. Nathan Hancart provided excellent research assistance.  This research is supported by grants from the National Science Foundation (Doval: SES-2131706; Skreta: SES-1851729). Vasiliki Skreta is grateful for generous financial support through the ERC consolidator grant 682417 ``Frontiers in design.''}}
\author{Laura Doval\thanks{Columbia University and CEPR. E-mail: \href{mailto:laura.doval@columbia.edu}{\texttt{laura.doval@columbia.edu}}}\and Vasiliki Skreta\thanks{University of Texas at Austin, University College London, and CEPR. E-mail: \href{mailto:vskreta@gmail.com}{\texttt{vskreta@gmail.com}}}}

\begin{document}
\pagenumbering{gobble}

\clearpage
\maketitle
\begin{abstract}
Product personalization opens the door to price discrimination.  A rich product line allows firms to better tailor products to consumers' tastes, but the mere choice of a product carries valuable information about consumers that can be leveraged for price discrimination. We study this trade-off in an upstream - downstream model, where a consumer buys a good of variable quality upstream, followed by an indivisible good downstream. The downstream firm’s use of the consumer’s purchase history for price discrimination introduces a novel distortion: The upstream firm offers a subset of the products that it would offer if, instead, it could jointly design its product line and downstream pricing. By controlling the degree of product personalization the upstream firm curbs ratcheting forces that result from the consumer facing downstream price discrimination.
\end{abstract}
\normalsize
\newpage
\pagenumbering{arabic}
\section{Introduction}\label{sec:intro}
\small
\begin{quote}
\textit{``[\dots] versioning has the benefit of reducing concerns about inequity that arise with personalized pricing, and big data may facilitate versioning strategies based on `mass customization' ''} 
\hfill  White House report on ``Big Data and Differential Pricing''
\end{quote}

\begin{quote}
\textit{``[\dots] there's a positive side to all that tracking companies do, too: it allows them to customize offers that customers do want.''}

\hfill Randall Rothenberg (Interactive Advertising Bureau)
\end{quote}
\normalsize

The trade-off between product personalization and price discrimination is at the center of the debate about the use of consumer data. Consumer data in the form of purchase histories is becoming increasingly available to firms\footnote{For instance, Google's Gmail keeps detailed data on consumers' purchases \href{https://www.cnbc.com/2019/05/17/google-gmail-tracks-purchase-history-how-to-delete-it.html}{(CNBC, 2019)}; brick and mortar stores track buying histories \href{https://abcnews.go.com/Business/supermarkets-introduce-personalized-pricing/story?id=21010246}{(ABCNews, 2013)}.} and used for personalized pricing.\footnote{Priceline acknowledges it ``personalizes search results based on a user's history of clicks and purchases'' \href{https://www.forbes.com/sites/kateashford/2014/10/28/lowest-price-study/?sh=5113ff0a309d}{(Forbes, 2014)}; Orbitz steers Mac users to pricier hotels \href{https://www.wsj.com/articles/SB10001424052702304458604577488822667325882}{(Wall Street Journal, 2012)}; auto dealerships tailor prices to  buyers' willingness to pay, using the way they dress and the car they currently drive \href{https://hbr.org/2017/10/how-retailers-use-personalized-prices-to-test-what-youre-willing-to-pay}{(Harvard Business Review, 2017)}; supermarkets peg prices to purchase histories \href{https://abcnews.go.com/Business/supermarkets-introduce-personalized-pricing/story?id=21010246}{(ABCNews, 2013)}.} Firms, however, also use consumer data for product personalization, allowing the firm to better meet consumers’ needs, which may result in Pareto improvements \citep{anderson2009price}.  As the opening quotes suggest, policy makers and industry practitioners note that rich product lines may compensate for the  the costs of the availability of consumer data in the form of price discrimination. 

Absent from this debate is that consumer data in the form of purchase histories is \emph{endogenously} determined. On the one hand, if the consumer is aware of price discriminating practices, the data is \emph{selected} as it reflects the consumer’s trade-off between a better product match and the costs of price discrimination.\footnote{A report on the impact of big data on differential pricing, \citet{obama2015bigdata}, states \emph{``[\dots] three broad trends suggest that concerns about big data and personalized pricing are not stifling consumer activity on the Internet [\dots]: (1) the rapid growth of electronic commerce, (2) the proliferation of consumer- empowering technologies, and (3) the slow uptake of privacy tools.''}} On the other hand, a firm designs the set the consumer chooses from and hence, how informative purchase histories may be about the consumer's preferences, taking into account the environment the firm and the consumer interact in. By selecting which products the consumer can choose from, the firm can control how much  information can be gleaned about the consumer from their interaction. 

In this article, we shed light on this debate by showing that the endogenous nature of consumer data in the form of purchase histories coupled with ratcheting forces can lead to narrow product lines. We study a dynamic mechanism design problem in the presence of limited commitment. We consider a canonical yet stylized model of upstream-downstream interaction, in which the upstream firm  faces the aforementioned trade-off. On the one hand, a rich product line allows the firm to better tailor the product to the consumer's tastes. On the other hand, a rich product line creates a richer purchase history, which can be exploited by a downstream firm (either the upstream firm itself or a third party) for price discrimination. In the spirit of the ratchet effect, the consumer demands upstream rents to be compensated for downstream rent extraction. Anticipating this, the product line and hence, the consumer's choice may be distorted.

In the model, a consumer interacts with an upstream and a downstream firm over two periods. In the first period, the upstream firm chooses its product line as in \cite{mussa1978monopoly}: It produces a good of variable quality \qualone\ at quadratic cost, whereas the buyer has private information about her willingness to pay for quality, indexed by $\type\sim U[0,1]$. (At the end of the introduction, we discuss the robustness of our results to removing the assumption that \type\ is uniformly distributed.) In the second period, a downstream firm sells an indivisible good at zero cost for which the buyer has a binary private value $\val\in\{\vl,\vh\}$, $\vl<\vh$. Valuations are correlated over time: the consumer's type \type\ also parametrizes the probability that the buyer's value for the second good is \vh. We assume that the downstream firm maximizes downstream profits, whereas the upstream firm's payoff is the sum of the upstream profits and a percentage $\commission\in[0,1]$ of downstream profits. When $\commission\in\{0,1\}$, we span the cases in which the firms are separate or the same entities. As we explain in \autoref{sec:model}, we can interpret the case $\commission\in(0,1)$ as the upstream firm obtaining a payment from the downstream firm for its use of upstream consumer data.

\paragraph{Preview of results:} As a benchmark, we derive the solution for the case in which the upstream firm can design both the upstream product line and the downstream allocation under commitment as would be the case if the firms were (vertically) integrated. We show that the product line is determined \emph{independently} of the \downstream\ allocation. This is intuitive: There is no payoff-relevant link between the \upstream\ and \downstream\ allocations and under commitment, the upstream firm internalizes the information externalities across periods. Thus, the product line coincides with that in the static analysis of \cite{mussa1978monopoly}: Because of decreasing returns to quality, the firm offers a \emph{complete} product line.  Importantly, the product line coincides with that in the first best \citep{anderson2015product}. Furthermore, the commitment solution features either no price discrimination or \emph{reverse} price discrimination, that is, the upstream firm offers the consumer a \emph{discount} for quality in the downstream good. 

We then characterize the upstream firm's optimal mechanism when the downstream firm observes the consumer's choice out of the product line \emph{before} choosing the period-2 mechanism. We refer to this as  \emph{limited commitment} because the downstream mechanism must be optimal given the downstream firm's information. In other words, the downstream firm does not internalize the upstream costs of using the consumer's purchase history for pricing. Anticipating the possibility of downstream price discrimination, the upstream firm offers a narrower product line than that in the commitment solution. By curtailing the range of products it offers to the consumer, the upstream firm obfuscates how much information can be gleaned about the consumer, thereby softening price discrimination downstream. 

The distortions to the product line depend on (i) the price the downstream firm would set absent consumer data and (ii) how much the upstream firm  cares about the downstream profits. The first determines the \emph{shape} of the product line by determining which products are used to convey the consumer's willingness to pay for the downstream good. Below we say that the downstream market is \emph{\upscale} or \emph{\mass} depending on whether \vh\ or \vl\ is the optimal price absent consumer information. The second determines the willingness of the upstream firm to share information with the downstream firm: After all, downstream profits are maximized by tailoring prices to the consumer's information. \autoref{fig:intro} depicts the consumer's choice out of the product line as a function of her type in the first best (black), second best (blue), and limited commitment (red).

\begin{figure}[t!]
\centering
\subfloat[Product line in \upscale\ market]{\scalebox{0.7}{%
\begin{tikzpicture}[thick,scale=3.5]
\draw[->](-0.25,0)--(2,0);
\draw[->](0,-0.25)--(0,2.1);
\node[label=below right:{type}] at (2,0){};
\node[label=left:{quality}] at (0,2.1){};
\node[label=below:{$\frac{1}{2}$}] at (1,0){};
\draw[thick,domain=0:2]plot (\x,\x){};
\draw[thick,blue,domain=0:1]plot (\x,0){};
\draw[thick,blue,domain=1:2]plot (\x,2*\x-2){};
\draw[thick,blue](0,0.01)--(1,0.01);
\node[label=below:{}] at (1.4,0){};
\draw[thick,red,domain=0:1.4]plot(\x,0){};
\draw[thick,red,domain=1.4:2]plot (\x,2*\x-2){};
\draw[-,dashed](1.4,0)--(1.4,0.8);
\pattern[pattern=north east lines,pattern color=red](-0.1,0.8)--(0.1,0.8)--(0.1,2)--(-0.1,2)--cycle;
\filldraw[red] (0,0) circle (0.75pt);
\end{tikzpicture}}\label{fig:premium}}
\subfloat[Product line in \mass\ market (low \commission)]{\scalebox{0.7}{%
\begin{tikzpicture}[thick,scale=3.5]
\draw[->](-0.25,0)--(2,0);
\draw[->](0,-0.25)--(0,2.1);
\node[label=below right:{type}] at (2,0){};
\node[label=left:{quality}] at (0,2.1){};
\node[label=below:{$\frac{1}{2}$}] at (1,0){};
\draw[thick,domain=0:2]plot (\x,\x){};
\draw[thick,blue,domain=0:1]plot (\x,0){};
\draw[thick,blue,domain=1:2]plot (\x,2*\x-2){};
\draw[-](-1.35,1.75)--(-1.25,1.75);
\node[label=right:{First best}] at (-1.3,1.75){};
\draw[-,blue](-1.35,1.65)--(-1.25,1.65);
\node[label=right:{Second best}] at (-1.3,1.65){};
\draw[-,red](-1.35,1.55)--(-1.25,1.55);
\node[label=right:{Limited commitment}] at (-1.3,1.55){};
\draw[thick,red,domain=0:1] plot (\x,0){};
\draw[thick,red,domain=1:1.190984] plot (\x,2*\x-2){};
\draw[thick,red,domain=1.190984:2] plot (\x,1){};
\pattern[pattern=north east lines,pattern color=red](-0.1,0)--(0.1,0)--(0.1,0.4)--(-0.1,0.4)--cycle;
\filldraw[red] (0,1) circle (0.75pt);
\end{tikzpicture}}\label{fig:mass-rents}}

\subfloat[Product line in \mass\ market (high \commission)]{\scalebox{0.7}{%
\begin{tikzpicture}[thick,scale=3.5]
\draw[->](-0.25,0)--(2,0);
\draw[->](0,-0.25)--(0,2.1);
\node[label=below right:{type}] at (2,0){};
\node[label=left:{quality}] at (0,2.1){};
\node[label=below:{$\frac{1}{2}$}] at (1,0){};
\draw[thick,domain=0:2]plot (\x,\x){};
\draw[thick,blue,domain=0:1]plot (\x,0){};
\draw[thick,blue,domain=1:2]plot (\x,2*\x-2){};
\draw[thick,red,domain=0:1] plot (\x,0){};
\draw[thick,red,domain=1:1.190984] plot (\x,2*\x-2){};
\draw[thick,red,domain=1.809016:2]plot (\x,2*\x-2){};
\draw[thick,red,domain=1.190984:1.809016] plot (\x,1){};
\pattern[pattern=north east lines,pattern color=red](-0.1,0)--(0.1,0)--(0.1,0.4)--(-0.1,0.4)--cycle;
\pattern[pattern=north east lines,pattern color=red](-0.1,1.618032)--(0.1,1.618032)--(0.1,2)--(-0.1,2)--cycle;
\filldraw[red] (0,1) circle (0.75pt);
\end{tikzpicture}\label{fig:mass-profits}}}
\caption{Product line and product choice. The product line under limited commitment is depicted in red on the $y$-axis.}\label{fig:intro}
\end{figure}

\autoref{proposition:delay-below-half} shows that in a \upscale\ market, the upstream firm offers a high-end product line to convey that the consumer has a high willingness to pay for quality and hence a high willingness to pay for the downstream good. In contrast to the commitment solution, low quality products are no longer offered, as \autoref{fig:premium} illustrates. Moreover, because only the most exclusive products remain, the decision to purchase the good of the lowest quality no longer implies that the consumer's utility for quality is low. This guarantees that the consumer faces no price discrimination downstream, having to pay a price of \vh\ to obtain the downstream good. Relative to the commitment solution, the consumer faces (weakly) higher downstream prices and higher upstream prices because of the narrow product line. As a result, each consumer type is worse off under limited commitment.

\autoref{proposition:nc-upper} shows that in a \mass\ market the optimal product line depends on how the upstream firm trades off upstream rents and downstream rent extraction. When \commission\ is low, if there is price discrimination, the upstream firm would get a small share of \downstream\ rent extraction, but would pay the costs of the rents the consumer demands. Thus, it is optimal to not allow for downstream price discrimination, which in this case corresponds to inducing a price of \vl\ with probability $1$. To do so, the upstream firm offers either one product or a two-tier product line: A \mass\ downstream market necessitates a product line with \mass-market appeal (see \autoref{fig:mass-rents}). As \commission\ increases, the upstream firm cares more about downstream profits, so that it internalizes the value of sharing consumer information for downstream pricing. The cost of offering a price of \vl\ is higher the higher is \type, as these consumer types are more likely to have high values in period 2. Thus, the upstream firm is willing to share this information. Because a high type consumer faces a high downstream price, there is no point in distorting their quality purchase and they obtain the same product as under commitment. The upstream firm then offers a three-tier product line with a distinct high-end range intended to compensate high type consumers for the downstream rent extraction, as \autoref{fig:mass-profits} illustrates. Because in a \mass\ market there are both downward and upward distortions in quality, not all consumer types are worse off under limited commitment.

\paragraph{Privacy as a remedy} One interpretation of our results is that absent a privacy policy that prevents the consumer's purchase history from being accessed by the downstream firm, the upstream firm then offers the consumer privacy through a coarser product line. We materialize this intuition in \autoref{sec:remedies}, where we consider the case in which the upstream firm designs both the product line \emph{and} the information available to the downstream firm. The comparison between this \emph{data intermediation} benchmark and the limited commitment solution separates the frictions introduced by optimal downstream pricing from those introduced by the observability of the purchase history. We show that the upstream firm offers the second best product line and offers the consumer \emph{full} privacy. In other words, a carefully designed privacy policy alleviates product line distortions, increases upstream profits, but its welfare effects depend on whether the downstream market is \premium\ or \mass.\footnote{This is consistent with Amazon's recent push to reduce the data available to third-party sellers about consumer's transactions with Amazon (see \href{https://www.modernretail.co/retailers/amazon-briefing-the-growing-customer-data-war/}{The growing customer data war}).}  We consider other remedies to limited commitment in \autoref{sec:remedies}.

\paragraph{Modeling choices} \label{page-model-choices}We conclude the introduction by discussing the role of our modeling choices; namely, the linear quadratic framework of \cite{mussa1978monopoly}, uniformly distributed types, and  the linearity in \type\ of the probability that the consumer's value is \vh.  Letting \priorf\ denote the consumer's type distribution and $p(\type)$ denote the probability that a consumer of type \type\ has value \vh, we show that the solutions to the commitment and data intermediation benchmarks are qualitatively the same if we instead considered $p(\type)=\priorf(\type)$.\footnote{As we discuss in \autoref{sec:commitment}, this is the assumption behind the optimality of no/reverse price discrimination in the commitment solution.} Under this assumption, the result that both benchmarks lead to complete product lines does not rely on the linear-quadratic framework of \cite{mussa1978monopoly}, but rather on the log-supermodularity of the upstream social surplus \citep{anderson2009price}.

The assumption that $p=\priorf$ and \type\ is uniformly distributed guarantees that the posterior mean of \type\ is a sufficient statistic for the upstream mechanism design problem under limited commitment.\footnote{In fact, the main qualitative features of our results extend to more general specifications of the period-1 interaction, as long as the consumer's utility is linear in \type.} Indeed, by leveraging the revelation principle under limited commitment and Markov environments in \cite{doval2022markov}, we construct the firm's optimal product line by marrying elements of mechanism design and information design. On the information design side, we rely on the techniques developed for continuum type spaces to transform the design of the product line, and hence how much information the firm learns about the consumer, into an information design problem  (e.g., \citealp{gentzkow2016rothschild,kolotilin2018optimal,dworczak2019simple,arieli2019optimal}). On the mechanism design side, we rely on the first order approach in dynamic mechanism design and dynamic public finance to characterize the solution to a relaxed problem and then provide conditions under which the firm can implement the solution to the relaxed problem  \citep{pavan2014dynamic,stantcheva2020dynamic}.

Whereas the characterization of the optimal product line under limited commitment under more general parametric specifications remains an open question, \autoref{proposition:ic-induces-gap} characterizes the product line distortions introduced by ratcheting forces when the upstream firm offers a menu of quality-transfer pairs under the assumption that $p$ is convex. Intuitively, a menu that induces no price discrimination must feature pooling at the bottom (if the downstream price is \vh) or at the top (if the downstream price is \vl). Importantly, we show that any incentive compatible menu that induces price discrimination downstream \emph{must} have a product line gap -- an interval of qualities that is not offered -- to compensate the consumer for the forgone \downstream\ rents.  These results rely on showing that incentive compatibility together with the optimality of downstream prices imply sorting of higher types into higher qualities.

\paragraph{Related Literature:} Our work contributes to the literatures on product line design (e.g., \citealp{mussa1978monopoly,itoh1983monopoly}) and  (intertemporal) price discrimination (e.g., \citealp{armstrong_2006,stokey1979intertemporal,hart1988contract,acquisti2005conditioning,fudenberg2006behavior}), which for the most part have proceeded on separate tracks. An exception is \cite{sun2014dynamic}, who studies a repeated version of the model in \cite{mussa1978monopoly}. \cite{sun2014dynamic} shows that offering a single variety may be optimal with binary values. Furthermore, when the firm chooses from a restricted class of mechanisms, he provides conditions under which a single variety is offered in the first period when types are drawn from a continuum: either it is optimal to offer a single variety under commitment, or the firm is patient so that it sacrifices product personalization today, in lieu of product personalization tomorrow. Whereas our article shares with \cite{sun2014dynamic} the observation that limited commitment limits varieties in the market, the results are not related otherwise: we do not restrict the set of mechanisms the firm offers to the consumer, but there is no product personalization in period $2$. For this reason, it can still be optimal to offer fully personalized products in period 1 to consumers on the high-end of the type distribution in our model.

We also contribute to the literatures on multiproduct competition and behavioral product line design (e.g., \citealp{johnson2003multiproduct,villas2004communication,ellison2005model,kamenica2008contextual,johnson2015properties,johnson2018determinants,xu2019product}).\footnote{Whereas in \cite{johnson2003multiproduct} product line pruning softens competition by increasing differentiation, \cite{zhang2011perils} shows that competition can also soften differentiation when price discrimination is possible. In \cite{zhang2011perils}, two firms  choose a location in a Hotelling line (a product) in the first period, anticipating that in the second period they can make price offers conditional on whether the consumer purchased from the firm or the rival. \cite{zhang2011perils} shows that both firms choose the same location in period one, making the decision to purchase uninformative.} \cite{johnson2018determinants} and \cite{ellison2005model} are the most relevant references. Starting from a benchmark of complete product lines absent firm competition, \cite{johnson2018determinants} characterize how cost and technological asymmetries drive product line pruning.  Our model starts from the same benchmark and shows  that intertemporal competition and the incentive to soften price discrimination lead to product line pruning.  \cite{ellison2005model} \label{page-ellison} shows that hidden upgrade prices (add-on pricing) may lead to product proliferation when competing against a horizontally differentiated rival, when absent competition the firm would only offer one product. Whereas in \cite{ellison2005model} there is strategic complementarity in the choice to engage in second-degree price discrimination, in our model, \downstream\ price discrimination makes it costly to engage in second-degree price discrimination \upstream.

By interpreting the upstream and downstream firm as different parties, our analysis contributes to the literature on downstream markets (e.g., \citealp{calzolari2006monopoly,calzolari2006optimality,argenziano2020information}). Unlike in \cite{calzolari2006monopoly,calzolari2006optimality} and consistent with the increasing availability of consumer data, we consider the case in which the upstream firm cannot prevent the downstream firm from observing the consumer's purchase. However, our data intermediation benchmark echoes the main result in \cite{calzolari2006optimality} that the upstream firm would prefer not to share information with the downstream firm.  In \cite{argenziano2020information}, a downstream firm observes the quantity purchased by a consumer upstream before setting prices. The authors study the effects of this \emph{data link} on upstream pricing decisions and evaluate the welfare implications of various privacy policies. Because there is no upstream product line design in their model, the ratchet effect only exacerbates downward distortions. Because these articles inherently embed a complex information feedback problem between consumer and firm's choices, like us, these articles rely on assumptions that reduce the dimension of the sufficient statistics needed to solve the problem (binary types and allocations in \citealp{calzolari2006optimality}; linear equilibrium and pricing in \citealp{argenziano2020information}).\footnote{\cite{calzolari2006optimality} rely on binary types and allocations to characterize the optimal mechanism when the conditions for the optimality of full privacy do not hold and hence, the optimal mechanism must deal with a complex information feedback problem.}

By showing the welfare implications of a carefully designed privacy policy, we relate to the works that study consumer privacy starting from the classic contributions of  \cite{taylor2004consumer} and \cite{calzolari2006optimality}.\footnote{\cite{cummings2015strange} study the impact of ad targeting in monopoly pricing in a two-period model with a continuum of types in the first period and binary types in the second period.} A series of recent articles study the (incentive compatible) use of consumer information in static models of second-degree price discrimination \citep{hidir2020privacy,ichihashi2020online,eilat2021bayesian}.

Finally, we contribute to the literature on dynamic mechanism design. From a methodological perspective, we contribute to the literature on limited commitment with continuum type spaces, which models mechanisms as menus \citep{skreta2006sequentially,deb2015dynamic,skreta2015optimal}. Whereas we rely on a more general class of mechanisms thanks to the revelation principle in \cite{doval2022markov} for Markovian environments such as those analyzed in this article, the optimal upstream mechanism can nevertheless be implemented by a menu. Conceptually, we contribute to the literature that studies conditions under which a designer with commitment power would release exogenously available information to an agent about her type, such as \cite{eso2007optimal} and \cite{li2017discriminatory}. Instead, we study a designer's incentives under limited commitment to disclose \emph{endogenous} information about a privately informed agent to a third party (or the designer's future self) that may use it for price discrimination.\label{page-dmd}

\paragraph{Organization:} The rest of the article proceeds as follows. \autoref{sec:model} describes the model and notation. \autoref{sec:benchmark} solves two benchmarks: \autoref{sec:pricing} characterizes the optimal \downstream\ mechanism as a function of the information gleaned from the \upstream\ interaction, whereas \autoref{sec:commitment} characterizes the optimal mechanism under commitment. \autoref{sec:lim com} derives the optimal mechanism under limited commitment and its welfare implications. \autoref{sec:remedies} discusses different remedies to the distortions introduced by limited commitment. \autoref{sec:conmr} concludes. All proofs are in  \autoref{appendix:mr}.

\section{Model}\label{sec:model}

A consumer interacts with an upstream and a downstream firm over two periods $t\in\{1,2\}$. The consumer is fully patient and thus, does not discount payoffs across periods.

In period $1$, the upstream firm (henceforth, \ufirm) produces a good of variable quality \qualone\ at quadratic costs, $\cost(\qualone)=\cost\qualone^2/2$. Upstream allocations are described by $(\qualone,\transferone)\in[0,\maxq]\times\mathbb{R}_+\equiv\allocations_1$, where \transferone\ denotes the payment from the consumer to the \ufirm.

In period $2$, the downstream firm (henceforth, \dfirm) produces an indivisible good at $0$ marginal cost. Period-$2$ allocations are described by $(\qual_2,\transfer_2)\in\{0,1\}\times\mathbb{R}_+\equiv\allocations_2$, where $\qual_2$ denotes whether the period-$2$ good is assigned to the consumer and $\transfer_2$ denotes the payment from the consumer to the \dfirm. 

\paragraph{Profits:} \Downstream\ profits are given by the period-2 payments, $\transfer_2$. Instead, upstream profits depend on the profits from the sale of the good of variable quality and \downstream\ profits. We assume that the \ufirm\ earns a portion $\commission\in[0,1]$ of the \dfirm's profits. Letting $(\qualone,\transferone)$ and $(\qual_2,\transfer_2)$ denote the \upstream\ and \downstream\ allocations, respectively, the \ufirm's profits are given by $\transferone-\cost(\qualone)+\commission\transfer_2.$

\paragraph{Consumer information and payoffs:} The consumer's valuation for each of the goods is her private information. In period $1$, if the consumer purchases a good of quality $\qualone$ and pays \transferone, her flow payoff is $u_1(\qualone,\transferone,\type)=\type \qualone-\transferone$, where $\type\in[\mint,\maxt]\equiv\Types$ denotes the consumer's type.  In period $2$, if she purchases the downstream good and pays $\transfer_2$, her flow payoff is $u_2(\qual_2,\transfer_2,\val)=\val\qual_2-\transfer_2$, where $\val\in\{\vl,\vh\}$, $0<\vl<\vh$. In what follows, we let \dv\ denote the difference $\vh-\vl$. \label{page-dv-model}

The consumer's period-1 type is distributed according to distribution \priorf\ on $[\mint,\maxt]$. We assume that \priorf\ has a density $\priorpdf>0$ and is such that the \emph{virtual values}, $\virtual(\priorf)=\type-(1-\priorf(\type))/\priorpdf(\type)$ are increasing in \type. In period $1$, the consumer does not know her valuation for the good in period $2$. Conditional on the consumer's type in period $1$ being \type, her valuation in period $2$ is \vh\ with probability $p(\type)= \priorf(\type)$. In particular, a consumer who values quality more is more likely to value the downstream good more, that is, $(\type,\val)$ are \emph{positively} correlated. The parametrization $p=\priorf$ ensures that $p$ is well-defined as a probability.\footnote{As we explain in \autoref{sec:commitment}, the parametrization also ensures that when $\commission=1$ the optimal dynamic mechanism under commitment features no price discrimination.} We assume that $p$, and hence \priorf, is Lipschitz continuous.

Finally, to characterize the optimal \upstream\ mechanism under limited commitment, we assume in \autoref{sec:lim com} that \priorf\ is the uniform distribution. As the analysis that follows makes clear, the main role of this assumption is to enable the application of the existing tools of information design with continuum type spaces, which have been developed exclusively for the case in which the sender and the receiver care only about the posterior mean (e.g., \citealp{gentzkow2016rothschild,kolotilin2018optimal,dworczak2019simple,arieli2019optimal}). However, as we discuss in that section, the economic forces underlying the distortions in the product line are more primitive and, as we discuss in \autoref{sec:conmr}, we expect that they will arise under more general assumptions.

\paragraph{Model interpretation} The above is a canonical model of dynamic consumer-firm interactions and thus admits different interpretations:

One interpretation is that the \ufirm\ and the \dfirm\ are the same firm
selling different goods over time to the consumer, as in the case of a car, followed by add-on features; a computer, followed by accessories; a hotel room, followed by amenities.\footnote{Because there are no consumption externalities across periods, our period-2 good is closer to an accessory than a prototypical add-on.} Under this interpretation, the parameter \commission\ can be interpreted as weights on the profits of the different sales $1/(1+\commission)$ on the upstream sale and $\commission/(1+\commission)$ on the downstream sale), or a discount factor. The most natural parametrization would be $\commission=1$.

Another interpretation is that the \ufirm\ and the \dfirm\ are different firms, in which case $\commission=0$ would be a natural parametrization. In that case, the \downstream\ interaction is merely an externality to the \ufirm\ through the consumer's ratchet effect. There are two ways to interpret the case $\commission>0$. First, the \dfirm\ may pay the \ufirm\ a commission for directing the consumer to the \dfirm. Second, the \dfirm\ may pay the \ufirm\ a fee for its use of consumer data in the form of the purchase history.\footnote{This is similar to the interpretation of the model in \cite{calzolari2006optimality}.} The largest price that the \dfirm\ would be willing to pay for the consumer's data is the difference in profits between accessing and not accessing this data.\footnote{In \autoref{sec:remedies}, we let  the \ufirm\ jointly design the product line and the data -- the \emph{cookies} -- available to the \dfirm. In that setting, the \ufirm\  releases the purchase history only if it is in the \ufirm's interest to do so.} 
 Letting \commission\ denote the \ufirm's bargaining power vis-\`a-vis the \dfirm, the \ufirm's total profits would then consist of the profits from the sales of the product line plus a percentage \commission\ of the difference in the \dfirm's profits. Because the \dfirm's profits without access to the purchase history do not depend on the upstream product line, it is without loss of generality to assume that the \ufirm's payoffs are as we specified.

\subsection{Mechanisms, timing, and solution concept}

We follow a mechanism design approach and thus, place no constraints on the mechanisms available to the firms. We consider constraints on the firms' ability to commit. Under \emph{(full) commitment}, we assume that the \ufirm\ can write long-term contracts with the consumer that, among other things, specify the terms of the \downstream\ interaction. In particular, the \downstream\ mechanism \emph{need not} be optimal given the information revealed by the consumer's choice out of the product line. Instead, under \emph{limited commitment}, we assume that the \ufirm\ can only commit to a short-term mechanism with the consumer and conditional on the outcome of that mechanism, the \dfirm\ offers the consumer a 
\downstream\ mechanism that is optimal given the information revealed by the consumer's interaction with the \upstream\ mechanism. The constraints on the \ufirm's ability to commit can be seen as a constraint on the completeness of contracts in this setting: Under full commitment, it is as if the firms are (vertically) integrated. Instead, under limited commitment, the \ufirm\ cannot contract on the \downstream\ allocation and hence designs its mechanism anticipating the effect that it may have on the \downstream\ allocation.

Given a sequence of mechanisms $\mechanism_1,\mechanism_2$ offered by the firms, \autoref{fig:timeline} summarizes the sequence of events that unfold on the consumer's side:\footnote{\autoref{appendix:mechanism} formally defines the mechanisms and the solution concept.}
\begin{figure}[ht!]
\centering\scalebox{0.9}{%
\begin{tikzpicture}[thick,scale=3.5]
\node(origin) at (0,0){};
\node[align=center](ctype) at (0,0.25){consumer\\ observes \type};
\draw[-](0,-0.1)--(0,0.1);
\node[align=center](ctype) at (0.75,-0.25){consumer faces\\ $\mechanism_1$};
\draw[-](0.75,-0.1)--(0.75,0.1);
\node[align=center](ctype) at (1.75,0.25){$(\qualone,\transferone)$ \\ is determined};
\draw[-](1.75,-0.1)--(1.75,0.1);
\node[align=center](ctype) at (2.45,-0.25){consumer \\ observes \val};
\draw[-](2.45,-0.1)--(2.45,0.1);
\node[align=center](ctype) at (3.2,0.25){consumer \\ faces $\mechanism_2(\cdot)$};
\draw[-](3.2,-0.1)--(3.2,0.1);
\node(end) at (4,0){};
\draw[->](origin)--(end);
\node(half) at (2,0){};
\end{tikzpicture}}
\caption{Timeline given a sequence of mechanisms $(\mechanism_1,\mechanism_2(\cdot))$}\label{fig:timeline}
\end{figure}

After observing her type \type, the consumer decides whether to participate in the upstream mechanism $\mechanism_1$. If she does not participate, she makes no payment to the upstream firm $(\transferone=0)$ and receives $\qualone=0$. Instead, if she participates in the mechanism, she submits a report, which determines, among other things, the consumer's allocation at the end of period $1$. The consumer then learns her value \val\ and faces the \downstream\ mechanism $\mechanism_2$, which can depend, among other things, on the consumer's participation decision and her period-1 allocation. Given the \downstream\ mechanism, the consumer decides whether to participate. If she does not participate, the no trade allocation, $(\qual_2,\transfer_2)=(0,0)$, obtains. Instead, if she participates, she submits a report and the \downstream\ mechanism determines the allocation.

Given a sequence of mechanisms faced by the consumer, we focus on consumer participation and reporting strategies that are sequentially rational. That is, the consumer's strategy is optimal in each period given her information and the sequence of mechanisms.

\section{Two benchmarks}\label{sec:benchmark}
\autoref{sec:benchmark} characterizes the solutions to two scenarios that help build intuition for the optimal mechanism under limited commitment. \autoref{sec:pricing} characterizes the optimal \downstream\ mechanism as a function of the \dfirm's beliefs about the consumer's period-1 type, which is useful to understand  the optimal mechanisms under commitment and limited commitment. \autoref{sec:commitment} then characterizes the \ufirm's optimal mechanism when it can design the \upstream\ and \downstream\ allocations under commitment. 

\subsection{\Downstream\ pricing without product-line design}\label{sec:pricing}
Consider the \dfirm's mechanism design problem. Standard arguments imply the optimal \downstream\ mechanism is a posted price.\footnote{This observation is immediate if the \dfirm\ offers a mechanism that elicits \val\ alone. The standard revelation principle, however, implies that the \downstream\ mechanism can elicit, a priori, both \type\ and \val. However, because \type\ is payoff irrelevant to the consumer, the \dfirm\ cannot elicit it. It follows that the optimal \downstream\ mechanism can only elicit \val. A formal proof of (a more general version of) this observation can be found in \cite{doval2022mechanism,doval2022markov}.\label{ftn:elicit}} Whether this posted price is \vl\ or \vh\ depends on the likelihood the firm assigns to the consumer's value being \vh. This likelihood, in turn, depends on the \dfirm's beliefs in period $2$ about the consumer's type, \type. 

Letting \posteriorf\ denote the \dfirm's belief  about \type\ in period $2$, the \dfirm\ assigns probability $\mathbb{E}_{\posteriorf}[p(\type)]\equiv\posteriorpmean$ to the consumer's value being \vh. Then, the optimal \downstream\ price is given by:
\begin{figure}[h!]
\centering
\begin{tikzpicture}[scale=1.25]
\node[label=below:{$0$}](origin) at (0,0){};
\node[label=below right:{$\posteriorpmean$}](end) at (8,-0.15){};
\node[label=below:{$\overline{\mu}=\frac{\vl}{\vh}$}](mu1) at (4,-0.1){};
\node[label=below:{}](origin2) at (0,-0.5){};
\node[label=below:{}](mu12) at (4,-0.5){};
\node[label=below:{}](end2) at (7.5,0.5){};
\node[label=below:{}](mu122) at (4,0.5){};
\node[label=below:{sell at $v_L$}](vl) at (2,-0.6){};
\node[label=above:{sell at $v_H$}](vh) at (6,0.6){};
\draw[|->,thick](origin)--(8,0);
\draw[|-),thick](origin2)--(mu12);
\draw[(-|,thick](mu122)--(end2);
\end{tikzpicture}
\caption{Optimal \downstream\ price as a function of $\posteriorpmean$}\label{fig:pricing}
\end{figure}

where $\delay=\vl/\vh$ is the belief about \vh\ at which the \dfirm\ is indifferent between selling at a price of \vh\ (obtaining revenue $\posteriorpmean\vh$) and selling at a price of \vl\ (obtaining revenue of \vl). \autoref{fig:pricing} illustrates two important themes for what follows. First, optimal \downstream\ pricing is sensitive to the information about \type, which gives rise to the possibility of price discrimination. Second, optimal \downstream\ pricing only depends on the posterior mean of $p(\type)$, \posteriorpmean. 

\paragraph{Downstream best response} For future reference, we define the \dfirm's best response correspondence via the probability that the \dfirm\ serves the consumer when her value is \vl. Formally, we let $\mathcal{Q}_2^*(\vl)$ denote the set of mappings $\pricetwo(\vl,\cdot):\Posteriors\mapsto[0,1]$ that satisfy the following: 
\begin{align}\label{eq:period-2-lc}\tag{D-BR}
q^{*}_2(\vl,\posteriorf)\left\{\begin{array}{ll}=1&\text{if }\posteriorpmean<\delay
\\\in[0,1]&\text{ if }\posteriorpmean=\delay
\\=0&\text{if }\delay<\posteriorpmean
\end{array}\right..
\end{align}
\paragraph{\Downstream\ pricing without consumer data} In what follows, the optimal downstream mechanism at the \emph{prior} mean of $p(\cdot)$ plays a role. This describes what the \dfirm\ would do in the absence of information and corresponds to $p_{\priorf}=1/2$. We say that the \downstream\ market is a \emph{\upscale} market when $\delay<p_{\priorf}=1/2$ because the optimal \downstream\ price is \vh. Instead, we say that the \downstream\ market is a \emph{\mass} market when $\delay>p_{\priorf}=1/2$ because the optimal \downstream\ price is \vl.

\subsection{Product-line design under commitment}\label{sec:commitment}
As our next benchmark, we consider the case in which the \ufirm\ can design the \upstream\ and \downstream\ allocations under full commitment. This allows us to study the distortions introduced to the product line when the \ufirm\ can internalize the information externalities across periods. Because of this, we draw an analogy with models of vertical integration. The contrast between the results in this section and those in \autoref{sec:lim com} is reminiscent of \emph{double marginalization}: when the \ufirm\ can internalize the information externalities across periods, the only distortions to the product line that remain are those coming from the consumer's adverse selection constraints.

Our model is a special case of the environments studied in \cite{pavan2014dynamic}, so we can rely on the standard revelation principle to characterize the \ufirm's optimal mechanism (see, e.g., \citealp{myerson1986multistage}). Without loss of generality, the \ufirm\ chooses a direct revelation mechanism
 \[\{(\qualone(\type),\transfer_1(\type),\qual_2(\type,\val),\transfer_2(\type,\val)):(\type,\val)\in\Types\times\{\vl,\vh\}\}, 
\] 
which specifies the allocation that the consumer receives in each period, as a function of the reports in each period. Importantly, when the consumer submits a type report, \typeb, in period $1$, she restricts the menu from which she chooses in period $2$, to $(q_2(\typeb,\cdot),\transfer_2(\typeb,\cdot))$.

A direct revelation mechanism determines the consumer's \upstream\ and \downstream\ payoffs as a function of her private information, $(\type,\val)$, and her reports, $(\typeb,\valb)$, as follows. Her \downstream\ payoff is given by
\begin{align}\label{eq:c-consumer-2}
u_2(\typeb,\valb)=\val\qual_2(\typeb,\valb)-\transfer_2(\typeb,\valb),
\end{align}
whereas her \upstream\ payoff is given by
\begin{align}\label{eq:c-consumer-1}
W_1^C(\typeb,\type)=\type\qualone(\typeb)-\transfer_1(\typeb)+p(\type)u_2(\typeb,\vh)+(1-p(\type))u_2(\typeb,\vl).
\end{align}
Let $U_1^C(\type)=W_1^C(\type,\type)$ denote the payoff from reporting \type\ truthfully in period 1.

The \upstream-profit maximizing mechanism then solves
\small
\begin{align}\label{eq:c-profit}\tag{C-OPT}
\max_{\qualone,\transferone,\qual_2,\transfer_2}&\int_\Types\left[\transfer_1(\type)-c(\qualone(\type))+\commission\left(p(\type)\transfer_2(\type,\vh)+(1-p(\type))\transfer_2(\type,\vl)\right)\right]\priorf(d\type)\\
\text{s.t.}&
(\forall\type\in\Types)U_1^C(\type)\geq0\label{eq:c-participation}\tag{C-PC}\\
&(\forall\type\in\Types)(\forall\typeb\in\Types)U_1^C(\type)\geq W_1^C(\typeb,\type)\label{eq:c-truthtelling-1}\tag{C-TT$_1$}\\
&(\forall\type\in\Types)(\forall\val,\valb\in\{\vl,\vh\}) u_2(\type,\val)\geq u_2(\type,\valb)\label{eq:c-truthtelling-2}.\tag{C-TT$_2$}
\end{align}
\normalsize
That is, the \ufirm's optimal mechanism must satisfy the following constraints. First, the consumer must find it optimal to participate \eqref{eq:c-participation}. Second, the consumer must find it optimal to report her type \type\ truthfully \eqref{eq:c-truthtelling-1}. Finally, for each type report, the consumer must find it optimal to truthfully report her value \eqref{eq:c-truthtelling-2}.  \autoref{eq:c-truthtelling-2} implies that the consumer does not benefit from deviating by first misreporting \type\ and then misreporting \val. 

\autoref{proposition:contractible} describes the optimal mechanism under commitment: 
\begin{proposition}[Vertical integration]\label{proposition:contractible} The following is the optimal mechanism when the \ufirm\ can design the product line and \downstream\ allocation under commitment:
\begin{enumerate}
\item \textbf{Product line:} The product line is given by $[0,\nicefrac{\maxt}{c}]$, with a type \type-consumer obtaining quality $\qualone(\type)=\max\{0,\nicefrac{\virtual(\priorf)}{c}\}$,
\item \textbf{Period 2:} The period-2 allocation $\qual_2(\type,\cdot)$ is as follows:
\begin{enumerate}
\item There is no distortion at the top, that is, for all $\type\in[\mint,\maxt]$, $\qual_2(\type,\vh)=1$,
\item Let $\type_*$ be such that $\priorf(\type_*)=\nicefrac{\max\{0,1-2\delay\}}{\left(2(1-\delay)-\commission\right)}$. Then, 
$\qual_2(\type,\vl)=\mathbbm{1}[\type\geq\type_*].$
\end{enumerate}
In particular, when $\delay\geq\nicefrac{1}{2}$, the consumer obtains the \downstream\ good with probability 1.
\end{enumerate}
\end{proposition}
See \autoref{appendix:app_virtual_c} for the proof.  \autoref{fig:commitment} describes the \upstream\ product line and the \downstream\ probability of trade with \vl\ as a function of \type\ for different parameter values under the assumption that \priorf\ is the uniform distribution.

\begin{figure}[ht!]
    \centering
\subfloat[$\commission=1,\delay\leq 1/2$]{\scalebox{0.65}{%
\begin{tikzpicture}[thick]
\node(origin) at (0,0){};
\node[label=below:{$\type$}](end) at (10,0){};
\node[label=below:{$\frac{1}{2}$}] at (5,0){};
\draw[->](origin)--(end);
\draw[<->,blue](5,0.5)--(10,0.5);
\node[label=above:{{\Large{$(\frac{2\type-1}{c},0)$}}}]at (7.5,0.5){};
\draw[<->,red](5,0.5)--(0,0.5);
\node[label=above:{{\Large{$(0,0)$}}}]at (2.5,0.5){};
\draw[-,dashed](5,0)--(5,0.5);
\end{tikzpicture}}\label{fig:commitment-vh}}\hfill
\subfloat[$\commission=1,\delay\geq1/2$]{\scalebox{0.65}{%
\begin{tikzpicture}[thick]
\node(origin) at (0,0){};
\node[label=below:{$\type$}](end) at (10,0){};
\node[label=below:{$\frac{1}{2}$}] at (5,0){};
\draw[->](origin)--(end);
\draw[<->,blue](5,0.5)--(10,0.5);
\node[label=above:{{\Large{$(\frac{2\type-1}{c},1)$}}}]at (7.5,0.5){};
\draw[<->,red](5,0.5)--(0,0.5);
\node[label=above:{{\Large{$(0,1)$}}}]at (2.5,0.5){};
\draw[-,dashed](5,0)--(5,0.5);
\end{tikzpicture}}\label{fig:commitment-vl}}

\subfloat[$\commission=0,\delay\leq 1/2$]{\scalebox{0.65}{%
\begin{tikzpicture}[thick]
\node(origin) at (0,0){};
\node[label=below:{$\type$}](end) at (10,0){};
\node[label=below:{$\frac{1}{2}$}] at (5,0){};
\node[label=below:{$\type_*$}] at (3,0){};
\draw[->](origin)--(end);
\draw[<->,blue](5,0.5)--(10,0.5);
\node[label=above:{{\Large{$(\frac{2\type-1}{c},1)$}}}]at (7.5,0.5){};
\draw[<->,red](3,0.5)--(0,0.5);
\node[label=above:{{\Large{$(0,0)$}}}]at (1.5,0.5){};
\draw[<->,red](3,0.5)--(5,0.5);
\node[label=above right:{{\Large{$(0,1)$}}}]at (3.2,0.5){};
\draw[-,dashed](3,0)--(3,0.5);
\draw[-,dashed](5,0)--(5,0.5);
\end{tikzpicture}}\label{fig:commitment-g-vh}}\hfill
\subfloat[$\commission=0,\delay\geq1/2$]{\scalebox{0.65}{%
\begin{tikzpicture}[thick]
\node(origin) at (0,0){};
\node[label=below:{$\type$}](end) at (10,0){};
\node[label=below:{$\frac{1}{2}$}] at (5,0){};
\draw[->](origin)--(end);
\draw[<->,blue](5,0.5)--(10,0.5);
\node[label=above:{{\Large{$(\frac{2\type-1}{c},1)$}}}]at (7.5,0.5){};
\draw[<->,red](5,0.5)--(0,0.5);
\node[label=above:{{\Large{$(0,1)$}}}]at (2.5,0.5){};
\draw[-,dashed](5,0)--(5,0.5);
\end{tikzpicture}}\label{fig:commitment-g-vl}}
\caption{\Upstream\ quality and \downstream\ probability of serving \vl\ as a function of \type\ under uniform distribution. The top two panels depict the solution when $\commission=1$ and the bottom two panels depict the solution when $\commission=0$.}\label{fig:commitment}
\end{figure}

Two features of the optimal mechanism under commitment are worth highlighting. First, the product line is determined \emph{independently} of the \downstream\ allocation. This is illustrated in \autoref{fig:commitment}, where across all panels the product line is the same. This is intuitive: There is no payoff-relevant link between the \upstream\ and \downstream\ allocations and the commitment solution internalizes the information externalities across periods. Thus, the product line coincides with that in the static analysis of \cite{mussa1978monopoly}: Because of decreasing returns to quality, the firm offers a \emph{complete} product line.\footnote{\cite{johnson2018determinants} refer to this as \emph{cost driven} second-degree price discrimination.}  Second, the commitment \downstream\ allocation features either \emph{no} price discrimination or  \emph{reverse} price discrimination, that is, consumer types that value quality more receive lower \downstream\ prices. Indeed, in a \mass\ \downstream\ market $(\delay>1/2)$, the consumer receives the good with probability $1$ \emph{independently} of her \upstream\ type. Instead, in a \upscale\ \downstream\ market, a type  $\type_*$ exists such that the consumer is excluded \downstream\ when her value is \vl\ as long as her \upstream\ type is below $\type_*$. In other words, in a \upscale\ market, the consumer receives a discount for quality on the \downstream\ good.\footnote{Airlines are a good example: A business ticket is more expensive than an economy one and provides access to free drinks and amenities that are available for a price to economy travelers.} 

To understand the pattern of \downstream\ prices when the \ufirm\ can jointly design the \upstream\ and \downstream\ allocations, note the following. The \downstream\ incentive constraints \eqref{eq:c-truthtelling-2} imply the consumer needs to obtain rents so that she truthfully reveals her value. These \downstream\ rents also determine the rents the \ufirm\ must leave the consumer in period 1. Indeed, when a consumer of type \type\ reports \typeb, her information rents are given by
\begin{align*}(\type-\typeb)\qualone(\typeb)+(p(\type)-p(\typeb))(u_2(\typeb,\vh)-u_2(\typeb,\vl)),\end{align*}
where the dependence of $u_2$ on \typeb\ represents the possibility of \downstream\ price discrimination as a function of the reported type in period $1$. That is, higher consumer types enjoy rents because they enjoy quality more and they are more likely to accrue \downstream\ rents. Thus, the \ufirm\ has two instruments to minimize the consumer's information rents: downward distortions in quality (familiar from \citealp{mussa1978monopoly}) and downward distortions in \downstream\ rents. Thus, if \downstream\ price discrimination exists, it is intuitive that it would be \emph{reversed}: By giving consumers who purchase high quality goods a discount on the \downstream\ good, the \ufirm\ makes purchasing low quality goods less attractive. As illustrated in \autoref{fig:commitment-g-vh}, when $\commission=0$ and \priorf\ is the uniform distribution, the consumer types who are excluded \downstream\ are also those who receive the lowest quality good upstream.

Whereas the above explains why price discrimination, if any, should be reverse, it does not explain (i) why it may be optimal to not price discriminate, (ii) the comparison between \delay\ and $1/2$ in determining the optimal mechanism, and  (iii) the role of \commission. To understand (i), note the following. Because the \ufirm\ contracts with the consumer in period 1, it can actually recoup part of the consumer's \downstream\ rents. Indeed, from the perspective of period 1, what matters for rents is (i) how \type\ determines the willingness to pay for quality and (ii) how informative \type\ is about \val\ (portion of \downstream\ rents that go to the consumer). In particular, the \ufirm\ can extract any portion of the consumer's downstream payoffs that is \emph{independent} of \type. Indeed, the \ufirm\ can always extract $u_2(\type,\vl)$ if there is no price discrimination downstream.

Now, if the downstream allocation is independent of \type, then the best the firm can do is to select the allocation as a function of the prior mean of $p(\type)$, which is $1/2$. Indeed, the \emph{ex ante} \downstream\ profit maximizing mechanism excludes the consumer when her value is \vl\ in a \upscale\ market and serves the consumer with probability $1$ in a \mass\ one. This helps explain why the comparison between \delay\ and the ex-ante mean of $p(\type)$ is one of the determinants of the \downstream\ allocation.

Finally, consider the role of \commission. When \commission\ is smaller than $1$, the \ufirm\ does not enjoy \downstream\ profits in their entirety, but pays the costs of those profits through the consumer's period-1 rents. Thus when $\commission$ is smaller than $1$, the mechanism trades off downstream profit maximization with downstream rent minimization. This effect comes through more forcefully when $\delay<1/2$: The incentive to diminish consumer rents together with not internalizing \downstream\ profit losses leads the \ufirm\ to offer \downstream\ discounts for the purchase of high quality goods in period 1.

\paragraph{No price discrimination when $\commission=1$} When $\commission=1$, we obtain the stark result that there is \emph{no} price discrimination regardless of the value of \delay\ (see Figures \ref{fig:commitment-vh} and \ref{fig:commitment-vl}). In particular, when $\commission=1$ and $\delay<1/2$, we have that $\type_*=1$, so in a \upscale\ market all consumer types face a price of \vh. This is a consequence of the assumption that $p(\type)=\priorf(\type)$. Under this assumption, the \emph{dynamic} virtual value of a consumer of type \type\ and value \vl\ is \emph{independent} of \type\ when $\commission=1$. Thus, the decision of whether to serve \vl\ is independent of \type.

\paragraph{Commitment vs limited commitment} It is immediate to see that the \downstream\ allocation in the commitment solution is not \downstream\ optimal given the information revealed about the consumer's type by her quality purchase \upstream. This is most easily seen when \priorf\ is the uniform distribution, where for consumer types above $1/2$ the quality provided at the end of period $1$ fully reveals the consumer's private information. For instance, consider the case in which $\delay>1/2$ and $\type>\delay$. For such a consumer, the optimal \downstream\ price is \vh, whereas the commitment solution allocates the good to the consumer also when her value is \vl.  In other words, when the \ufirm\ can jointly design the product line and the \downstream\ allocation under commitment, it can ignore the information revealed by the product line when choosing the \downstream\ price.

\section{Product-line design with limited commitment}\label{sec:lim com}
\autoref{sec:lim com} studies the properties of the \ufirm-profit maximizing mechanism when the \downstream\ mechanism is chosen based on the information provided by the consumer's choice out of the \ufirm's product line. \autoref{sec:mechanisms} applies the revelation principle in \cite{doval2022markov} to derive a constrained optimization problem, \ref{eq:nc-opt}, the solution to which characterizes the upstream-profit maximizing mechanism under limited commitment. We use this program to illustrate the forces that lead the \ufirm\ to offer a reduced product line relative to the commitment solution (see \autoref{proposition:ic-induces-gap}). \autoref{sec:lim-com-mech} characterizes the solution to the \emph{relaxed} version of  \ref{eq:nc-opt} under the assumption that \priorf\ is the uniform distribution. When the solution to the relaxed program solves  \ref{eq:nc-opt}, we show that it can be implemented with the consumer choosing from a menu of quality-transfer pairs.

\subsection{The \upstream\ mechanism design problem}\label{sec:mechanisms}

We set up in this section a constrained optimization problem, \ref{eq:nc-opt}, the solution to which characterizes the \ufirm's optimal mechanism, and in particular, product line, under limited commitment. 

\paragraph{Direct Blackwell mechanisms} Our model is a special case of the environment studied in \cite{doval2022markov}, so we can rely on the revelation principle in that article to characterize the \ufirm's maximum payoff under limited commitment. Without loss of generality, the \ufirm\ chooses a \emph{direct Blackwell mechanism}, which consists of two mappings 
\begin{align*}
\beta:\Types\mapsto\Delta\left(\Posteriors\right),\;\alpha:\Posteriors\mapsto\Delta(\allocations_1),
\end{align*}
that specify for each consumer type \type, a Blackwell experiment $\beta(\cdot|\type)\in\Delta\left(\Posteriors\right)$ and for each realized posterior, \posteriorf, a distribution over allocations $\alpha(\cdot|\posteriorf)$. Conditional on participating in a direct Blackwell mechanism, the consumer \emph{privately} submits a type report, \typeb. This determines the distribution $\beta(\cdot|\typeb)$ from which a posterior \posteriorf\ is drawn. In turn, this determines the distribution $\alpha(\cdot|\posteriorf)$ from which an allocation $(\qualone,\transferone)\in\allocations_1$ is drawn. Whereas the consumer's type report is unobserved to the firms, the realized posterior and allocation are publicly observed.

Finally, \autoref{lemma:deterministic-q1} shows that it is without loss of generality to consider direct Blackwell mechanisms in which each posterior \posteriorf\ is mapped to one quality-transfer pair, $(\qualone(\posteriorf),\transferone(\posteriorf))$.

\paragraph{Downstream information and pricing} The Blackwell experiment $\beta$ summarizes the information that the \dfirm\ obtains from observing the consumer participate in the mechanism and her choice out of the upstream product line. Indeed, by the revelation principle in \cite{doval2022markov}, it is without loss of generality to assume that when the mechanism outputs \posteriorf, then \posteriorf\ is the \downstream\ belief about the consumer's type. \autoref{sec:pricing} characterized the optimal \downstream\ mechanism as a function of the \downstream\ information about the consumer's type, which is summarized by the \dfirm's best response correspondence, $\mathcal{Q}_2^*(\vl)$.

Fix a \dfirm's best response, $\pricetwo(\vl,\cdot)\in\mathcal{Q}_2^*(\vl)$, and a posterior \posteriorf. The \dfirm's profits as a function of \type\ and the \downstream\ belief \posteriorf\ are given by
\begin{align}
\profits_D^L(\type,\posteriorf\mid\pricetwo)=\pricetwo(\vl,\posteriorf)\vl+(1-\pricetwo(\vl,\posteriorf))p(\type)\vh.
\end{align}

\paragraph{\Upstream\ mechanism design} A direct Blackwell mechanism together with the downstream best response determine the consumer's \upstream\ payoff as a function of her private information \type\ and her report \typeb\ as follows:
\begin{align}\label{eq:nc-deviant}
W_1^L(\typeb,\type)=\int_{\Posteriors}\left[\type\qualone(\posteriorf)-\transferone(\posteriorf)+p(\type)\pricetwo(\vl,\posteriorf)\dv\right]\beta(d\posteriorf|\typeb),
\end{align}
where $\pricetwo(\vl,\posteriorf)$ is defined in \autoref{eq:period-2-lc}. To see how \autoref{eq:nc-deviant} obtains, note that in period $2$, the consumer makes a positive payoff only when her valuation is \vh\ and the \dfirm\ sells the good at a price of \vl, in which case, she earns $\vh-\vl\equiv\dv$.  Let $U_1^L(\type)=W_1^L(\type,\type)$ denote the payoff from truthfully reporting \type.\label{page-dv}

Theorem 1 in \cite{doval2022markov} implies that the \ufirm's optimal mechanism solves the following constrained optimization problem:
\begin{align}\label{eq:nc-opt}\tag{L-OPT}
&\max_{\beta,\qualone,\transferone,\pricetwo\in\mathcal{Q}_2^*(\vl)}\int_\Types\int_{\Posteriors}\left[\transfer_1(\posteriorf)-c(\qualone(\posteriorf))+\gamma\profits_D^L(\type,\posteriorf\mid\pricetwo)\right]\beta(d\posteriorf|\type)\priorf(d\type)\\
\text{s.t.}\;&
(\forall\type\in\Types)U_1^L(\type)\geq0\label{eq:nc-participation}\tag{L-PC}\\
&(\forall\type\in\Types)(\forall\typeb\in\Types)U_1^L(\type)\geq W_1^L(\typeb,\type)\label{eq:nc-truthtelling-1}\tag{L-TT$_1$}\\
&(\forall\measurablet\subseteq\Types)(\forall\measurablem\subseteq\Posteriors)\int_{\measurablet}\beta(\measurablem|\type)\priorf(d\type)=\int_{\Types}\int_{\measurablem}\posteriorf(\measurablet)\beta(d\posteriorf|\type)\priorf(d\type).\label{eq:nc-bp}\tag{BP}
\end{align}
\normalsize
That is, it is without loss of generality to restrict attention to \upstream\ mechanisms such that the consumer participates with probability $1$ \eqref{eq:nc-participation}, truthfully reports her type \eqref{eq:nc-truthtelling-1}, and  the Blackwell experiment $\beta$ satisfies a Bayes' plausibility constraint \eqref{eq:nc-bp}, which we explain below. To understand the right hand side of \autoref{eq:nc-participation}, note the following. Because without loss of generality the consumer participates in the mechanism, then non-participation is an off-path event. Thus, Bayes' rule does not pin down the \dfirm's beliefs about \type\ conditional on not participating. In particular, the \dfirm\ can assign probability $1$ to the consumer's type being $\type=1$ upon non-participation. Thus, offering a price of \vh\ in period $2$ is optimal. It follows that if the consumer does not participate in the period-$1$ mechanism, the consumer's payoff is $0$. Finally, the Bayes' plausibility constraint \ref{eq:nc-bp} states that whenever the mechanism outputs a belief \posteriorf, this is the belief the \dfirm\ has about the consumer's type. 

\paragraph{Comparison with \ref{eq:c-profit}:} We can always interpret the program \ref{eq:nc-opt} as one in which the \ufirm\ chooses the upstream and downstream allocations as the \ufirm\ does in \ref{eq:c-profit}, but subject to additional constraints. As we explain next, both constraints push in the direction of product line pruning.

\emph{Sequential rationality.} The first constraint is that the \downstream\ allocation must satisfy the \downstream\ \emph{sequential rationality} constraint, summarized by $\pricetwo(\vl,\cdot)$. The sequential rationality constraint captures that in order to maximize profits the \dfirm\ leverages the data revealed from the \upstream\ purchase for price discrimination. In other words, relative to the commitment solution, the \ufirm\ chooses the product line taking into account how the information revealed through the product line affects the \dfirm's optimal price in period $2$.  As the analysis in \autoref{sec:lim-com-mech} illustrates, this constraint alone may induce product line pruning. In other words, product line pruning may obtain precisely to \emph{avoid} price discrimination \downstream\ (as is the case in Propositions \ref{proposition:delay-below-half} and \ref{proposition:nc-upper}, \cref{itm:low-commission}). By offering a coarser product line than would be optimal, the \ufirm\ induces more pooling of consumer types, which in turn obfuscates the information available \downstream\ for price discrimination.

\emph{Truthtelling.} Contrary to \ref{eq:c-profit}, the consumer's truthtelling constraint, \ref{eq:nc-truthtelling-1}, is now intertwined with the \downstream-sequential rationality constraints: Consumer data available to the \dfirm\ (and hence, \downstream\ prices) must be consistent with the consumer's choices by the Bayes' plausibility constraint, \ref{eq:nc-bp}.  At the same time, the consumer's willingness to report truthfully depends on the downstream prices by \ref{eq:nc-truthtelling-1}. Thus, the \upstream\ product line design can no longer be separated from that of the data that is generated about the consumer \downstream. If this data is used for price discrimination, the consumer will demand to be compensated \upstream\ for the lost \downstream\ rents. As the analysis in \autoref{sec:lim-com-mech} illustrates, this compensation takes the form of a product line \emph{gap} (see \autoref{proposition:nc-upper}, \cref{itm:high-commission}). Whereas the analysis in \autoref{sec:lim-com-mech} relies on the assumption that \type\ is uniformly distributed, \autoref{proposition:ic-induces-gap} below shows that the property that price discrimination induces a product line gap is more fundamental and holds whenever the \ufirm's optimal mechanism takes the form of a \emph{menu}.

\paragraph{Menu mechanisms} Looking ahead, the \ufirm's optimal mechanism in \autoref{sec:lim-com-mech} can be implemented as a menu. That is, each consumer type is mapped to one posterior belief, \posteriorf\ and hence, each consumer type is mapped to one pair $(\qualone(\type),\transferone(\type))$. Blackwell mechanisms are richer than menus, as they allow the \ufirm\ to obfuscate \downstream\ information by assigning a consumer type \type\ to different posterior beliefs. Despite this, the optimal mechanism in \autoref{sec:lim-com-mech} can be implemented as a menu, so that the analysis that follows allows us to interpret the results in that section without assuming that \priorf\ is the uniform distribution.

To a menu mechanism $\menu=\{(\qualone(\type),\transferone(\type)):\type\in\Types\}$ we can associate a family of posterior beliefs $\{\posteriorf^\type\in\Posteriors:\type\in\Types\}$ such that $\posteriorf^\type$ is the \dfirm's belief about the consumer's period-1 type conditional on observing the \upstream\ allocation $(\qualone(\type),\transferone(\type))$. This, in turn, determines the \dfirm's best response $\pricetwo(\vl,\posteriorf^\type)$. We say that the menu $\menu$ is \emph{incentive compatible} if given the \dfirm's best response, the \upstream\ truthtelling constraints, \ref{eq:nc-truthtelling-1} hold. We say that an incentive compatible menu \emph{induces price discrimination} if two types, \type\ and \typeb, exist such that $\pricetwo(\vl,\posteriorf^{\type})\neq\pricetwo(\vl,\posteriorf^{\typeb})$. Finally, we say that there is a \emph{product line gap} if $\{\qualone(\type):\type\in\Types\}$ is not an interval.

\autoref{proposition:ic-induces-gap} summarizes the properties of incentive compatible menus:
\begin{proposition}[Price discrimination induces a product line gap]\label{proposition:ic-induces-gap}
Suppose \menu\ is incentive compatible. Furthermore, suppose $p^\prime(\type)$ is (weakly) increasing and bounded below by $\gap>0$. Then, the following hold:
\begin{enumerate}
\item\label{itm:q-beliefs}\Downstream\ beliefs track qualities, that is, \downstream\ prices depend on period-1 qualities but not on period-1 prices,
\item\label{itm:up-prices}\Downstream\  prices are monotone in the quality purchased,
\item\label{itm:gap} If \menu\ induces price discrimination, then there is a product line gap. Formally, if \qualone\ is followed by \vl\ and \qualoneb\ is followed by \vh, then $\qualone+\gap\dv\leq\qualoneb$.
\end{enumerate}
\end{proposition}
The proof is in \autoref{appendix:upgrades}. Whereas \autoref{proposition:ic-induces-gap} does not depend on the definition of $p=\priorf$, note that when $p=\priorf$ and $\priorf$ is the uniform distribution the assumption on $p^\prime$ automatically holds with $\gap=1$. 

\autoref{proposition:ic-induces-gap} illustrates how the consumer's incentives lead to product line pruning: An incentive compatible menu that induces price discrimination \emph{must} have a product line gap to compensate the consumer for the forgone \downstream\ rents. Key to this result is the observation that higher types sort into higher qualities in an incentive compatible menu. Thus, \downstream\ prices are increasing in the quality purchased (\cref{itm:up-prices}) and in fact, independent of the \upstream\ prices (\autoref{itm:q-beliefs}). As a consequence, if \qualone\ is followed by \vl\ and \qualoneb\ is followed by \vh, then $\qualone\leq\qualoneb$. \Cref{itm:gap} qualifies this statement by describing the extent to which \qualoneb\ must differ from \qualone\ if they are followed by different prices: the quality difference must account for the forgone downstream rents after the purchase of \qualoneb. Finally, note that \autoref{proposition:ic-induces-gap} also has implications for the product line absent downstream price discrimination: Indeed, \cref{itm:up-prices} implies that there must be pooling at the bottom (top) whenever the downstream price is \vh\ (\vl).

Having understood the forces that lead to product line pruning, we now turn to the characterization of the \ufirm-profit maximizing mechanism in \autoref{sec:lim-com-mech}.

\subsection{Product line design as data design}\label{sec:lim-com-mech}
\autoref{sec:lim-com-mech} characterizes the \ufirm-profit maximizing mechanism under limited commitment when \priorf\ is the uniform distribution relying on the first order approach in dynamic mechanism design and public finance \citep{pavan2014dynamic,stantcheva2020dynamic}. We do so by a combination of mechanism design and information design tools, which reflects the underlying theme of the article: When designing its product line -- a typical mechanism design problem-- the \ufirm\ is also designing the data on the basis of which \downstream\ prices are determined -- a typical information design problem. 

\paragraph{Relaxed program} As is standard in the literature in dynamic mechanism design, we achieve this characterization by studying the solution to a relaxed version of \ref{eq:nc-opt}, which only involves the consumer's downward looking incentive constraints. In \autoref{appendix:limited-commitment}, we show how to obtain an envelope representation of the consumer's utility, $U_1^L(\type)$, which we use to replace the transfers out of the \ufirm's profits. This representation, in turn, allows us to express the \ufirm's expected profit in terms of virtual values and to reduce the \upstream\ mechanism design problem to the problem of choosing two objects: the data available \downstream, in the form of a Bayes' plausible distribution over posteriors, $\bsplit\in\Delta\left(\Posteriors\right)$, and for each posterior, a quality level, $\qualone(\posteriorf)$. The \ufirm\ chooses these objects to maximize the virtual surplus subject to a \emph{monotonicity} constraint, requiring that the consumer's marginal utility
\begin{align}\label{eq:mon}\tag{MON}
U_1^{L{^\prime}}(\type)=\int_{\Posteriors}\left[\qualone(\posteriorf)+p^\prime(\type)\dv\pricetwo(\vl,\posteriorf)\right]\beta(d\posteriorf|\type), \text{ is increasing in \type.}
\end{align}
The \emph{relaxed} program corresponds to maximizing the virtual surplus with respect to the posterior distribution \bsplit\ and the product line $\qualone(\cdot)$ ignoring the monotonicity constraint.

\label{page-rel-mon}To understand the role of the constraint \ref{eq:mon} and the implications of ignoring it in the analysis that follows, consider again a menu mechanism, where each consumer type \type\ is mapped to one posterior distribution, $\posteriorf^\type$. In that case, \ref{eq:mon} is equivalent to the requirement that $\qualone(\posteriorf^\type)+p^\prime(\type)\dv\pricetwo(\vl,\posteriorf^\type)$ is increasing in \type. As we showed in \autoref{proposition:ic-induces-gap}, the posterior mean of $p$, \posteriorpmean, is increasing in \type, and thus $\pricetwo(\vl,\posteriorf^\type)$ has a \emph{cutoff} structure, so that \ref{eq:mon} can be written as\footnote{If $\hat{\type}=\mint$, then the \downstream\ price is  always \vh. Instead, if $\hat{\type}=\maxt$, then the \downstream\ price is always \vl.}
\begin{align}\label{eq:menu-mon}
U{_1^L}^\prime(\type)=\left\{\begin{array}{ll}\qualone(\posteriorf^\type)+p^\prime(\type)\dv&\text{if }\type\leq\hat{\type}\\
\qualone(\posteriorf^\type)&\text{otherwise}\end{array}\right.,\text{ is increasing in \type}.
\end{align}

In other words, the monotonicity constraint is precisely the product line gap constraint in \autoref{proposition:ic-induces-gap}. Thus, the solution to the relaxed program may lead to a product line that is pruned less than what the monotonicity constraint would dictate (see the discussion before \autoref{proposition:nc-upper}). Despite this, as the results that follow highlight, the \ufirm\ nevertheless offers a coarser product line because the relaxed program still captures the \downstream\ sequential rationality constraints and the consumer's downward-looking truthtelling constraints, all of which push towards product line pruning relative to the commitment solution.

\paragraph{Posterior means} In the relaxed problem, the assumption that \priorf\ is the uniform distribution yields the result that the virtual surplus is a function of a low-dimensional sufficient statistic of the consumer data: the posterior mean of \type. Thus, we can solve this problem with the information design tools for continuum type spaces, which deal exclusively with the case in which the receiver's action and the sender's payoff are a function of the posterior mean (e.g., \citealp{gentzkow2016rothschild,kolotilin2018optimal,dworczak2019simple}). 

In what follows, we describe the solution to the relaxed problem (Propositions \ref{proposition:delay-below-half} and \ref{proposition:nc-upper}) and show that under a wide range of parameter configurations, the solution to the relaxed problem is a solution to \ref{eq:nc-opt} (Corollaries \ref{corollary:delay-below-half} and \ref{corollary:nc-upper}). The optimal mechanism turns out to be a compromise between the \ufirm's desire to offer a rich product line and at the same time discipline \downstream\ price discrimination. Except when $\delay\leq1/4$, the optimal mechanism distorts quality provision to discipline the revelation of information about \type\ across periods. How the product line is distorted and whether price discrimination arises depends on \delay, which determines the \downstream\ price in the absence of \upstream\ information, and on \commission, which determines how much the \ufirm\ cares about downstream profits $(\commission)$ and downstream consumer payoffs $(1-\commission)$. The sections that follow describe the \ufirm's optimal mechanism depending on whether it faces a \upscale\ or \mass\ downstream market.

\subsubsection{Low-end bundling: \Upscale\ downstream market begets a \upscale\ product line}\label{sec:premium}
Absent further consumer data,  the \dfirm\ would serve the consumer only when her value is \vh\ in a \upscale\ \downstream\ market. Instead, except when $\commission=1$,  the \ufirm\ prefers to give the consumer a discount for quality, setting a \downstream\ price equal to \vh\ for $\type\leq\type_*$ and \vl, otherwise. Unfortunately, there is no way to provide consumer data that would make it optimal to offer the discount for quality \downstream.\footnote{Formally, reverse price discrimination in the commitment solution obtains from the constraint that \vl\ does not imitate \vh\ binding in the commitment solution for high consumer types. The optimality of the \dfirm's mechanism implies that only the constraint that \vh\ does not imitate \vl\ binds, except when both \vh\ and \vl\ receive the same allocation.} As a consequence, the \ufirm\ designs the product line so as to sustain high downstream prices whenever possible. Thus, in a \upscale\ downstream market, both the product line and the \downstream\ allocation may be distorted relative to the commitment solution.

\autoref{fig:delay-below-half} describes the product line and downstream pricing distortions in the case of a \upscale\ downstream market. 
\begin{figure}[ht!]
\subfloat[No price discrimination]{\scalebox{0.7}{%
\begin{tikzpicture}[thick,scale=3.5]
\draw[->](-0.25,0)--(2,0);
\draw[->](0,-0.25)--(0,2);
\node[label=below right:{$\theta$ }] at (2,0){};
\node[label=left:{$\qualone$}] at (0,2){};
\node[label=below:{$\frac{1}{2}$}] at (1,0){};
\draw[thick,domain=0:2]plot (\x,\x){};
\draw[thick,blue,domain=0:1]plot (\x,0){};
\draw[thick,blue,domain=1:2]plot (\x,2*\x-2){};
\draw[thick,blue](0,0.01)--(1,0.01);
\node[label=below:{}] at (1.4,0){};
\draw[<->,thick,red](0,-0.2)--(1.4,-0.2){};
\node[label=below:{$\mean^*$}] at (1.4,0){};
\draw[<->,thick,red](0,-0.2)--(1.4,-0.2){};
\node[label=below:{$\mathbb{E}[\theta|\theta<m^*]=\overline{\mu}$}] at (0.7,-0.2){};
\draw[thick,red,domain=0:1.4]plot(\x,0){};
\draw[thick,red,domain=1.4:2]plot (\x,2*\x-2){};
\draw[-,dashed](1.4,0)--(1.4,0.8);
\end{tikzpicture}}\label{fig:delay-above-quarter}}
\hfill\subfloat[Price discrimination in the low-end]{
\scalebox{0.7}{%
\begin{tikzpicture}[thick,scale=3.5]
\draw[->](-0.25,0)--(2,0);
\draw[->](0,-0.25)--(0,2);
\node[label=below right:{$\theta$ }] at (2,0){};
\node[label=left:{$\qualone$}] at (0,2){};
\node[label=below:{$\frac{1}{2}$}] at (1,0){};
\draw[thick,domain=0:2]plot (\x,\x){};
\draw[thick,blue,domain=0:1]plot (\x,0){};
\draw[thick,blue,domain=1:2]plot (\x,2*\x-2){};
\draw[thick,blue](0,0.01)--(1,0.01);
\draw[-](-1.35,1.75)--(-1.25,1.75);
\node[label=right:{First best}] at (-1.25,1.75){};
\draw[-,blue](-1.35,1.65)--(-1.25,1.65);
\node[label=right:{Second best}] at (-1.25,1.65){};
\draw[-,red](-1.35,1.55)--(-1.25,1.55);
\node[label=right:{Limited Commitment}] at (-1.25,1.55){};
\node[label=below:{}] at (1.4,0){};
\draw[<->,thick,red](0,0.1)--(0.6,0.1){};
\node[label=above right:{$p_2=\vl$}] at (0.2,0.1){};
\node[label=below:{$\mean^*$}] at (1.4,-0.01){};
\node[label=below:{$\mean_*$}] at (0.6,0){};
\draw[<->,thick,red](0.6,-0.2)--(1.4,-0.2){};
\node[label=below:{$\mathbb{E}[\theta|\theta<m^*]=\overline{\mu}$}] at (1,-0.2){};
\draw[thick,red,domain=0:1.4]plot(\x,0){};
\draw[thick,red,domain=1.4:2]plot (\x,2*\x-2){};
\draw[-,dashed](1.4,0)--(1.4,0.8);
\end{tikzpicture}}
\label{fig:delay-quarter-c-small}}
\caption{Product choice in \upscale\ downstream market in first best (black), commitment (blue), and limited commitment (red)}\label{fig:delay-below-half}
\end{figure}
In a \premium\ market, the solution to the relaxed problem differs in whether it prevents (\autoref{fig:delay-above-quarter}) or allows for (\autoref{fig:delay-quarter-c-small}) for price discrimination. However, the product line has the same qualitative features in both cases: The \ufirm\ provides a high-end product line to convey to the \dfirm\ that the consumer's valuation for the \downstream\ good is high. That is, the \ufirm\ bundles a series of low quality upgrades into a minimum quality good that is of high enough quality that the consumer's purchase history does not necessarily convey that the consumer's value for quality is low.
%

\autoref{fig:delay-above-quarter} illustrates the case in which the \ufirm\ does not allow for \downstream\ price discrimination. Thus, the minimum quality good (the one that corresponds to type $\mean^*$ in the figure) is sufficiently high that all consumer types face a price of \vh\ downstream, regardless of their purchase history. The solution in \autoref{fig:delay-above-quarter} obtains either when $\delay\leq 1/4$ or  when $\delay\geq 1/4$ and the cost of personalization is high. In the first case, the \ufirm\ offers a complete product line: By observing that the consumer buys the lowest quality good, the \dfirm\ assigns probability $\mathbb{E}[\type|\type\leq1/2]=1/4>\delay$ that the consumer's value is \vh, and hence is willing to offer a high price.\footnote{However, limited commitment shapes the way information is disclosed relative to the commitment solution. Indeed, when $\delay\leq1/4$ there is a sense in which the commitment solution reveals ``too much'' information: consumer types below $1/2$ reveal \type\ to the mechanism which is used neither for \upstream\ product personalization nor for \downstream\ price discrimination.} In the second case, the high costs of personalization imply that the \ufirm\ does not find it worth it to pay the cost of downstream rents as well. As $\delay$ goes above $1/4$, the product line is pruned at the bottom to guarantee no price discrimination.

However, the closer \delay\ is to $1/2$ the more high-end the product line needs to be to prevent price discrimination in period $2$. Thus, when $\delay$ is high or the cost of product personalization is low (see \autoref{proposition:delay-below-half} below), the \ufirm\ would prefer to disclose more than just whether the consumer purchased a good of the lowest quality in period $1$. \autoref{fig:delay-quarter-c-small} illustrates the solution to the relaxed problem in this case: The \ufirm\ separates the lowest consumer types that buy $\qualone=0$ (i.e., those in $[0,\mean_*)$) from the low-to-middle consumer types that buy $\qualone=0$, (i.e., those in $[\mean_*,\mean^*]$). The \dfirm\ then offers the former a price of \vl\ and the latter a price of \vh. In turn, this allows the \ufirm\ not to sacrifice product personalization for the high consumer types in period $1$ (i.e., those above $\mean^*$), because the \ufirm\ no longer needs to pool them with the lowest types to keep a price of \vh\ in period $2$. \autoref{proposition:ic-induces-gap} implies that in this case the solution to the relaxed program is not a solution to \ref{eq:nc-opt}: After all, the solution to the relaxed program is a menu mechanism that induces price discrimination \emph{and} there is no product line gap to compensate consumer types in $[\mean_*,\mean^*]$ for the forgone rents.

\autoref{proposition:delay-below-half} describes the solution to the relaxed program in a \upscale\ market and \autoref{corollary:delay-below-half} describes when the solution to the relaxed program solves \ref{eq:nc-opt}. In what follows, to simplify notation we denote the product \cost\vh\ by \costb.

\begin{proposition}[\Upscale\ \downstream\ market]\label{proposition:delay-below-half}
Suppose $\priorf$ is the uniform distribution. Let 
\[l_0(\delay,\commission,\costb)=(1-\delay)-\commission\delay-\frac{(4\delay-1)^2}{2\costb}.\]
The solution to the relaxed program is as follows:
\begin{enumerate}

\item\label{itm:low-delay-2} If $\delay\leq1/4$ or $\delay\in(\nicefrac{1}{4},\nicefrac{1}{2}]$ and $l_0(\delay,\commission,\costb)\geq0$
\begin{enumerate}
\item \textbf{Product line:} There is product line pruning at the bottom, that is, the product line is given by $[\max\{0,\nicefrac{4\delay-1}{c}\},\nicefrac{1}{c}]$, with a type \type-\consumer\ being assigned quality equal to $\nicefrac{2\type-1}{c}$ whenever $\type\geq\max\{2\delay,1/2\}$ and $0$ otherwise,
\item \textbf{Period-2 pricing:} There is no price discrimination \downstream: the price is \vh\ for all consumer types.
\end{enumerate}
In particular, when $\delay\leq 1/4$, the product line coincides with that in \autoref{proposition:contractible}.
\item\label{itm:low-delay-3} If $\delay\in[\nicefrac{1}{4},\nicefrac{1}{2}]$ and $l_0(\delay,\commission,\costb)\leq0$, let $\mean_*,\mean^*$ be such that $\delay=\mathbb{E}\left[\type|\type\in[\mean_*,\mean^*]\right]$ and  \autoref{eq:goal} in \autoref{appendix:lim-com-mech} holds.\footnote{Among the pairs $\mean_*,\mean^*$ that satisfy $\delay=\mathbb{E}\left[\type|\type\in[\mean_*,\mean^*]\right]$, \autoref{eq:goal-3} identifies the one that is part of the solution to the relaxed program.}
Then, we have the following
\begin{enumerate}
\item \textbf{Product line:} There is product line pruning at the bottom,  that is, the product line is given by $[\nicefrac{2\mean^*-1}{c},\nicefrac{1}{c}]$, with a type \type-\consumer\ being assigned quality equal to $\nicefrac{2\type-1}{c}$ whenever $\type\geq \mean^*$ and $0$ otherwise,
\item \textbf{Period-2 pricing:} There is price discrimination \downstream: Consumer types below $\mean_*$ face a price of \vl\ and consumer types above $\mean_*$ face a price of \vh.
\end{enumerate}
\end{enumerate}
\end{proposition}
\begin{corollary}\label{corollary:delay-below-half} Under the assumptions of \autoref{proposition:delay-below-half}, the solution to the relaxed program solves 
\ref{eq:nc-opt} in case \ref{itm:low-delay-2} of \autoref{proposition:delay-below-half}. 
\end{corollary}

Note that in case \ref{itm:low-delay-2} there is product line pruning and no price discrimination. As anticipated in \autoref{sec:mechanisms}, the \downstream\ sequential rationality constraint is enough to introduce distortions in the product line. In this case, the product line is pruned at the bottom relative to the commitment solution. By doing this, the \ufirm\ forces low \type\ consumers to pool and hence, sustain high \downstream\ prices.

Under the conditions of case \ref{itm:low-delay-2} in \autoref{proposition:delay-below-half}, both the \ufirm\ and the consumer lose from the \ufirm's inability to control \downstream\ actions. Relative to the commitment solution,  the consumer faces either the same \downstream\ price ($\type\leq\type_*$) or a higher price ($\type>\type_*$). Moreover, in period $1$, a consumer with type below $\mean^*$ receives the lowest quality good, whereas a consumer with type above $\mean^*$ faces higher prices (see \autoref{fig:delay-above-quarter}). Indeed, by pruning products from the product line, the \ufirm\ gives the consumer fewer opportunities to self-select in period $1$. Therefore, the \ufirm\ needs to leave less rents to the consumer in period $1$, and hence charges higher prices. In other words, in \premium\ markets, limited commitment exacerbates downward distortions. Despite the possibility of charging higher prices, the \ufirm\ is clearly worse off because it cannot implement the commitment solution. \autoref{corollary:w-delay-below-half} summarizes this discussion:

\begin{corollary}[Consumer welfare and social surplus in \upscale\ market]\label{corollary:w-delay-below-half}
Suppose the conditions in case \ref{itm:low-delay-2} in \autoref{proposition:delay-below-half} hold. Then, all consumer types are worse off under limited commitment, that is, $U_1^C(\type)\geq U_1^L(\type)$ for all $\type\in[0,1]$, and upstream profits are lower. Thus, upstream social surplus is lower under limited commitment.
\end{corollary}

\subsubsection{Bundling and pruning: \Mass\ downstream market begets a product line gap}\label{sec:mass}
Recall that in a \mass\ \downstream\ market, the \dfirm\ would serve the consumer with probability $1$ absent \upstream\ consumer data, which is also  optimal in the commitment solution. One way in which the \ufirm\ can induce low \downstream\ prices is to produce a \emph{\mass} product line, consisting of low to middle-range quality products. The absence of high-end products forces high \type\ consumers to buy mid-range products, thereby convincing the \dfirm\ that it is facing a \mass\ market downstream.  Low \downstream\ prices, in turn, allow the \ufirm\ to recoup part of the downstream rents, while at the same time avoiding ratcheting forces. This, however, comes at the cost of not offering the consumer personalized products when her type is high. As we describe below, the solution in a \mass\ \downstream\ market depends on the weight the \ufirm\ attaches to  the cost of revealing consumer data $(1-\commission)$.%

\paragraph{Mid-range bundling and high-end pruning} \autoref{fig:delay-above-half-low-gamma} illustrates the optimal product line for low values of \commission\ (see \autoref{proposition:nc-upper}).  The possibility of \downstream\ price discrimination implies the consumer demands rents up front from the \ufirm. When \commission\ is low, this is very costly to the \ufirm, so that the optimal product line induces low prices downstream. This is accomplished in two ways. First, the \ufirm\ offers no high-end products (see Figure \ref{fig:delay-above-half-low-gamma}): The perception of a \mass\ downstream market is sustained by a \mass\ product line. Second, when the \ufirm\ provides a range of low-quality products as in \autoref{fig:delay-above-half-low-gamma-2}, it creates a distinct middle-quality product, by bundling a bunch of intermediate-level upgrades to just one upgrade. This allows the \ufirm\ to engage in some second-degree price discrimination, while at the same time preventing \downstream\ price discrimination. Because there is no price discrimination, the solution to the relaxed program is a solution to \ref{eq:nc-opt} when \commission\ is low.

\begin{figure}[h!]
\centering
\subfloat[$\delay\in[1/2,3/4)$]{
\scalebox{0.7}{%
\begin{tikzpicture}[thick,scale=3.5]
\draw[->](-0.25,0)--(2,0);
\draw[->](0,-0.25)--(0,2);
\node[label=below right:{ \type\ }] at (2,0){};
\node[label=left:{$\qualone$}] at (0,2){};
\node[label=below:{$\frac{1}{2}$}] at (1,0){};
\draw[thick,domain=0:2]plot (\x,\x){};
\draw[thick,blue,domain=0:1]plot (\x,0){};
\draw[thick,blue,domain=1:2]plot (\x,2*\x-2){};
\node[label=below:{$\mean_*$}] at (0.9,0){};
\draw[<->,thick,red](0.9,-0.2)--(2,-0.2){};
\node[label=below:{$\mathbb{E}[\type|\mean_*\leq\type]=\delay$}] at (1.5,-0.2){};
\draw[thick,red,domain=0:0.9] plot (\x,0){};
\draw[thick,red,domain=0.9:2] plot (\x,0.2){};
\draw[-,dashed](1.3,0)--(1.3,0.6);
\draw[-,dashed](0.9,0)--(0.9,0.2);
\end{tikzpicture}}
\label{fig:delay-above-half-low-gamma-1}}
\subfloat[$\delay\geq 3/4$]{
\scalebox{0.7}{%
\begin{tikzpicture}[thick,scale=3.5]
\draw[->](-0.25,0)--(2,0);
\draw[->](0,-0.25)--(0,2);
\node[label=below right:{ \type\ }] at (2,0){};
\node[label=left:{$q$}] at (0,2){};
\node[label=below:{$\frac{1}{2}$}] at (1,0){};
\draw[thick,domain=0:2]plot (\x,\x){};
\draw[thick,blue,domain=0:1]plot (\x,0){};
\draw[thick,blue,domain=1:2]plot (\x,2*\x-2){};
\draw[-](-1.35,1.75)--(-1.3,1.75);
\node[label=right:{First best}] at (-1.3,1.75){};
\draw[-,blue](-1.35,1.65)--(-1.3,1.65);
\node[label=right:{Second best}] at (-1.3,1.65){};
\draw[-,red](-1.35,1.55)--(-1.3,1.55);
\node[label=right:{Limited commitment}] at (-1.3,1.55){};
\draw[thick,red,domain=0:1] plot (\x,0){};
\draw[thick,red,domain=1:1.190984] plot (\x,2*\x-2){};
\draw[thick,red,domain=1.190984:2] plot (\x,1){};
\node[label=below:{$\mean_*$}] at (1.190984,0){};
\draw[<->,thick,red](1.190984,-0.2)--(2,-0.2){};
\node[label=below:{$\mathbb{E}[\type|\mean_*\leq\type]=\delay$}] at (1.6,-0.2){};
\end{tikzpicture}}\label{fig:delay-above-half-low-gamma-2}}
\caption{Product choice in a \mass\ market and ``low'' \commission\ in first best (black), commitment (blue), and limited commitment (red)}\label{fig:delay-above-half-low-gamma}
\end{figure}
In this case, the limited commitment solution always features a distortion at the top, whereas sometimes it does not feature distortions at the bottom (\autoref{fig:delay-above-half-low-gamma-2}). 

\paragraph{Two upgrades: mid-range bundling and high-end price discrimination} As \commission\ increases, the incentive cost induced by \downstream\ rent extraction is compensated by \downstream\ profits, which can be more effectively maximized with access to more detailed consumer data.  The \ufirm\ then offers a three-tier product line, with a range of low and high-end quality products, and a mid-range quality product as illustrated in \autoref{fig:delay-above-half-high-gamma}.\footnote{Three-tier product lines are ubiquitous, with the easiest example being economy, business, first in airlines, or silver, gold, platinum in loyalty programs. For instance, Netflix has basic, standard, and premium subscriptions; Mailchimp has new business, growing business, and pro marketer; Slickplan has basic, premium, and unlimited. } The consumer now faces \downstream\ price discrimination:   Whereas consumer types who upgrade from low to mid-quality products face a price of \vl, those that upgrade from the middle to high-quality products face a price of \vh. This has two effects on the product line:  First, because consumer types who buy high-end products face \downstream\ price discrimination, the \ufirm\ then offers them personalized products, which fully reveal their data to the \dfirm. Second, as anticipated in \autoref{proposition:ic-induces-gap}, there is a product line gap between middle and high quality products to compensate for the forgone \downstream\ rents.

\begin{figure}[h!]
\centering
\subfloat[$\delay\in[1/2,3/4)$]{
\scalebox{0.7}{%
\begin{tikzpicture}[thick,scale=3.5]
\draw[->](-0.25,0)--(2,0);
\draw[->](0,-0.25)--(0,2);
\node[label=below right:{ \type\ }] at (2,0){};
\node[label=left:{$\qualone$}] at (0,2){};
\node[label=below:{$\frac{1}{2}$}] at (1,0){};
\draw[thick,domain=0:2]plot (\x,\x){};
\draw[thick,blue,domain=0:1]plot (\x,0){};
\draw[thick,blue,domain=1:2]plot (\x,2*\x-2){};
\draw[<->,thick,red](1.3,0.1)--(2,0.1);
\node[label=above:{$p_2=\vh$}] at (1.65,0.1){};
\node[label=below:{$\mean_*$}] at (0.9,0){};
\node[label=below:{$\mean^*$}] at (1.3,0){};
\draw[<->,thick,red](0.9,-0.2)--(1.3,-0.2){};
\node[label=below:{$\mathbb{E}[\type|\mean_*\leq\type\leq\mean^*]=\delay$}] at (1.1,-0.2){};
\draw[thick,red,domain=0:0.9] plot (\x,0){};
\draw[thick,red,domain=0.9:1.3] plot (\x,0.2){};
\draw[thick,red,domain=1.3:2]plot (\x,2*\x-2){};
\draw[-,dashed](1.3,0)--(1.3,0.6);
\draw[-,dashed](0.9,0)--(0.9,0.2);
\end{tikzpicture}}
\label{fig:delay-intermediate}}
\subfloat[$\delay\geq3/4$]{
\scalebox{0.7}{%
\begin{tikzpicture}[thick,scale=3.5]
\draw[->](-0.25,0)--(2,0);
\draw[->](0,-0.25)--(0,2);
\node[label=below right:{ \type\ }] at (2,0){};
\node[label=left:{$\qualone$}] at (0,2){};
\node[label=below:{$\frac{1}{2}$}] at (1,0){};
\draw[thick,domain=0:2]plot (\x,\x){};
\draw[thick,blue,domain=0:1]plot (\x,0){};
\draw[thick,blue,domain=1:2]plot (\x,2*\x-2){};
\draw[-](-1.35,1.75)--(-1.25,1.75);
\node[label=right:{First best}] at (-1.25,1.75){};
\draw[-,blue](-1.35,1.65)--(-1.25,1.65);
\node[label=right:{Second best}] at (-1.25,1.65){};
\draw[-,red](-1.35,1.55)--(-1.25,1.55);
\node[label=right:{Limited Commitment}] at (-1.25,1.55){};
\draw[thick,red,domain=0:1] plot (\x,0){};
\draw[thick,red,domain=1:1.190984] plot (\x,2*\x-2){};
\draw[thick,red,domain=1.809016:2]plot (\x,2*\x-2){};
\draw[thick,red,domain=1.190984:1.809016] plot (\x,1){};
\draw[<->,thick,red](1.809016,0.1)--(2,0.1);
\node[label=above:{$p_2=\vh$}] at (1.94,0.1){};
\node[label=below:{$\mean_*$}] at (1.190984,0){};
\node[label=below:{$\mean^*$}] at (1.809016,0){};
\draw[<->,thick,red](1.190984,-0.2)--(1.809016,-0.2){};
\node[label=below:{$\mathbb{E}[\type|\mean_*\leq\type\leq\mean^*]=\delay$}] at (1.5,-0.2){};
\end{tikzpicture}}
\label{fig:delay-high}}
\caption{Product choice in a \mass\ market, ``high'' \commission\  in first best (black), commitment (blue), and limited commitment (red)}\label{fig:delay-above-half-high-gamma}
\end{figure}

\autoref{proposition:nc-upper} summarizes the above discussion:
\begin{proposition}[\Mass\ \downstream\ market]\label{proposition:nc-upper} Assume \priorf\ is the uniform distribution and $\delay\geq1/2$. Let 
\begin{align*}
l_1(\delay,\commission,\costb)&=(1-\delay)(1-\commission)+\frac{(2\delay-1)^2}{\costb}-\frac{1}{2\costb},\\
l_2(\delay,\commission,\costb)&= 1-\commission-\frac{4}{\costb}(1-\delay).
\end{align*}
The solution to the relaxed program is as follows:
\begin{enumerate}
\item\label{itm:low-commission} If either (i) $\delay\leq 3/4$ and $l_1\geq0$, or (ii) $\delay\geq3/4$ and $l_2\geq0$, then
\begin{enumerate}
\item \textbf{Product line:} The \upstream\ product line is pruned at the top and is given by
\begin{align*}
\left[0,\frac{2\min\{2\delay-1,\nicefrac{1}{2}\}-1}{2c}\right]\cup\left\{\frac{2\delay-1}{c}\right\},
\end{align*}
so that the \ufirm\ offers only one quality whenever $\delay\leq3/4$,
\item \textbf{Period-2 pricing:} There is no price discrimination \downstream: all consumer types face a price of \vl.
\end{enumerate}
\item\label{itm:high-commission} Instead, if either (i) $\delay\leq3/4$ and $l_1\leq0$, or (ii) $\delay\geq3/4$ and $l_2\leq0$, then  let $\mean_*,\mean^*$ be such that $\delay=\mathbb{E}\left[\type|\type\in[\mean_*,\mean^*]\right]$ and \autoref{eq:goal-3} in \autoref{appendix:lim-com-mech} holds.\footnote{Among the pairs $\mean_*,\mean^*$ that satisfy $\delay=\mathbb{E}\left[\type|\type\in[\mean_*,\mean^*]\right]$, \autoref{eq:goal-3} identifies the one that is part of the solution to the relaxed program.} We have the following:

\begin{enumerate}
\item \textbf{Product line:} There is a gap in the \upstream\ product line, which is given by
\begin{align*}
\left[0,\frac{\max\{2\mean_*-1,0\}}{c}\right]\cup\left\{\frac{2\delay-1}{c}\right\}\cup\left[\frac{2\mean^*-1}{c},\frac{1}{c}\right],
\end{align*}
\item \textbf{Period-2 pricing:} There is price discrimination at the high end of the product line: Consumer types below $\mean^*$ face a price of \vl, and above $\mean^*$ face a price of \vh.
\end{enumerate}
\end{enumerate}
\end{proposition}
The solution to the relaxed program features price discrimination in case \ref{itm:high-commission} in \autoref{proposition:nc-upper}. Because the relaxed program ignores the monotonicity constraint, the product line gap due to \downstream\ price discrimination is not necessarily enough to compensate the consumer for the forgone \downstream\ rents, $\type\dv\propto\type(1-\delay)$. When \delay\ is small, these upfront rents are ``tempting'' for low consumer types in period $1$, who may now wish to report that they value quality in period $1$ more than they actually do.\footnote{The above logic  is reminiscent of the ``take the money and run'' strategy in \cite{laffont1988dynamics}.} In other words, the solution to the relaxed problem may fail to satisfy the monotonicity constraint for low values of \delay. 

\autoref{corollary:nc-upper} provides conditions under which the solution to the relaxed problem satisfies the monotonicity constraints in case \ref{itm:high-commission} in \autoref{proposition:nc-upper}. Whereas we provide the full set of conditions in the appendix, we state them only for the cases of $\commission=0$ and $\commission=1$ for clarity.
\begin{corollary}\label{corollary:nc-upper} Under the assumptions of \autoref{proposition:nc-upper}, the solution to the relaxed problem is a solution to \ref{eq:nc-opt} in case \ref{itm:low-commission} and whenever the following holds in case \ref{itm:high-commission}: Either $\commission=0$, or $\commission=1$ and $\delay\leq1/2+\costb/4$. 
\end{corollary}
\autoref{fig:monotonicity} in the appendix illustrates the tuples $(\delay,\costb)$ for which the monotonicity constraint holds for $\commission=0$ and $\commission=1$ across \premium\ and \mass\ downstream markets.

In contrast to a \upscale\ downstream market, the implications of limited commitment for consumer welfare and social surplus are more nuanced in a \mass\ market. The reason is that in a \mass\ market there are both upward and downward distortions in quality and less exclusion than in the commitment solution. Serving lower consumer types has the benefit of making purchase histories less informative, at the cost of giving more consumer rents upstream in the form of more quality provision than in the commitment solution. That is, some consumer types benefit from receiving higher quality products than in the commitment solution (as is the case for types $[\mean_*,\delay]$ in \autoref{fig:delay-above-half-high-gamma}), whereas those consumer types who receive lower quality products face lower prices than in the commitment solution (as is the case for types $[\delay,\mean^*]$ in \autoref{fig:delay-above-half-high-gamma}). It follows that not all consumer types are worse off under limited commitment: Intuitively, high consumer types are those that may prefer the commitment solution because they face either the largest quality distortions or price discrimination. Under our parametric assumptions, we obtain that average consumer welfare may be higher under limited commitment, but there are also instances in which each consumer type may be (weakly) better off under limited commitment.

\begin{corollary}[Consumer welfare in \mass\ market]\label{corollary:w-delay-above-half}
In a \mass\ market, limited commitment has heterogeneous impact on consumer welfare. Indeed, the following hold:
\begin{enumerate}
\item If  $\delay\geq 3/4$ and $l_2\geq0$, then $U_1^L(\type)\geq U_1^C(\type)$ for all types, with the inequality being strict for $\type>\mean_*$,
\item In all other cases, there is an intermediate range of consumer types who strictly prefer the allocation under limited commitment to that under commitment, whereas high consumer types have the opposite preference.
\end{enumerate}
\end{corollary}
The proof of \autoref{corollary:w-delay-above-half} provides more detail about the welfare comparison across the commitment and limited commitment allocations.

We close \autoref{sec:lim com} by discussing the implementation of the optimal mechanism and the difficulties in incorporating the monotonicity constraint:

\paragraph{Menu implementation:} The results in Propositions \ref{proposition:delay-below-half} and \ref{proposition:nc-upper}  show that the \ufirm\ distorts its product line relative to the commitment solution in an attempt to obfuscate how much the consumer's choice out of the product line reveals information about her preferences for quality.  Indeed, as described above, the optimal product line sorts the different consumer types in (sometimes multiple) separation and pooling intervals. Despite this, whenever the monotonicity constraint \ref{eq:mon} holds, the \ufirm's optimal mechanism has a \emph{simple} implementation. Indeed, the \ufirm\ can offer the consumer a menu of qualities and payments, such that what the \dfirm\ learns from observing the consumer's choice from the menu coincides with the information that is induced by the optimal mechanism. 

\paragraph{Incorporating the monotonicity constraint} As discussed, the monotonicity condition \ref{eq:mon} ensures that the quality upgrade between products that induce low prices and those that induce high prices compensates the consumer for the forgone rents. However, incorporating the monotonicity constraint into the upstream mechanism design problem is not without difficulty. First, we would turn the product line design problem into a constrained information design one, as in \cite{doval2022constrained}. Unfortunately, the tools developed for these kinds of problems do not readily extend to continuum type spaces. Second, we can no longer rely on the existing tools for continuum type spaces as we would lose the property that the virtual surplus depends only on the posterior mean of \type.\label{page-monotonicity}

\section{Policy implications}\label{sec:remedies}
We briefly discuss different remedies to the product line distortions introduced by limited commitment, whose feasibility may depend on the context:

\paragraph{Bundling}  One interpretation of the commitment solution is that the sale of the upstream good is \emph{bundled} with that of the downstream good. From this perspective, bundling the upstream and downstream allocation and pricing decisions restores commitment and eliminates the distortions in the upstream product line and downstream pricing. 

\paragraph{\Downstream\ competition}  As long as there are no exclusivity clauses or compatibility requirements with the upstream good, price competition in the \downstream\ market may also help eliminate product line distortions. To see this, suppose that at least two firms can offer the period-2 good at $0$ marginal cost and they compete in prices. In this case, there is no price discrimination in the \downstream\ market \emph{regardless} of how informative purchase histories are. This implies that the \ufirm\ can implement the commitment product line. This enhances consumer welfare in a \upscale\ downstream market and has ambiguous welfare effects in a \mass\ downstream market. 
 Having said this, exclusivity clauses or compatibility with the upstream good may do away with the benefits of \downstream\ competition.\label{page-bertrand}

\paragraph{Privacy design} Recall that one possible interpretation of the \ufirm\ earning downstream profits is that the downstream firm pays a fee for its use of the consumer data stemming from the purchase history. Given the distortions documented thus far, it is natural to ask whether the upstream firm wants to share information with the downstream firm and if so, whether it would be in the form of the consumer's purchase history. 

Motivated by this we characterize the upstream profit-maximizing mechanism when the \ufirm\ can design the product line together with the data available to the \dfirm.\footnote{This is analogous to the setting in \cite{calzolari2006optimality}, but without perfectly persistent types.} Importantly, the \dfirm's only source of consumer data is that from the \ufirm. That is, the \dfirm\ no longer observes the consumer's purchase out of the product line. Thus, the \ufirm\ acts as a \emph{data intermediary} between the consumer and the \dfirm.

In this data intermediation benchmark, the \ufirm's optimal mechanism features no product line distortions, full consumer privacy, and no price discrimination. \autoref{proposition:data-intermediary} states this formally:

\begin{proposition}[\Upstream\ data design]\label{proposition:data-intermediary}
Under \upstream\ product and data design, the upstream profit maximizing mechanism is as follows:
\begin{enumerate}
\item\textbf{Product line:} The product line is given by $[0,\nicefrac{\maxt}{c}]$, with a type \type-consumer obtaining quality $\qualone(\type)=\max\{0,\nicefrac{\virtual(\priorf)}{c}\}$,
\item\textbf{Period 2:} There is no downstream price discrimination:
\begin{enumerate}
\item In a \upscale\ market, all consumer types face a  price of \vh,
\item In a \mass\ market, all consumer types face a price of \vl.
\end{enumerate}
\end{enumerate}
\end{proposition}
The proof is in \autoref{appendix:data-design}. When the \ufirm\ designs the data the \dfirm\ has access to, it preserves full consumer privacy. Thus, the \dfirm\ cannot engage in behavior-based price discrimination. Because the \ufirm\ can conceal consumer choices, it offers a complete product line. In other words, the \ufirm\ utilizes the consumer data to engage in second-degree price discrimination and does not share this data downstream to be used for price discrimination.

We note the following three features of the optimal mechanism in \autoref{proposition:data-intermediary}. First, relative to the commitment solution, the \ufirm\ loses the ability to engage in reverse price discrimination. This explains why this feature is absent under limited commitment as well. Second, recall the interpretation of the \ufirm's profits as arising from the \dfirm\ paying for access to the consumer's purchase history. Because no value of \commission\ exists such that the \ufirm\ shares consumer data with the \dfirm, it follows that the \ufirm\ would not sell consumer data to the \dfirm, even if the \ufirm\ could extract the entirety of the downstream profits.\footnote{\autoref{proposition:data-intermediary} echoes the main theorem in \cite{calzolari2006optimality}. In a persistent-type setting, they provide sufficient conditions under which  the upstream firm may choose not to share consumer information downstream. Whereas we consider non-perfectly persistent types, the conditions under which privacy is optimal in their setting also hold in our model. Namely, the consumer's and \ufirm's payoffs are additively separable in the period-1 and period-2 allocations,  and there is positive correlation in the valuations. In \cite{calzolari2006optimality}, positive correlation is not in the statistical sense, but rather the mechanism design sense: It means that the set of binding incentive constraints is the same in the upstream and downstream interaction.} Finally, relative to the limited commitment solution, the \ufirm\ benefits from a carefully designed privacy policy, but whether this benefits the consumer depends on whether the downstream market is \upscale\ or \mass. In particular, by implementing the full product line, the \ufirm\ excludes more consumers relative to the limited commitment solution in a \mass\ market. This illustrates the nuanced effects of privacy policies: They may neither harm firms, nor benefit consumers. Moreover, because the welfare effects of limited commitment are not uniform across consumer types, it is not immediate that giving the consumer the choice to reveal her purchase history to the \dfirm\ would be a solution.

\section{Concluding remarks}\label{sec:conmr}

Our results provide an additional perspective on the availability of consumer information and its use for price discrimination (\citealp{bergemann2021information}). Economists usually highlight that the consumer data available to the firms is endogenous because it is the result of consumer choices \emph{given} the choice sets offered by the firm. Our results complement this view by highlighting another way in which this data is endogenous: the choice set the consumer faces is \emph{chosen} by the firm. It is precisely because we study the \emph{joint} determination of the firm's product line together with the consumer's behavior within the set of products offered by the firm (and its informational impact on period-$2$ pricing) that we uncover a distortion in the product line on top of the one driven by rent extraction. This perspective puts at the forefront a concern in empirical work about the endogeneity of the set of options consumers face (e.g., \citealp{ivaldi1994competition,miravete2002estimating,luo2018structural}), making salient the possibility that this endogeneity may be pervasive in settings where firms and consumers interact repeatedly over time. 

Our work opens up several avenues for future research. First, the strength of the ratchet effect, and hence the extent to which the \ufirm\ may prune its product line, depends on how forward looking consumers are. Considering the effects of having consumers of different levels of sophistication in the \ufirm's optimal mechanism would be interesting. Second, one interpretation of our results is that the \ufirm\ optimally responds to ratcheting forces by choosing a mechanism that is less ``history dependent.'' Thus, our framework and results could be used as a first step to understand the costs and benefits of making procurement contracts sensitive to past performance. In a setting with moral hazard, \cite{decarolis2016past} illustrate the benefits of taking into account past performance in awarding procurement contracts. Instead, our results suggest that in settings with adverse selection procurement contracts should be less sensitive to past performance.

Finally, whereas our analysis relies on a number of parametric assumptions whose only role is to allow us to apply the existing methodology of information design for continuum type spaces, the economic force underlying the optimal product line is likely to extend to more general settings. Indeed, an interpretation of our model is that, faced with the dynamic inconsistency of the commitment solution, the \ufirm\ prefers to acquire less information about the consumer as a (self-)disciplining device, as in \cite{carrillo2000strategic}. Even if it is natural to conjecture that this economic force extends to more general settings, showing this formally requires extending the existing toolkit of information design with continuum type spaces. Because it will enable a deeper exploration of the issues raised in this article and open the analysis of new problems, we see this extension as a fruitful avenue for further research.

\appendix
\section{Omitted proofs}\label{appendix:mr}
\begin{remark}\label{remark:technical}
Throughout, we make the following technical assumptions. Unless noted otherwise, all spaces are Polish spaces; we endow them with their Borel $\sigma$-algebra. Second, product spaces are endowed with their product $\sigma$-algebra. Third, for a Polish space $X$, we let $\Delta(X)$ denote the set of Borel probability measures over $X$, endowed with the weak$^*$ topology. Thus, $\Delta(X)$ is also Polish \citep{aliprantis2013infinite}. Finally, for any two measurable spaces $X$ and $Y$ , a mapping $\varphi:X\mapsto\Delta(Y)$ is a \emph{transition probability} from $X$ to $Y$ if, for any measurable $C\subseteq Y$, $\varphi(C|x)\equiv\varphi(x)(C)$ is a measurable real valued function of $x\in X$.
\end{remark}
\subsection{Proofs of \autoref{sec:commitment}}\label{appendix:app_virtual_c}
\begin{proof}[Proof of \autoref{proposition:contractible}]
Let $\Delta_\val u_2(\type)=u_2(\type,\vh)-u_2(\type,\vl)$. \autoref{eq:c-truthtelling-2} implies a monotonicity condition for $\Delta_\val u_2(\type)$, that we use later on:
\begin{align}\label{eq:c-transfer-bounds}
\dv \qual_2(\type,\vl)\leq\Delta_\val u_2(\type)\leq\dv q_2(\type,\vh).
\end{align}
Furthermore, because $p(\cdot)$ is differentiable, we can apply the envelope theorem in \cite{milgrom2002envelope} to obtain the following envelope condition from  Equation \ref{eq:c-truthtelling-1}:
\begin{align}\label{eq:c-envelope}\tag{C-E}
U{_1^C}^{\prime}(\type)=\qualone(\type)+p^\prime(\type)\Delta_\val u_2(\type),
\end{align}
\autoref{eq:c-envelope} delivers the following expression for the period-1 transfers
\begin{align*}
\transfer_1(\type)=\type\qualone(\type)+p(\type)\Delta_\val u_2(\type)+u_2(\type,\vl)-\int_{\mint}^\type\left(\qualone(s)+p^\prime(s)\Delta_\val u(s)\right)ds.
\end{align*}
Replacing this in \autoref{eq:c-profit}, we obtain:
\begin{align}\label{eq:c-virtuals}\tag{C-VS}
\profits_U^C=\int_\Types\left[\qualone(\type)\virtual(\priorf)-c(\qualone(\type))+\commission\profits_D^C(\type)+(1-\commission)\welfare_D^C(\type)\right]\priorf(d\type),
\end{align}
where $\profits_D^C(\type)$ denotes the downstream profits adjusted by the dynamic information rents
\begin{align}\label{eq:c-profits-2}
\profits_D^C(\type)=p(\type)\vh\qual_2(\type,\vh)+(1-p(\type))\qual_2(\type,\vl)-p^\prime(\type)\frac{(1-\priorf(\type))}{\priorpdf(\type)}\Delta_vu_2(\type),
\end{align}
whereas $\welfare_D^C(\type)$ is a measure of the \consumer's downstream payoffs in terms of the dynamic rents:
\begin{align}\label{eq:c-welfare-2}
\welfare_D^C(\type)=u_2(\type,\vl)+\left(p(\type)-\frac{1-\priorf(\type)}{\priorpdf(\type)}p^\prime(\type)\right)\Delta_vu_2(\type).
\end{align}
We now show that the mechanism described in \autoref{proposition:contractible} maximizes the objective in 
\ref{eq:c-virtuals}. Because it satisfies all constraints, it follows that it is the optimal mechanism. 

Pointwise maximization of the objective in \ref{eq:c-virtuals} with respect to \qualone(\type) delivers that the product line described in \autoref{proposition:contractible} is optimal.

We now show the properties of the period-2 allocation. Note that the term multiplying $\Delta_\val u_2(\type)$ in \ref{eq:c-virtuals} is given by
\begin{align}
\left[(1-\commission)p(\type)-p^\prime(\type)\frac{1-\priorf(\type)}{\priorpdf(\type)}\right]=(2-\commission)\priorf(\type)-1,
\end{align}
and note that this is increasing in $\type$, negative for $\type=\mint$ and positive for $\type=\maxt$. Define $\type_\commission$ to be such that $\priorf(\type_\commission)=\nicefrac{1}{(2-\commission)}$. Applying \autoref{eq:c-transfer-bounds}, we obtain the following implication on $\Delta_\val u_2(\type)$:
\begin{align}
\Delta_\val u_2(\type)&=\left\{\begin{array}{cc}\dv\qual_2(\type,\vl)&\text{ if }\type\leq\type_\commission\\
\dv\qual_2(\type,\vh)&\text{ if }\type>\type_\commission\end{array}\right.,
\end{align}
which we can replace in \autoref{eq:c-virtuals}. For $\type\leq\type_\commission$, we obtain that
\begin{align}\label{eq:c-below-type-g}
&\commission\profits_D^C(\type)+(1-\commission)\welfare_D^C(\type)\\
&=\commission p(\type)\vh\qual_2(\type,\vh)+\qual_2(\type,\vl)\vh\left(2\delay-1+\priorf(\type)(2-2\delay-\commission)\right)-(1-\commission)\transfer_2(\type,\vl).\nonumber
\end{align}
Clearly, $\qual_2(\type,\vh)=1$ and $\transfer_2(\type,\vl)=0$. When $\delay<\nicefrac{1}{2}$, a threshold type $\type_*\leq\type_\commission$ exists such that $\priorf(\type_*)=\nicefrac{(1-2\delay)}{(2(1-\delay)-\commission)}$ and it is optimal to set $\qual_2(\type,\vl)=\mathbbm{1}[\type\geq\type_*]$.

Instead, when $\delay\geq\nicefrac{1}{2}$, the term multiplying $\qual_2(\type,\vl)$ is (weakly) positive for all $\type\leq\type_\commission$, in which case, $\qual_2(\type,\vl)=\mathbbm{1}[\type\geq0]$.

Consider now $\type>\type_\commission$. We have that
\begin{align}\label{eq:c-above-type-g}
&\commission\profits_D^C(\type)+(1-\commission)\welfare_D^C(\type)=\qual_2(\type,\vl)\vl\left(\commission(1-\priorf(\type))+1-\commission\right)-(1-\commission)\transfer_2(\type,\vl)\nonumber\\
&=\qual_2(\type,\vh)\vh\left[\priorf(\type)(\commission(2-\delay)+2(1-\commission)(1-\delay))-(1-\delay)\right].
\end{align}
In this case, $\qual_2(\type,\vl)=1$ and $\transfer_2(\type,\vl)=0$ for all $\type>\type_\commission$. Furthermore, the term multiplying $\qual_2(\type,\vh)$ is increasing in \type\ and positive for $\type\geq\type_\commission$ so that $\qual_2(\type,\vh)=1$ for all $\type>\type_\commission$.
\end{proof}
\begin{observation}[Implementability]\label{observation:c-implementable}
The solution to the relaxed program can be implemented. Indeed, it is possible to show that the monotonicity constraint ($U_1^{C^{\prime}}(\type)$ increasing in \type) is satisfied.
\end{observation}
\subsection{Proofs of \autoref{sec:lim com}}\label{appendix:limited-commitment}
\subsubsection{Proofs of \autoref{sec:mechanisms}}
Given a mechanism $\langle\beta,\alpha,\transfer_1\rangle$, denote the average quality conditional on \posteriorf\ by $\qualone(\posteriorf)=\int_0^{Q}\qualone\alpha(d\qualone|\posteriorf)$. \autoref{lemma:deterministic-q1} shows that it is always payoff-improving for the \ufirm\ (and payoff irrelevant for the \consumer) to have the mechanism induce  $\qualone(\posteriorf)$ with probability 1:
\begin{lemma}[Deterministic \qualone\ conditional on \posteriorf]\label{lemma:deterministic-q1}Let $\mechanism=\langle\beta,\alpha,\transfer_1\rangle$ denote a mechanism that satisfies \ref{eq:nc-participation}, \ref{eq:nc-truthtelling-1}, \ref{eq:nc-bp}. Then, $\mechanism^\prime=\langle\beta,\alpha^\prime,\transfer_1\rangle$ such that $\alpha^\prime(\cdot|\posteriorf)$ assigns probability $1$ to $\qualone(\posteriorf)$  satisfies \ref{eq:nc-participation}, \ref{eq:nc-truthtelling-1}, \ref{eq:nc-bp} and is preferred by the \ufirm.
\end{lemma}
\begin{proof}[Proof of \autoref{lemma:deterministic-q1}]
The proof follows from the expressions for upstream profits in \ref{eq:nc-opt} and consumer payoffs in \autoref{eq:nc-deviant}. Because \downstream\ information does not depend on the allocation rule $\alpha$, $\mechanism^\prime$ induces the same \downstream\ response as \mechanism. Note that the \ufirm's profits from \mechanism\ are given by:\small
\begin{align*}
&\int_\Types\int_{\Posteriors}\int_0^Q\left[\transfer_1(\posteriorf)-c(\qualone)+\commission\left(\pricetwo(\vl,\posteriorf)\vl+(1-\pricetwo(\vl,\posteriorf))p(\type)\dv\right)\right]\alpha(d\qualone|\posteriorf)\beta(d\posteriorf|\type)\priorf(d\type)\\
&=\int_\Types\int_{\Posteriors}\left[\transfer_1(\posteriorf)-\int_0^Qc(\qualone)\alpha(d\qualone|\posteriorf)+\commission\left(\pricetwo(\vl,\posteriorf)\vl+(1-\pricetwo(\vl,\posteriorf))p(\type)\dv\right)\right]\beta(d\posteriorf|\type)\priorf(d\type)\\
&\leq\int_\Types\int_{\Posteriors}\left[\transfer_1(\posteriorf)-c(\qualone(\posteriorf))+\commission\left(\pricetwo(\vl,\posteriorf)\vl+(1-\pricetwo(\vl,\posteriorf))p(\type)\dv\right)\right]\beta(d\posteriorf|\type)\priorf(d\type),
\end{align*}\normalsize
where the latter are the profits from $\mechanism^\prime$. The second equality uses that conditional on \posteriorf\ the only element that depends on \qualone\ are the costs and the inequality in the third line follows from convexity of the costs (and it is strict whenever $\mathrm{supp}\;\alpha(\cdot|\posteriorf)$ has at least two elements. The consumer's payoffs from reporting \typeb\ when her type is \type\ under \mechanism\ are given by:
\begin{align*}
&\int_{\Posteriors}\int_0^Q\left[\type\qualone-\transfer(\posteriorf)+p(\type)\dv\pricetwo(\vl,\posteriorf)\right]\alpha(d\qualone|\posteriorf)\beta(d\posteriorf|\typeb)\\
&=\int_{\Posteriors}\left[\type\qualone(\posteriorf)-\transfer(\posteriorf)+p(\type)\dv\pricetwo(\vl,\posteriorf)\right]\beta(d\posteriorf|\typeb),
\end{align*}
where the latter are the payoffs from reporting \typeb\ under $\mechanism^\prime$. The result follows.
\end{proof}
\subsubsection{Proof of \autoref{proposition:ic-induces-gap}}\label{appendix:upgrades}
The proof of \autoref{proposition:ic-induces-gap} follows from three lemmas, which we state and prove below.
Given the menu $\{(\qualone(\type),\transferone(\type)):\type\in\Types\}$, recall that $\posteriorf^\type\equiv\posteriorf^{(\qualone(\type),\transferone(\type))}$ denotes the \downstream\ belief when $(\qualone(\type),\transferone(\type))$ is the choice out of the menu.
\begin{lemma}[Downstream beliefs track qualities]\label{lemma:beliefs-quality}
Let \menu\ be incentive compatible. Let $\type<\typeb$ and suppose that $\qualone(\type)=\qualone(\typeb)$. Then, $\pricetwo(\vl,\posteriorf^{\type})=\pricetwo(\vl,\posteriorf^{\typeb})$. 
\end{lemma}
\begin{proof}[Proof of \autoref{lemma:beliefs-quality}]
Towards a contradiction, suppose that $\pricetwo(\vl,\posteriorf^{(\qualone(\type),\transferone(\type))})\neq\\\pricetwo(\vl,\posteriorf^{(\qualone(\typeb),\transferone(\typeb))})$.
 Then, it must be that case that $\transferone(\type)\neq \transferone(\typeb)$ as \downstream\ prices are different after the allocations of \type\ and \typeb. Because the menu is incentive compatible, the following holds:
\begin{align*}
(p(\typeb)-p(\type))\dv \pricetwo(\vl,\posteriorf^{\typeb})\geq \transferone(\typeb)-\transferone(\type)\geq (p(\typeb)-p(\type))\dv \pricetwo(\vl,\posteriorf^\type),
\end{align*}
which means that $\pricetwo(\vl,\posteriorf^{\typeb})=1>0=\pricetwo(\vl,\posteriorf^\type)$. Therefore, we must have
\begin{align}\label{eq:key}
\mathbb{E}\left[p|\posteriorf^\type\right]\geq\delay\geq\mathbb{E}\left[p|\posteriorf^{\typeb}\right].
\end{align}
There are two possibilities:
\begin{enumerate}
\item\label{itm:down} $(\exists\type_+<\typeb)$ such that $(\qualone(\type_+),\transferone(\type_+))=(\qualone(\typeb),\transferone(\typeb))$, or
\item\label{itm:up} $(\exists\type^+>\type)$ such that $(\qualone(\type^+),\transferone(\type^+))=(\qualone(\type),\transferone(\type))$.
\end{enumerate}
Otherwise, simultaneous violation of both of the above implies that $\mathbb{E}[p|\posteriorf^{\type}]<\mathbb{E}[p|\posteriorf^{\typeb}]$, a contradiction.

Suppose that \autoref{itm:down} holds. Incentive compatibility implies that \emph{any} such $\type_+$ must satisfy that $\type_+>\type$. Hence, $\mathbb{E}[p|\posteriorf^{\typeb}]\geq p(\type)$. It follows that \autoref{itm:up} must also hold. However, incentive compatibility implies that \emph{any} such $\type^+$ must satisfy that $\type^+<\typeb$. In fact, any such $\type^+$ must be less than any $\type_+<\typeb$ that receives \typeb's allocation. It follows that \autoref{eq:key} cannot hold. 
\end{proof}
Because of \autoref{lemma:beliefs-quality}, below we write $\mathbb{E}[p|\qualone(\type)]$ instead of $\mathbb{E}[p|\posteriorf^\type]$.
\begin{lemma}[Prices are monotone in qualities]\label{lemma:monotonicity-1}
An incentive compatible menu cannot have two qualities $\qualone<\qualoneb$ such that $\pricetwo(\vl,\qualone)=0<\pricetwo(\vl,\qualoneb)=1$.
\end{lemma}
\begin{proof}[Proof of \autoref{lemma:monotonicity-1}]
Towards a contradiction, suppose this is the case. Let $\typeb$ denote the \consumer\ type that chooses $\qualoneb$ and let \type\ denote the \consumer\ type that chooses $\qualone$. Note that we must have that $\mathbb{E}[p|\qualoneb]\leq\delay\leq\mathbb{E}[p|\qualone]$. Now, incentive compatibility implies that
\begin{align}\label{eq:prop-2-monotonicity}
(\typeb-\type)(\qualoneb-\qualone)+(p(\typeb)-p(\type))\dv\geq0.
\end{align}
Note that we must have that $\typeb>\type$. Otherwise, because $\qualone^\prime>\qualone$ and $p$ is increasing, \autoref{eq:monotonicity} cannot hold. Then, as in the proof of \autoref{lemma:beliefs-quality}, we must have that either
\begin{enumerate}
\item\label{itm:down-2} $(\exists\type_+<\typeb)$ such that $\qualone(\type_+)=\qualoneb$,
\item\label{itm:up-2} $(\exists\type^+>\type)$ such that $\qualone(\type^+)=\qualone$.
\end{enumerate}
Suppose that \autoref{itm:down-2} holds. Note that for any such $\type_+$, incentive compatibility implies that $\type<\type_+$. Thus, we have that $\mathbb{E}[p|\qualoneb]\geq p(\type)$ and hence $\mathbb{E}[p|\qualone]\geq p(\type)$. 

There are two possibilities. If $\type$ is the only type that purchases $\qualone$, then we have $p(\type)=\mathbb{E}[p|\qualone]\geq\delay\geq\mathbb{E}[p|\qualoneb]\geq p(\type)$, contradicting that $\pricetwo(\vl,\qualone)\neq \pricetwo(\vl,\qualoneb)$. Otherwise, there exists $\type^+>\type$ that satisfies \autoref{itm:up-2}. Incentive compatibility then implies that $\type^+<\typeb$. As before, it must be that $\type^+<\type_+$ for any $\type_+$ that satisfies \autoref{itm:down-2}. It follows that we cannot have that $\mathbb{E}[p|\qualoneb]\leq\mathbb{E}[p|\qualone]$. 
\end{proof}
We have the following corollary:
\begin{corollary}[Cutoff]
Let $(\qualone,\transferone,\pricetwo)$ be an incentive compatible menu. Then, there exists a cutoff quality $\qualone^\star$ such that 
\begin{enumerate}
\item $\pricetwo(\vl,\qualone)=1$ if $\qualone<\qualone^\star$,
\item $\pricetwo(\vl,\qualone)=0$ if $\qualone>\qualone^\star$.
\end{enumerate}
\end{corollary}
\begin{lemma}[Price discrimination induces gap]\label{lemma:gap}
If $\qualone$ and $\qualoneb$ are part of an incentive compatible menu and $\qualone<\qualoneb$ and $\pricetwo(\vl,\qualone)=1>\pricetwo(\vl,\qualoneb)=0$, then $\qualoneb\geq \qualone+\gap\dv$.
\end{lemma}

\begin{proof}[Proof of \autoref{lemma:gap}]
Towards a contradiction, suppose that $\qualoneb>\qualone$, $\pricetwo(\vl,\qualoneb)=0<\pricetwo(\vl,\qualone)=1$, but $\qualoneb<\qualone+\dv$. Let \typeb\ denote the type that receives $\qualoneb$ and \type\ denote the type that receives $\qualone$. Suppose that $\qualoneb<\qualone+\gap\dv$. Incentive compatibility implies that:
\begin{align}\label{eq:ic-gap-2}
(\typeb-\type)(\qualoneb-\qualone)-(p(\typeb)-p(\type))\dv\geq0.
\end{align}
 We want to show that it cannot be that $\typeb>\type$. Indeed, suppose that $\typeb>\type$, then
 \begin{align}\label{eq:sc-key}
&(\typeb-\type)(\qualoneb-\qualone)-(p(\typeb)-p(\type))\dv<(\typeb-\type)\gap\dv-(p(\typeb)-p(\type))\dv\nonumber\\
&=(\typeb-\type)\dv\left(\gap-\frac{p(\typeb)-p(\type)}{\typeb-\type}\right)\leq0,
 \end{align}
where the first inequality follows from $\qualoneb<\qualone+\gap\dv$ and the last one follows from convexity of $p$ and the assumption that $p^\prime\geq\gap$. Thus, if \autoref{eq:ic-gap-2} holds, then it must be that $\typeb<\type$. 

To complete the proof of \autoref{lemma:gap}, we need to show that all types $\typec<\typeb$ prefer $(\qualoneb,\pricetwo(\vl,\qualoneb))$ to $(\qualone,\pricetwo(\vl,\qualone))$ and all types $\typec>\type$ have the opposite preference. 
Let $\typec<\typeb$. \typec\ prefers $(\qualoneb,\pricetwo(\vl,\qualoneb))$ over $(\qualone,\pricetwo(\vl,\qualone))$ if the following holds:
\begin{align*}
\typec (\qualoneb-\qualone)-p(\typec)\dv\geq \transfer_1^\prime-\transfer_1,
\end{align*}
where $\transfer_1$ is the payment associated to \qualone\ and $\transfer_1^\prime$ is the payment associated to \qualoneb. Note that we have that
\begin{align*}
\typeb (\qualoneb-\qualone)-p(\typeb)\dv\geq \transfer_1^\prime-\transfer_1,
\end{align*}
so that it would suffice that we show 
\begin{align*}
\typec (\qualoneb-\qualone)-p(\typec)\dv\geq \typeb (\qualoneb-\qualone)-p(\typeb)\dv\Leftrightarrow (\typeb-\typec)(\qualoneb-\qualone)-(p(\typeb)-p(\typec))\dv\leq 0.
\end{align*}
Note that this follows from $\qualoneb<\qualone+\gap\dv$ as in \autoref{eq:sc-key}. Similar steps show that $\typec>\type$ prefers $(\qualone,\pricetwo(\vl,\qualone))$ to $(\qualoneb,\pricetwo(\vl,\qualoneb))$.

Because all types $\typec<\typeb$ prefer $\qualoneb$ (and the implied period-2 price) to $\qualone$ and all types $\typec>\type$ prefer $\qualone$ (and the implied period-2 price) to $\qualoneb$, it follows that in order for the condition on the period-2 prices to hold, we must have that either there is $\type^+>\typeb$ that chooses $\qualoneb$ or $\type_+<\type$ that chooses $\qualone$. It must be that $\typeb<\type^+<\type_+<\type$ for any such types. Indeed, for $\typec\in[\typeb,\type]$ write
\begin{align*}
u(\typec)=\typec \qualoneb-\transferone^\prime+\max\{0,\typec(\qualone-\qualoneb)+p(\typec)\dv+\transferone-\transferone^\prime\}.
\end{align*}
Note that the second term is continuous and increasing in \typec. By assumption it is positive for $\typec=\type$ and it is zero or negative when $\typec=\typeb$. Thus, there is $\type^\star\in[\typeb,\type]$ such that for \type\ above $\type^\star$, $\qualone$ is the preferred quality and for $\type<\type^\star$, $\qualoneb$ is the preferred quality. We conclude again that it cannot be the case that $\mathbb{E}[p|\qualone^\prime]\geq\delay\geq\mathbb{E}[p|\qualone]$. 
\end{proof}

\subsubsection{Proofs of \autoref{sec:lim-com-mech}}\label{appendix:lim-com-mech}
\paragraph{Envelope representation of payoffs:}
We first argue that the consumer's period-1 payoff defined in \autoref{sec:mechanisms} is Lipschitz continuous, and hence almost everywhere differentiable. To see this, consider the payoff from the following deviation: The consumer with type \type\ reports \typeb\ and then follows the strategy of \typeb\ in period $2$. Her payoff would then be given by $W_1^L(\typeb,\type)$ as defined in \autoref{eq:nc-truthtelling-1}. 

The optimality of truthtelling implies 
\begin{align*}
U_1^L(\type)=\max_{\typeb\in\Types}W_1^L(\typeb,\type).
\end{align*}
We now establish that the family $\{W_1^L(\typeb,\cdot):\typeb\in\Types\}$ is equi-Lipschitz continuous. Let \type\ and \typec\ be such that $\type\neq\typec$, and consider\small
\begin{align*}
&|W_1^L(\typeb,\type)-W_1^L(\typeb,\typec)|=\left|\int_{\Posteriors}((\type-\typec)\qualone(\posteriorf)+(p(\type)-p(\typec))\dv \pricetwo(\vl,\posteriorf))\beta(d\posteriorf|\typeb)\right|\\
&\leq |\type-\typec|\int_{\Posteriors}\left[\qualone(\posteriorf)+\left|\frac{p(\type)-p(\typec)}{\type-\typec}\right|\dv \pricetwo(\vl,\posteriorf)\right]\beta(d\posteriorf|\typeb)\leq |\type-\typec|(\maxq+K\dv),
\end{align*}\normalsize
where $K$ is the Lipschitz constant of $p$ and $\maxq$ is the bound on quality. We conclude that $U_1^L$ is Lipschitz continuous because it is the max over a family of equi-Lipschitz continuous functions. Moreover, at any point of differentiability of $U_1^L(\cdot)$, we have
\begin{align}\label{eq:monotonicity}\tag{L-E}
U{_1^L}^\prime(\type)=\int_{\Posteriors}\left[\qualone(\posteriorf)+p^\prime(\type)\dv \pricetwo(\vl,\posteriorf)\right]\beta(d\posteriorf|\type).
\end{align}
Incentive compatibility implies $U{_1^L}^\prime(\type)$ is nondecreasing. \autoref{eq:monotonicity} implies \small
\begin{align}\label{eq:transfers}
&\int_{\Types}\int_{\Posteriors}\transfer_1(\posteriorf)\beta(d\posteriorf|\type)\priorf(d\type)=\int_{\Types}\int_{\Posteriors}\left[\type \qualone(\posteriorf)+p(\type)\dv \pricetwo(\vl,\posteriorf)\right]\beta(d\posteriorf|\type)\priorf(d\type)\nonumber\\
&-\int_{\Types}\int_{0}^\type\left(\int_{\Posteriors}\left[\qualone(\posteriorf)+p^\prime(\type)\dv \pricetwo(\vl,\posteriorf)\right]\beta(d\posteriorf|u)\right)du\priorf(d\type).
\end{align}\normalsize
\paragraph{Virtual surplus:} Replacing \autoref{eq:transfers} in \ref{eq:nc-opt} and integrating by parts, we obtain the virtual surplus representation of the \ufirm's profits. Denote by $\bsplit$ the distribution on $\Types\times\Posteriors$ defined as $P(\measurablet\times\measurablem)=\int_{\measurablet}\beta(\measurablem|\type)\priorf(d\type)$, for all measurable subsets $\measurablet,\measurablem$ of $\Types$ and $\Posteriors$. Letting $\bsplit$ denote its marginal on \Posteriors, Proposition 3.6 in \cite{crauel2002random} delivers the virtual surplus representation of the \ufirm's problem:\small
\begin{align}\label{eq:nc-virtuals}\tag{NC-VS}
\max_{\qualone(\cdot),\bsplit}&\int_{\Posteriors}\int_\Types\left[\virtual(\priorf)\qualone(\posteriorf)-c(\qualone(\posteriorf))+\commission\profits_D^L(\type,\posteriorf)+(1-\commission)\welfare_D^L(\type,\posteriorf)\right]\posteriorf(d\type)\tau(d\posteriorf)\\
\text{s.t.}\;&\bsplit\text{ is Bayes' plausible given }\priorf\tag{BP}\\
&U_1^{L{^\prime}}(\type)=\int_{\Posteriors}\left[\qualone(\posteriorf)+p^\prime(\type)\dv \pricetwo(\vl,\posteriorf)\right]\beta(d\posteriorf|\type)\text{ is increasing in }\type.\label{eq:monotonicity-app}\tag{MON}
\end{align}\normalsize
In the above expression, $\profits_D^L$ are the \downstream\ profits adjusted by \emph{dynamic} rents:
\begin{align*}
\profits_D^L(\type,\posteriorf)&=p(\type)\vh(1-\pricetwo(\vl,\posteriorf))\\
&+\pricetwo(\vl,\posteriorf)\left[p(\type)\vh+(1-p(\type))\left(\vl-\frac{1-\priorf(\type)}{\priorpdf(\type)}\frac{p^\prime(\type)}{1-p(\type)}\dv\right)\right],
\end{align*}
and $W_D^L$ is a measure of \downstream\ consumer welfare, adjusted by \emph{dynamic} rents:
\begin{align*}
\welfare_D^L(\type,\posteriorf)&=\dv\pricetwo(\vl,\posteriorf)\left(p(\type)-\frac{1-\priorf(\type)}{\priorpdf(\type)}p^\prime(\type)\right).
\end{align*}
\textbf{Relaxed problem:} Ignoring the monotonicity constraint, we can choose $\qualone(\posteriorf)$ to maximize pointwise the integrand in \autoref{eq:nc-virtuals}, in which case, we obtain 
\begin{align}\label{eq:rep}
\qualone(\posteriorf)=\max\left\{0,\frac{2\posteriormean-1}{c}\right\}\equiv \qualone(\posteriormean).
\end{align}
Replacing \autoref{eq:rep} in the objective in \ref{eq:nc-virtuals} leads to the following expression for the objective function in \ref{eq:nc-virtuals}:
\small
\begin{align}\label{eq:firm-mean-app}
&\int_{\Posteriors}\frac{(\max\{2\posteriormean -1,0\})^2}{2c}\bsplit(d\posteriorf)+(1-\commission)\int_{\Posteriors}\pricetwo(\vl,\posteriorf)\dv(2\posteriormean-1)\bsplit(d\posteriorf)\\
&+\commission\int_{\Posteriors}\left[(1-\qual_2^*(\vl,\posteriorf))\vh\posteriormean+\qual_2^*(\vl,\posteriorf)\left(\vh\posteriormean+(\vl-\dv)(1-\posteriormean)\right)\right]\bsplit(d\posteriorf).\nonumber
\end{align}
\normalsize
Thus, the virtual \upstream\ profits are a function of the posterior mean of \posteriorf, and we can pool together all distributions \posteriorf\ which induce the same posterior mean.

\paragraph{Product line design as information design:} Thus, letting \mean\ denote the posterior mean, the firm's payoff can be written as:
\begin{align*}
\int_0^1\Revenue(\mean)\mpc(d\mean),
\end{align*}
where $\mpc$ is a distribution on $\Types$ that is dominated by \priorf\ in the convex order, and in a slight abuse of notation, the function \Revenue\ corresponds to the virtual surplus as a function of the posterior mean \mean\ and is defined as follows. When $\delay<\nicefrac{1}{2}$,  we have
\begin{align}\label{eq:revenue-mean-lower}
\Revenue(\mean)&=\left\{\begin{array}{ll}
\vh\left((2\mean-1)(1-\delay)+\commission\delay\right)&\text{ if }\mean<\delay\\
\vh\commission \mean&\text{ if }\delay\leq \mean<\nicefrac{1}{2}\\
\vh\left(\commission \mean+\frac{(2m-1)^2}{2\cost\vh}\right)&\text{ if }\mean\geq\nicefrac{1}{2}
\end{array}\right..
\end{align}
Instead, when $\delay>\nicefrac{1}{2}$, we have
\begin{align}\label{eq:revenue-mean-upper}
\Revenue(\mean)&=\left\{\begin{array}{ll}
\vh\left((2m-1)(1-\delay)+\commission\delay\right)&\text{ if }m<\nicefrac{1}{2}\\
\vh\left((2m-1)(1-\delay)+\commission\delay+\frac{(2m-1)^2}{2\cost\vh}\right)&\text{ if }\nicefrac{1}{2}\leq m\leq\delay\\
\vh\left(\commission m+\frac{(2m-1)^2}{2\cost\vh}\right)&\text{ if }m>\delay
\end{array}\right..
\end{align}
The above expressions show that the value of \vh\ does not matter, so in what follows we write $\costb\equiv \cost\vh$. The relaxed problem then reduces to 
\begin{align}\label{eq:relaxed}
\max_{G:\priorf\succ_{cx}G}\int_0^1\profits(\mean)\mpc(d\mean).
\end{align}
Because $\profits$ satisfies the conditions of \cite{dworczak2019simple}, their results imply that $\mpc^*$ is a solution to  \autoref{eq:relaxed} if and only if a convex function \price\ exists such that (i) $\mathbbm{E}_{\mpc^*}\price(\mean)=\mathbb{E}_{\priorf}\price(\mean)$, (ii) $\price\geq \profits$, and (iii) $\text{supp }\mpc^*\subseteq\{\mean:\price(\mean)=\profits(\mean)\}$. The proof of Propositions \ref{proposition:delay-below-half} and \ref{proposition:nc-upper} proceeds as follows. In each case, we prove that the induced distribution over posterior means is optimal by constructing a convex function $\price$ and verifying that the above conditions hold. We then use \autoref{eq:transfers} to construct the transfers and provide conditions under which the solution satisfies that $U^\prime(\type)$ is nondecreasing.

\begin{proof}[Proof of \autoref{proposition:delay-below-half}] The solution when $\delay\leq1/4$ follows from the proof of \autoref{proposition:data-intermediary}. Suppose that $\delay>1/4$. Let $\mean^*=2\delay$ and note that $\delay=\mathbb{E}[\type|\type\leq \mean^*]$. When $l_1(\cdot)\geq0$, the distribution over posteriors is as follows:
\begin{align*}
\mpc(\mean)=\left\{\begin{array}{ll}0&\text{ if }m<\delay\\ \priorf(\mean^*) & \text{if }\delay\leq m<\mean^*\\
\priorf(m)&\text{ otherwise }\end{array}\right..
\end{align*}
The price function that supports this as a solution is 
\begin{align*}
\price(m)=\left\{\begin{array}{ll}\Revenue(\delay)+\frac{\Revenue(\mean^*)-\Revenue(\delay)}{\mean^*-\delay}(m-\delay)&\text{ if }m\leq \mean^*
\\
\Revenue(\mean)&\text{ otherwise }\end{array}\right..
\end{align*}
Note that we only need to check that $\price(0)\geq \Revenue(0)$. This is the case if and only if
\begin{align*}
\price(0)&=2 \Revenue(\delay)-\Revenue(\mean^*)=\vh\left(-\frac{(4\delay-1)^2}{2\costb}\right)\geq \Revenue(0)=\vh(\commission\delay-(1-\delay))\\
&\Leftrightarrow\frac{(4\delay-1)^2}{2\costb}\leq(1-\delay)-\commission\delay,
\end{align*}
which corresponds to $l_0\geq0$.

Instead, when $l_0\leq0$, let $\mean_*,\mean^*$ be such that $\mathbb{E}[\type|\type\in[\mean_*,\mean^*]]=\delay$. In this case, the distribution over posteriors is as follows:
\begin{align*}
\mpc(\mean)=\left\{\begin{array}{ll}\priorf(m)&\text{ if }0\leq m<\mean_*\text{ or }\mean^*<m\leq1\\ 
\priorf(\mean_*)&\text{ if }\mean_*\leq m<\delay\\
\priorf(\mean^*) & \text{if }\delay\leq m\leq \mean^*\end{array}\right..
\end{align*}
The price function that supports this as a solution is 
\begin{align*}
\price(m)=\left\{\begin{array}{ll}\Revenue(\mean_*)+\frac{\Revenue(\mean^*)-\Revenue(\mean_*)}{\mean^*-\mean_*}(m-\mean_*)&\text{ if }\mean_*\leq m \leq \mean^*
\\
\Revenue(\mean)&\text{ otherwise }\end{array}\right..
\end{align*}
Assume first that $\mean_*,\mean^*$ exist. We construct a convex function $\price(\mean)$ that satisfies the necessary properties. Define $\price:[0,1]\mapsto\mathbb{R}$ as follows:
\begin{align}\label{eq:price-below-half}
    \price(\mean)=\left\{\begin{array}{cc}\Revenue(\mean)&\text{if }\mean\notin[\mean_*,\mean^*]\\
    \Revenue(\mean_*)+\frac{\Revenue(\mean^*)-\Revenue(\mean_*)}{\mean^*-\mean_*}(\mean-\mean_*)&\text{otherwise }\end{array}\right..
\end{align}
Clearly, $\price$ is convex and $\price(\mean)\geq \Revenue(\mean)$. To see that $\mathbb{E}_\mpc[\price]=\mathbb{E}_{\priorf}[\price]$, note that
\begin{align*}
    \mathbb{E}_{\priorf}[\price(\mean)]-\mathbb{E}_{\mpc}[\price(\mean)]=\int_{\mean_*}^{\mean^*}\price(\mean)d\mean-(\mean^*-\mean_*)\price(\delay).
\end{align*}
Now, because $\price$ is linear on $[\mean_*,\mean^*]$, we have that
\begin{align*}
    \int_{\mean_*}^{\mean^*}\price(\mean)d\mean=(\mean^*-\mean_*)\price\left(\frac{\mean_*+\mean^*}{2}\right)=(\mean^*-\mean_*)\price(\delay).
\end{align*}
To finish the proof, we show $\mean_*,\mean^*$ exist so that $\price(\delay)=\Revenue(\delay)$. The goal is to find $\mean_*,\mean^*$ such that 
\begin{align}\label{eq:goal}
\frac{\mean_*+\mean^*}{2}=\delay&   &\Revenue(\delay)= \Revenue(\mean_*)+\frac{\Revenue(\mean^*)-\Revenue(\mean_*)}{\mean^*-\mean_*}(\delay-\mean_*).
\end{align}
Note that the second condition is equivalent to
\begin{align}\label{eq:cond-x-below-half}
   \Revenue(\mean^*)&= \Revenue(\mean_*)+\frac{\Revenue(\delay)-\Revenue(\mean_*)}{\delay-\mean_*}(\mean^*-\mean_*).
\end{align}
Verifying that $\mean_*,\mean^*$ exist that satisfy the above equations is equivalent to showing that the quadratic equation
\begin{align*}
h_0(x)\equiv\frac{2}{\costb}x^2-x\left(\frac{2}{\costb}+2(1-\delay)-\commission\right)+\frac{1}{2\costb}+4\delay(1-\delay)-(1-\delay)-\commission\delay=0,
\end{align*}
has a solution $\mean^*\in[1/2,1]$. It is easy to verify that $h_0(x)$ achieves its minimum at $\frac{1}{2}+(1-\commission+1-2\delay)(\costb/4)\geq1/2$. So we need to show that $h_0(2\delay)>0$. This ensures that $\mean_*=2\delay-\mean^*>0$. The latter condition is equivalent to $l_0\leq0$, which completes the proof.
\end{proof}
\begin{proof}[Proof of \autoref{corollary:delay-below-half}]The mechanism described in \autoref{proposition:delay-below-half} when $\delay\leq1/4$ or $\delay\geq 1/4$ and $l_0\geq0$ implies that in the solution to the relaxed problem types below $\mean^*$ are excluded in period $1$, whereas types above $\mean^*$ receive $(2\type-1)/c$ and pay $\frac{\type^2-\mean^*(1-\mean^*)}{c}$. It is immediate to show  $U{_1^L}^\prime(\type)$ is monotone. Instead, when $\delay\geq1/4$ and $l_0\leq0$, $U{_1^L}^\prime(\type)$ is not monotone, so the solution to the relaxed program cannot be implemented.
\end{proof}
\begin{proof}[Proof of \autoref{corollary:w-delay-below-half}]
Relying on the envelope representation of the consumer's payoffs for the commitment \eqref{eq:c-envelope} and limited commitment \eqref{eq:monotonicity} solutions, we obtain the consumer's payoffs under the assumptions of \autoref{proposition:delay-below-half}, case \ref{itm:low-delay-2}. Consumer payoffs under limited commitment are
\begin{align*}
U_1^L(\type)&=\left\{\begin{array}{ll}0&\text{ if }\type\leq\min\{2\delay,1/2\}\\
\int_{\max\{1/2,2\delay\}}^\type\frac{2x-1}{c}dx&\text{otherwise}\end{array}\right..
\end{align*}
Instead, consumer payoffs under commitment depend on whether $\type_*\geq 1/2$ (top) or $\type_*\leq 1/2$ (bottom):\small
\begin{align*}
U_1^C(\type)&=\left\{\begin{array}{ll}0&\text{ if }\type\leq1/2\\
\int_{1/2}^\type\frac{2x-1}{c}dx&\text{ if }\type\in[1/2,\type_*]\\
\int_{1/2}^\type\frac{2x-1}{c}dx+(\type-\type_*)\dv&\text{ if }\type\geq\type_*
\end{array}\right., \\
U_1^C(\type)&=\left\{\begin{array}{ll}0&\text{ if }\type\leq\type_*\\
(\type-\type_*)\dv&\text{ if }\type\in[\type_*,1/2]\\
\int_{1/2}^\type\frac{2x-1}{c}dx+(\type-\type_*)\dv&\text{ if }\type\geq1/2
\end{array}\right..
\end{align*}\normalsize
Tedious, but straightforward algebra, verifies that $U_1^C(\type)\geq U_1^L(\type)$, strictly so whenever (i) $\commission<1$ and $\type>0$, or (ii) $\commission=1$, $\delay>1/4$, and $\type>1/2$.
\end{proof}

\begin{proof}[Proof of \autoref{proposition:nc-upper}, \cref{itm:low-commission}] Define $\mean_*$ to be such that $\delay=\mathbb{E}[\type|\type\geq \mean_*]$. That is, $\mean_*=2\delay-1$. Note that whether $\mean_*$ is above or below $\nicefrac{1}{2}$ depends on whether $\delay$ is above or below $\nicefrac{3}{4}$. In both cases, the posterior distribution is the same:
\begin{align*}
\mpc(\mean)&=\left\{\begin{array}{ll}\priorf(m)&\text{ if }0\leq m<\mean_*\\
\priorf(\mean_*)&\text{ if }\mean_*\leq m\leq\delay\\
1&\text{ otherwise}\end{array}\right.,
\end{align*}
and it is supported by the following convex function:
\begin{align*}
\price(\mean)=\left\{\begin{array}{ll}\Revenue(\mean)&\text{ if }0\leq m\leq \mean_*\\
\Revenue(\mean_*)+\frac{\Revenue(\delay)-\Revenue(\mean_*)}{\delay-\mean_*}(m-\mean_*)&\text{ if }\mean_*\leq m \leq 1
\end{array}\right..
\end{align*}
The only remaining step is to check that $\price(\mean)\geq\Revenue(\mean)$ for  all \mean. Indeed, it suffices to check that this happens at $\mean=1$.

Suppose first that $\nicefrac{1}{2}\leq\delay\leq\nicefrac{3}{4}$ so that $\mean_*=2\delay-1\leq\nicefrac{1}{2}$. Then,
\begin{align*}
\price(1)&=2\Revenue(\delay)-\Revenue(\mean_*)=(1-\delay)+\commission\delay+\frac{(2\delay-1)^2}{2\costb}\\
\Revenue(1)&=\commission+\frac{1}{2\costb}.
\end{align*}
Then, $\price(1)\geq\Revenue(1)$ if and only if 
\begin{align*}
(1-\delay)(1-\commission)+\left[\frac{(2\delay-1)^2}{2\costb}-\frac{1}{2\costb}\right]\geq0,
\end{align*}
yielding the condition that $l_2\geq0$.

Consider now $\delay\in[\nicefrac{3}{4},1]$, in which case $\mean_*\geq\nicefrac{1}{2}$. Then, 
\begin{align*}
\price(1)&=2\Revenue(\delay)-\Revenue(\mean_*)=(1-\delay)+\commission\delay+\frac{(2\delay-1)^2}{2\costb}-\frac{(2(2\delay-1)-1)^2}{2\costb}\\
\Revenue(1)&=\commission+\frac{1}{2\costb}.
\end{align*}
Then, $\price(1)\geq\Revenue(1)$ if and only if 
\begin{align*}
(1-\delay)\left[1-\commission-\frac{4}{\costb}(1-\delay)\right]\geq0,
\end{align*}
yielding the condition that $l_3\geq0$.
\end{proof}
\begin{proof}[Proof of \autoref{proposition:nc-upper}, \cref{itm:high-commission}]
Let $\mean_*,\mean^*$ be such that $\mathbb{E}[\type|\type\in[\mean_*,\mean^*]]=\delay$. In this case, the distribution over posteriors is as follows:
\begin{align*}
\mpc(\mean)=\left\{\begin{array}{ll}\priorf(m)&\text{ if }0\leq m<\mean_*\text{ or }\mean^*<m\leq1\\ 
\priorf(\mean_*)&\text{ if }\mean_*\leq m<\delay\\
\priorf(\mean^*) & \text{if }\delay\leq m\leq \mean^*\end{array}\right..
\end{align*}
The price function that supports this as a solution is 
\begin{align*}
\price(m)=\left\{\begin{array}{ll}\Revenue(\mean_*)+\frac{\Revenue(\mean^*)-\Revenue(\mean_*)}{\mean^*-\mean_*}(m-\mean_*)&\text{ if }\mean_*\leq m \leq \mean^*
\\
\Revenue(\mean)&\text{ otherwise }\end{array}\right..
\end{align*}
Similar to the steps in the proof of part \ref{itm:low-delay-3} of \autoref{proposition:delay-below-half}, it suffices to check that $\mean_*,\mean^*$ can be chosen so that $\price(\delay)=\Revenue(\delay)$. The goal is to find $\mean_*,\mean^*$ such that 
\begin{align}\label{eq:goal-3}
\frac{\mean_*+\mean^*}{2}=\delay&   &\Revenue(\delay)= \Revenue(\mean_*)+\frac{\Revenue(\mean^*)-\Revenue(\mean_*)}{\mean^*-\mean_*}(\delay-\mean_*).
\end{align}
As in the proof of \cref{itm:low-commission} in \autoref{proposition:nc-upper}, we consider two cases. Suppose first that $\mean_*\leq0.5$. Then, finding a solution to \autoref{eq:goal-3} is equivalent to finding a solution $\mean^*\in[\delay,1]$ to the following quadratic equation:
\begin{align*}
h_1(x)=\frac{2}{\costb}x^2+x\left(\commission-2(1-\delay)-\frac{2}{\costb}\right)-\left(\frac{4\delay^2-4\delay}{\costb}+\frac{1}{2\costb}+\delay(1+\commission)-1\right).
\end{align*}
Now, because
\begin{align}
h_1(\delay)=-\frac{(2\delay-1)^2}{2\costb}+(1-\delay)(1-2\delay)<0,
\end{align}
we need to check that $h_1(2\delay-1/2)\leq0$ and $h_1(1)\geq0$. The first ensures that $\mean_*=2\delay-\mean^*\leq1/2$ and the latter ensures that $\mean^*\leq1$. Note that because $\mean^*\in[2\delay-1/2,1]$, it must be the case that $\delay\leq 3/4$.

Now, $h_1(1)\geq0$ is equivalent to the condition that
\[l_1(\cdot)=(1-\commission)(1-\delay)+\frac{(2\delay-1)^2}{\costb}-\frac{1}{2\costb}\leq0.\]
Instead, $h_1(2\delay-1/2)\leq0$ is equivalent to 
\[\delay\leq\frac{2+(4-\commission)\costb}{4(1+\costb)}\in\left[\frac{1}{2},\frac{4-\commission}{4}\right].\]
Suppose now that $\mean_*\in[0.5,\delay)$. Then, finding a solution to \autoref{eq:goal-3} is equivalent to finding a solution $\mean_*\in[\nicefrac{1}{2},\delay]$ to the following quadratic equation:
\begin{align*}
h_2(x)&=\frac{4}{\costb}x^2-x\left(\frac{8}{\costb}\delay+\commission-2(1-\delay)\right)+\delay\left(\frac{4\delay}{\costb}+\commission\right)-(1-\delay)(4\delay-1).
\end{align*}
Now, because
\[h_2(\delay)=(1-\delay)(1-2\delay)<0,\]
we need to check that $h_2(1/2)\geq0$ and $h_2(2\delay-1)\geq0$. The first ensures that $1/2\leq\mean_*$ and the second ensures that $\mean^*=2\delay-\mean_*\leq1$. Now, when $\delay\leq 3/4$, $2\delay-1\leq1/2$ so that $h_2(1/2)\geq0$ implies $h_2(2\delay-1)\geq0$. The condition $h_2(1/2)\geq0$ requires that
\[\delay\geq \frac{2+(4-\commission)\costb}{4(1+\costb)}.\]
Thus, we have that
\begin{align*}
\frac{2+(4-\commission)\costb}{4(1+\costb)}\leq\delay\leq\frac{3}{4}.
\end{align*}
This implies that $(1-\commission)\costb\leq 1$. As a consequence, we obtain that
\begin{align*}
2(1-\delay)(1-\commission)\costb+2(2\delay-1)^2-1\leq2(1-\delay)+2(2\delay-1)^2-1=(2\delay-1)(4\delay-3)\leq0,
\end{align*}
which is the remaining condition in \cref{itm:high-commission} in \autoref{proposition:nc-upper} when $\delay\leq3/4$.
Instead, when $\delay>3/4$, $h_2(2\delay-1)\geq0$ requires that
\begin{align*}
(1-\delay)\left[\frac{4}{\costb}(1-\delay)-(1-\commission)\right]\geq0.
\end{align*}
\end{proof}
\begin{proof}[Proof of \autoref{corollary:nc-upper}]
Standard results imply that if $U{_1^L}^\prime$ is monotone, then transfers exist that implement the allocation that solves the relaxed program. This immediately holds under the conditions of case \ref{itm:low-commission} in \autoref{proposition:nc-upper}. Consider then the solution to the relaxed problem under the conditions of case \ref{itm:high-commission}. Note that when $\mean_*\leq 1/2$, we have that 
\begin{align*}
U{_1^L}^\prime(\type)&=\left\{\begin{array}{ll}\dv&\text{if }\type<\mean_*\\
    \frac{2\delay-1}{c}+\dv&\text{if }\mean_*\leq \type\leq\mean^*\\
    \frac{2\type-1}{c}&\text{otherwise}\end{array}\right.,
\end{align*}
whereas when $\mean_*\geq1/2$, we have that 
\begin{align*}
U{_1^L}^\prime(\type)&=\left\{\begin{array}{ll}\dv&\text{if }\type\leq1/2\\
    \frac{2\type-1}{c}+\dv&\text{ if }1/2\leq\type<\mean_*\\
    \frac{2\delay-1}{c}+\dv&\text{if }\mean_*\leq \type\leq\mean^*\\
    \frac{2\type-1}{c}&\text{otherwise}\end{array}\right..
\end{align*}
Thus, in both cases monotonicity is satisfied if and only if
\begin{align}\label{eq:mean-mon}
\mean^*\geq\delay+\frac{(1-\delay)\costb}{2}.
\end{align}
\paragraph{Case 1: $\mean_*\leq1/2$} Recall that in this case the conditions of \autoref{proposition:nc-upper} boil down to $\delay\leq\min\{3/4, \frac{2+(4-\commission)\costb}{4(1+\costb)}\}$ and $l_1(\delay,\commission,\costb)\leq0$. 

\autoref{eq:mean-mon} holds if $h_1(\delay+(1-\delay)\costb/2)\leq0$ and $\delay+(1-\delay)\costb/2\leq1$. The latter holds if $\costb\leq2$.  This former holds if and only if:
\[\delay(1-\delay)\leq\frac{1+(1-\commission)\costb^2}{4+\costb^2}.\]
When $\commission=1$, the above equation is inconsistent with $\delay\leq \frac{2+(4-\commission)\costb}{4(1+\costb)}$. Instead, when $\commission=0$, the above equation is always satisfied.

\paragraph{Case 2: $\mean_*\geq 1/2$} Because $h_2(\cdot)$ is expressed in terms of $\mean_*$, \autoref{eq:mean-mon} is equivalent to $\mean_*\leq\delay-(1-\delay)\costb/2$. Thus, in this case we need that
\begin{align*}
\delay-\frac{\costb(1-\delay)}{2}\geq\frac{1}{2}&\Leftrightarrow 1-2\delay+\costb(1-\delay)\leq0\\
h_2(\delay-(1-\delay)\costb/2)\leq0&\Leftrightarrow (1-\delay)\left(1-2\delay+\frac{\commission\costb}{2}\right)\leq0.
\end{align*}

Recall that when $\mean_*\geq1/2$,  the conditions of \autoref{proposition:nc-upper} can be split into two groups:

\emph{Case 2a:} In this case, $\delay\in[\frac{2+(4-\commission)\costb}{4(1+\costb)},3/4]$ (As we showed before, this implies that $l_1(\delay, \commission,\costb)\leq0$). We thus have the following three equations:
\begin{align*}
1-2\delay+2\costb(1-\delay)&\leq0,&
1-2\delay+(1-\delay)\costb&\leq0, &
1-2\delay+\costb\commission/2&\leq0,
\end{align*}
where the first is the condition that defines Case 2a. We thus obtain that when $\commission=0$, monotonicity always holds. Instead, when $\commission=1$, the condition $\delay\in[\frac{2+(4-\commission)\costb}{4(1+\costb)},3/4]$  implies $\delay=3/4$, which is part of the next case.

\emph{Case 2b:} In this case, $\delay\geq 3/4$ and $(4/\costb)(1-\delay)-(1-\commission)\geq0$ and the monotonicity constraint implies that 
\begin{align*}
1-2\delay+(1-\delay)\costb&\leq0,
1-2\delay+\costb\commission/2\leq0.
\end{align*}
When $\commission=0$, the conditions are implied by $\costb\leq 4(1-\delay)$, so that monotonicity always holds. Instead, when $\commission=1$, we obtain the condition that $1-2\delay+\costb/2\leq0$. 
\end{proof}
\autoref{fig:monotonicity} shows the configurations of $(\delay,\costb)$ such that the monotonicity constraint holds.
\begin{figure}
\centering
\subfloat[$\commission=0$]{\scalebox{0.8}{%
\begin{tikzpicture}
    \begin{axis}[axis lines=middle,
    legend cell align={left},
    legend style={draw=none,font=\scriptsize,anchor=north, at={(-0.3,0.9)}},
            xmin=0,xmax=1,
        ymin=0,ymax=2,
        xtick={0,0.25,0.5,0.75},
        ytick={0.5,1,1.5},
        xlabel=$\delay$,ylabel=$\costb$,
        every axis x label/.style={at={(current axis.right of origin)},below right=2mm},
every axis y label/.style={at={(current axis.north west)},left=5mm}]
        \addplot [blue,thick,domain=0.25:0.495,name path=D,mark=$g(c)$]  {(16*x^2-8*x+1 )/(2-2*x)};
                \addlegendentry{$l_0(\delay,0,\costb)=0$}
   \addplot [mark=none,dashed,name path=C,forget plot] coordinates {(1/2, 0) (1/2, 2)};
\addplot [gray!50,forget plot] fill between [
of=D and C,
soft clip={domain=0.25:1}, ];    
   \addplot [mark=none,dashed,name path=E,forget plot] coordinates {(0.75, 0) (0.75, 2)};
\end{axis}
    \end{tikzpicture}}\label{fig:monotonicity-0}}
\subfloat[$\commission=1$]{\scalebox{0.8}{%
\begin{tikzpicture}
    \begin{axis}[axis lines=middle,
    legend cell align={left},
    legend style={draw=none,font=\scriptsize,anchor=north, at={(1.2,0.9)}},
            xmin=0,xmax=1,
        ymin=0,ymax=2,
        xtick={0,0.25,0.5,0.75},
        ytick={0.5,1,1.5},
        xlabel=$\delay$,ylabel=$\costb$,
        every axis x label/.style={at={(current axis.right of origin)},below right=2mm},
every axis y label/.style={at={(current axis.north west)},left=5mm}]
        \addplot [blue,thick,domain=0.25:0.495,name path=D,mark=$g(c)$]  {(16*x^2-8*x+1 )/(2-4*x)};
                \addlegendentry{$l_0(\delay,1,\costb)=0$}
        \addplot[red,thick,name path=B,domain=0.5:1]  {4*x - 2 };
        \addlegendentry{$\delay=0.5+\costb/4$}
   \addplot [mark=none,dashed,name path=C] coordinates {(1/2, 0) (1/2, 2)};
\addplot [gray!50] fill between [
of=D and B,
soft clip={domain=0.25:1}, ];    
 \addplot [mark=none,dashed,name path=E,forget plot] coordinates {(0.75, 0) (0.75, 2)};
\end{axis}
    \end{tikzpicture}}\label{fig:monotonicity-1}}
    \caption{Shaded gray area shows parameter values for which the solution to the relaxed problem does not satisfy monotonicity}\label{fig:monotonicity}
\end{figure}

\begin{proof}[Proof of \autoref{corollary:w-delay-above-half}]
In a \mass\ \downstream\ market, consumer's payoffs under commitment are as follows:
\begin{align*}
U_1^C(\type)&=\type\dv+\mathbbm{1}[\type\geq1/2]\int_{1/2}^\type\frac{2x-1}{c}dx,
\end{align*}
whereas average consumer welfare is given by
\begin{align*}
AW_C=\int_0^1U_1^C(\type)d\type=\vh\left(\frac{(1-\delay)}{2}+\frac{1}{24\costb}\right).
\end{align*}
Instead, the payoffs under limited commitment depend on \delay\ and \commission. For ease of exposition, we refer to the cases by the figures that illustrate them. 

\paragraph{Case 1: Product line is as in \autoref{fig:delay-above-half-low-gamma-1}} Relative to the commitment solution, there is less exclusion (all types above $2\delay-1\leq 1/2$ are served) and only one quality besides the outside option one is provided ($(2\delay-1)/c$). Like in the commitment solution, there is no downstream price discrimination. Consumer payoffs are given by:
\begin{align*}
U_1^L(\type)=\type\dv+\mathbbm{1}[\type\geq2\delay-1](\type-(2\delay-1))\frac{2\delay-1}{c}.
\end{align*}
It is easy to see that $\type\leq 2\delay-1$ are indifferent and $\type\in[2\delay-1,1/2]$ prefer limited commitment to the commitment solution. Consider now $\type\geq1/2$, then
\begin{align*}
U_1^C(\type)-U_1^L(\type)=-\left(\frac{1}{2}-(2\delay-1)\right)\frac{2\delay-1}{c}+\frac{1}{c}\left(\type-\frac{1}{2}\right)\left(\type+\frac{1}{2}-2\delay\right). 
\end{align*}
This difference is increasing in $\type$ for $\type\geq\delay$, negative for $\type=\delay$, and positive for $\type=1$. Instead, average consumer welfare under limited commitment is given by
\begin{align*}
AW_L=\dv+\frac{2\delay-1}{c}\int_{2\delay-1}^1\left[\type-(2\delay-1)\right]d\type=\dv+\frac{2\delay-1}{c}2(1-\delay)^2.
\end{align*}
\autoref{fig:avg-welfare-1} illustrates the difference $\costb\left(AW_C-AW_L\right)$ as a function of \delay. When \delay\ is close to $1/2$, the limited commitment solution is close to full exclusion -- even if it serves a larger number of types. As \delay\ grows, the upward distortion in quality kicks in so that average consumer welfare is larger under limited commitment.
\begin{figure}[th!]
\centering
\scalebox{0.8}{%
\begin{tikzpicture}
\begin{axis}[axis lines=middle,
    legend cell align={left},
    legend style={draw=none,font=\footnotesize,anchor=north, at={(1.3,1)}},ticks=none,
            xmin=0.5,xmax=0.8,
        ymin=-0.05,ymax=0.05,
        xtick={0.5,0.75},
        xlabel=$\delay$,ylabel={},
        every axis x label/.style={at={(current axis.right of origin)},below right=2mm},
every axis y label/.style={at={(current axis.north west)},left=5mm}]
        \addplot[blue,name path=A,domain=0.5:0.75,mark=$AW_C$]  {(49/24)-8*x+10*x^2-4*x^3};
\end{axis}
\end{tikzpicture}}
\caption{Difference in average welfare under commitment and limited commitment as a function of \delay\ for $\delay\in[0.5,0.75]$.}\label{fig:avg-welfare-1}
\end{figure}

\paragraph{Case 2: Product line is as in \autoref{fig:delay-above-half-low-gamma-2}:} The product line coincides with that in the commitment solution for consumer types in $[1/2,2\delay-1]$. Consumer payoffs under limited commitment are given by
\begin{align*}
U_1^L(\type)=\type\dv+\mathbbm{1}[\type\geq1/2]\int_{1/2}^{\min\{2\delay-1,\type\}}\frac{2x-1}{c}dx+\mathbbm{1}[\type\geq 2\delay-1](\type-(2\delay-1))\frac{2\delay-1}{c}.
\end{align*}
It is immediate that $U_1^C(\type)=U_1^L(\type)$ whenever $\type\leq 2\delay-1$. Instead, for $\type\geq 2\delay-1$, we have that \type\ is worse off under the commitment solution. Indeed, 
\begin{align*}
c\left(U_1^C(\type)-U_1^L(\type)\right)=\int_{2\delay-1}^{\type}2(x-\delay)dx=(\type-(2\delay-1))(\type-1)\leq0,
\end{align*}
where the difference is strict whenever $\type\in(2\delay-1,1)$. There are two reasons for this comparison: First, consumer types in $[2\delay-1,\delay]$ obtain a good of a higher quality than in the commitment solution. Second, whereas consumer types in $[\delay,1]$ obtain a good of lower quality, they also pay less for it, as the most the \ufirm\ can extract is what the consumer with type $2\delay-1$ pays for the good. 

\paragraph{Case 3: Product line is as in \autoref{fig:delay-intermediate}:} Relative to the commitment solution, there is less exclusion (all types above $\mean_*\leq 1/2$ are served) and there is price discrimination for types in the high-end of the product line ($\type\geq\mean^*$). Consumer payoffs are given by:\small
\begin{align*}
U_1^L(\type)=\max\{\type,\mean^*\}\dv+\mathbbm{1}[\type\geq\mean_*](\min\{\mean^*,\type\}-\mean_*)\frac{2\delay-1}{c}+\mathbbm{1}[\type\geq \mean^*]\int_{\mean^*}^{\type}\frac{2x-1}{c}dx.
\end{align*}\normalsize
Clearly, consumer types below $\mean_*$ are indifferent and consumer types in $[\mean_*,1/2]$ are better off under limited commitment. Consider now a consumer with type in $[\mean_*,\mean^*]$. In this case,
\begin{align*}\small
U_1^C(\type)-U_1^L(\type)=-\left(\frac{1}{2}-\mean_*\right)\left(\frac{2\delay-1}{c}\right)+\frac{2}{c}\left(\type-\frac{1}{2}\right)\left(\frac{1}{2}\left(\type+\frac{1}{2}\right)-\delay\right).
\end{align*}\normalsize
Instead, for $\type\geq\mean^*$, we have that:\small
\begin{align*}
U_1^C(\type)-U_1^L(\type)=(\type-\mean^*)\dv+\frac{1}{c}\left(\mean^*-\frac{1}{2}\right)\left(\mean^*+\frac{1}{2}-2\delay\right)-\left(\frac{1}{2}-\mean_*\right)\left(\frac{2\delay-1}{c}\right),
\end{align*}\normalsize
which is increasing in $\type$ and positive for $\type=1$: $U_1(1)=(1-\mean^*)\dv+(1/c)(\mean^*+1/2-2\delay)^2>0$.

\paragraph{Case 4: Product line is as in \autoref{fig:delay-high}:} Exclusion is the same as in the commitment solution, but there is price discrimination on the high-end of the product line ($\type\geq\mean^*$). Consumer payoffs are given by\small
\begin{align*}
U_1^L(\type)&=\max\{\type,\mean^*\}\dv+\mathbbm{1}[\type\geq\mean_*](\min\{\mean^*,\type\}-\mean_*)\frac{2\delay-1}{c}\\
&+\mathbbm{1}[\type\geq 1/2]\int_{1/2}^{\min\{\type,\mean_*\}}\frac{2x-1}{c}dx+\mathbbm{1}[\type\geq \mean^*]\int_{\mean^*}^{\type}\frac{2x-1}{c}dx.
\end{align*}\normalsize
It is immediate that consumer types $\type\leq\mean_*$ are indifferent between the commitment and limited commitment allocations. Consider now $\type\in[\mean_*,\mean^*]$. We have that
\begin{align*}
c\left(U_1^C(\type)-U_1^L(\type)\right)=(\type-\mean_*)(\type-\mean^*)\leq0.
\end{align*}
As in case 2, this obtains because types below $\delay$ obtain higher qualities and types above $\delay$ pay less for the lower-quality good that they obtain. Finally, consider now $\type\geq\mean^*$. In this case, 
\begin{align*}
U_1^C(\type)-U_1^L(\type)=(\type-\mean^*)\dv\geq0.
\end{align*}
Relative to the commitment solution, these types receive the same (average) rents upstream, but now they face price discrimination downstream.

\end{proof}
\subsection{Proof of \autoref{proposition:data-intermediary}}\label{appendix:data-design}
 
When the \ufirm\ can design both the product line and the information available to the \dfirm\, it is without loss of generality to restrict attention to  mechanisms $\varphi:\Types\mapsto\Delta(\allocations_1\times\{\vl,\vh\})$ that assign to each type \type\ a lottery over allocations $(\qualone,\transferone)\in\allocations_1$ and price recommendations for the \dfirm, $\rec\in\{\vl,\vh\}$ \citep{myerson1982optimal}.\footnote{These are the only recommendations \dfirm\ would find optimal to follow.} Furthermore, it is without loss to restrict attention to mechanisms such that the consumer participates and truthfully reports her type and the \dfirm\ \emph{obeys} the received recommendation. Thus, when the consumer's type is \type\ and the \dfirm\ receives recommendation $\rec$, \downstream\ profits are given by:
 \begin{align}
\profits_D^M(\type,\transfer_2)=\mathbbm{1}[\rec=\vl]\vl+(1-\mathbbm{1}[\rec=\vl])p(\type)\vh.
\end{align}
Furthermore, the consumer's payoff when her type is \type\ and she reports \typeb\ is\small
\begin{align}
W_1^M(\typeb,\type)=\int_{\allocations_1\times\{\vl,\vh\}}\left[\type\qualone(\posteriorf)-\transferone(\posteriorf)+p(\type)\mathbbm{1}[\rec=\vl]\dv\right]\varphi(d(\qualone,\transferone,\transfer_2)|\typeb).\notag
\end{align}\normalsize
Let $U_1^M(\type)\equiv W_1^M(\type,\type)$ denote the payoff from truthtelling.
 The \ufirm-profit maximizing mechanism solves
 \small
\begin{align}\label{eq:m-profit}\tag{M-OPT}
\max_{\varphi:\Types\mapsto\Delta(\allocations_1\times\{\vl,\vh\})}&\int_\Types\int_{\allocations_1\times\{\vl,\vh\}}\left[\transfer_1(\type)-c(\qualone(\type))+\commission\profits_D^M(\type,\rec)\right]\varphi(d(\qualone,\transferone,\rec)|\type)\priorf(d\type)\\
\text{s.t.}&
(\forall\type\in\Types)U_1^M(\type)\geq0\label{eq:m-participation}\tag{M-PC}\\
&(\forall\type\in\Types)(\forall\typeb\in\Types)U_1^M(\type)\geq W_1^M(\typeb,\type)\label{eq:m-truthtelling-1}\tag{M-TT$_1$}\\
&\int_{\Types\times\allocations_1\times\{\vl\}}\left(\vl-p(\type)\vh\right)\varphi(d(\qualone,\transferone,\transfer_2)|\type)\priorf(d\type)\geq0\label{eq:m-obedience-l}.\tag{M-OB$_{\vl}$}\\
&\int_{\Types\times\allocations_1\times\{\vh\}}\left(p(\type)\vh-\vl\right)\varphi(d(\qualone,\transferone,\transfer_2)|\type)\priorf(d\type)\geq0\label{eq:m-obedience-h}.\tag{M-OB$_{\vh}$}
\end{align}\normalsize
The last two constraints are the \dfirm's obedience constraints. It is possible to show that \ref{eq:m-profit} is equivalent to maximizing\small
\begin{align}\label{eq:m-virtuals}\tag{M-VS}
&\max_{\varphi}\int\left[\virtual(\priorf)\qualone-c(\qualone)+\commission\profits_D^M(\type,\rec)+(1-\commission)\welfare_D^M(\type,\rec)\right]\varphi(d(\qualone,\rec)|\type)\priorf(d\type)\\
\text{s.t.}\;&\text{\ref{eq:m-obedience-l}, \ref{eq:m-obedience-h}}, \text{ and }\notag\\
&U{_1^M}^\prime(\type)=\int_{[0,\maxq]\times\{\vl,\vh\}}\left[\qualone+p^\prime(\type)\dv\mathbbm{1}[\rec=\vl]\right]\varphi(d(\qualone,\rec)|\type)\text{ is increasing in }\type.\notag
\end{align}\normalsize
In the above expression, $\profits_D^M$ are the \downstream\ profits adjusted by \emph{dynamic} rents:
\begin{align*}
\profits_D^M(\type,\transfer_2)&=p(\type)\vh(1-\mathbbm{1}[\rec=\vl])\\
&+\mathbbm{1}[\rec=\vl]\left[p(\type)\vh+(1-p(\type))\left(\vl-\frac{1-\priorf(\type)}{\priorpdf(\type)}\frac{p^\prime(\type)}{1-p(\type)}\dv\right)\right],
\end{align*}
and $W_D^M$ is a measure of \downstream\ consumer welfare, adjusted by \emph{dynamic} rents:
\begin{align*}
\welfare_D^M(\type,\rec)&=\dv\mathbbm{1}[\rec=\vl]\left(p(\type)-\frac{1-\priorf(\type)}{\priorpdf(\type)}p^\prime(\type)\right).
\end{align*}
Note that the objective in \ref{eq:m-virtuals} would be maximized by placing probability $1$ on $\qualone=\virtual(\priorf)/c$ conditional on any recommendation to the \dfirm. This can be achieved without affecting the obedience constraints \ref{eq:m-obedience-l} and \ref{eq:m-obedience-h}. 

Thus, the maximization problem \ref{eq:m-virtuals} boils down to finding a mapping $\chi:\Types\mapsto\Delta(\{\vl,\vh\})$ to maximize:
\begin{align}
\int_\Types\sum_{\rec\in\{\vl,\vh\}}\left((1-\commission)\welfare_D^M(\type,\rec)+\commission\profits_D^M(\type,\rec)\right)\chi(\rec|\type)\priorf(d\type),
\end{align}
subject to \ref{eq:m-obedience-l} and \ref{eq:m-obedience-h}. Standard results imply that this problem is equivalent to finding a Bayes' plausible posterior distribution $\bsplit\in\Delta\Posteriors$ and a selection \pricetwo\ from the \dfirm's best response correspondence to solve the above problem. Replacing our parametric assumptions shows that the objective function only depends on the posterior mean of $p(\cdot)$. Thus, we need to solve:\small
\begin{align}\label{eq:m-posterior}
\max_{\mpc:U\succ_{cx} \mpc}\int_0^1\left[(1-\pricetwo(\vl,p))\commission p\vh+\pricetwo(\vl,p)\left(2(1-\delay)p+\delay(1+\commission)-1\right)\right]\mpc(dp),
\end{align}\normalsize
where $U\succ_{cx}\mpc$ states that the uniform distribution (which is the prior distribution of $p=\priorf$) dominates \mpc\ in the convex order. The result that there is no price discrimination in the solution to \ref{eq:m-virtuals} follows from establishing that the concavification of the integrand in \autoref{eq:m-posterior} evaluated at the prior coincides with the integrand evaluated at the prior. In what follows, denote the integrand in \autoref{eq:m-posterior} by $\profits_U^M(p)$. Note that it is piecewise linear in $p$.

When $\delay<1/2$, $\profits_U^M$ satisfies the following: (i) There is a jump up at $p=\delay$, (ii) the concavification of $\profits_U^M(\cdot)$ coincides with $\commission p \vh$ for $p\geq\delay$. The reason for the latter is that $\commission p\vh\geq 2(1-\delay)p+\delay(1+\commission)-1$ for all $p\in[0,\delay]$. Instead, when $\delay>1/2$, $\profits_U^M$ satisfies the following: (i) There is a jump down at $p=\delay$, (ii) the concavification of $\profits_U^M(\cdot)$ coincides with $2p(1-\delay)+\delay(1+\commission)-1$ for $p\leq\delay$. The reason for the latter is that $\commission p\vh\leq 2(1-\delay)p+\delay(1+\commission)-1$ for all $p\in[\delay,1]$. In both cases, the results in \cite{arieli2019optimal} imply that no information revelation is optimal.

\section{Mechanism-selection game and solution concept}\label{appendix:mechanism}
We describe here the mechanisms available to the \ufirm\ and \dfirm\ in the data intermediation and limited commitment settings. We refer the reader to \cite{doval2022mechanism} for a discussion of mechanism-selection games with a continuum of types and the appropriate solution concept. Here, we take this discussion as given and merely specify the firms' action sets and the sequential rationality requirements.

\paragraph{Mechanisms:} The firm which operates in period $t\in\{1,2\}$ can choose a mechanism $\mechanism_t$, defined as follows. The mechanism $\mechanism_t$ is a tuple $\left(M_t,S_t,\varphi_t\right)$, where $M_t$ is a set of input messages, $S_t$ is a set of output messages, and $\varphi:M_t\mapsto\Delta(S_t\times A_t)$ is a mapping associating to each input message $m\in M_t$ a lottery over output messages, $s_t\in S_t$ and allocations, $a_t\in A_t$. Recall from \autoref{sec:model} that $A_1=[0,\maxq]\times\mathbb{R}_+$ and $A_2=\{0,1\}\times\mathbb{R}_+$.

Given a mechanism $\mechanism_t$, the consumer chooses whether to participate. If she does not participate, then the allocation is no trade. If she participates, then she \emph{privately} sends an input message $m\in M_t$ into the mechanism, which determines the distribution $\varphi_t(\cdot|m)$ from which the output message and the allocation are drawn. The firm designing the mechanism and the consumer observe the output message and the allocation.

\paragraph{Information available to \dfirm\ and consumer:} The \dfirm\ observes the output message from the \ufirm's upstream mechanism and the consumer's participation decision both in the data intermediation and the limited commitment case.\footnote{\cite{calzolari2006optimality} make the same assumption in their sequential common agency problem.} In the case of limited commitment, the \dfirm\ also observes the allocation that the consumer obtains in the upstream interaction. 

It follows that in the data intermediation setting the \dfirm\ chooses a mechanism for each output message, $\mechanism_2(s_1)$, whereas in the limited commitment setting the \dfirm\ chooses a mechanism for each output message $s_1$ and each allocation $(\qualone,\transferone)$.

\paragraph{\dfirm\ sequential rationality:} Given an upstream mechanism,  $\mechanism_1$, and a family of downstream mechanisms $\mechanism_2(\cdot)$, the consumer faces an extensive form game. The consumer's participation and reporting strategy together with the upstream mechanism determine the \dfirm's beliefs as a function of what the \dfirm\ observes via Bayes' rule. We assume that the \dfirm\ chooses $\mechanism_2$ to maximize revenue conditional on its beliefs. 

\paragraph{\ufirm\ optimality:} Given the consumer's strategy and the downstream mechanism(s), we assume that the upstream mechanism maximizes upstream profits. 

{\singlespacing{
\bibliographystyle{ecta}
\bibliography{pld}}}

\begin{thebibliography}{53}
\newcommand{\enquote}[1]{``#1''}
\expandafter\ifx\csname natexlab\endcsname\relax\def\natexlab#1{#1}\fi

\bibitem[\protect\citeauthoryear{Acquisti and Varian}{Acquisti and
  Varian}{2005}]{acquisti2005conditioning}
\textsc{Acquisti, A. and H.~R. Varian} (2005): \enquote{Conditioning prices on
  purchase history,} \emph{Marketing Science}, 24, 367--381.

\bibitem[\protect\citeauthoryear{Aliprantis and Border}{Aliprantis and
  Border}{2013}]{aliprantis2013infinite}
\textsc{Aliprantis, C. and K.~Border} (2013): \emph{Infinite Dimensional
  Analysis: A Hitchhiker's Guide}, Springer-Verlag Berlin and Heidelberg GmbH
  \& Company KG.

\bibitem[\protect\citeauthoryear{Anderson and Dana}{Anderson and
  Dana}{2009}]{anderson2009price}
\textsc{Anderson, E.~T. and J.~D. Dana} (2009): \enquote{When is price
  discrimination profitable?} \emph{Management Science}, 55, 980--989.

\bibitem[\protect\citeauthoryear{Anderson and Celik}{Anderson and
  Celik}{2015}]{anderson2015product}
\textsc{Anderson, S.~P. and L.~Celik} (2015): \enquote{Product line design,}
  \emph{Journal of Economic Theory}, 157, 517--526.

\bibitem[\protect\citeauthoryear{Argenziano and Bonatti}{Argenziano and
  Bonatti}{2020}]{argenziano2020information}
\textsc{Argenziano, R. and A.~Bonatti} (2020): \enquote{Information Revelation
  and Privacy Protection,} \emph{Working paper}.

\bibitem[\protect\citeauthoryear{Arieli, Babichenko, Smorodinsky, and
  Yamashita}{Arieli et~al.}{2023}]{arieli2019optimal}
\textsc{Arieli, I., Y.~Babichenko, R.~Smorodinsky, and T.~Yamashita} (2023):
  \enquote{Optimal persuasion via bi-pooling,} \emph{Theoretical Economics},
  18, 15--36.

\bibitem[\protect\citeauthoryear{Armstrong}{Armstrong}{2006}]{armstrong_2006}
\textsc{Armstrong, M.} (2006): \emph{Recent Developments in the Economics of
  Price Discrimination}, Cambridge University Press, vol.~2 of
  \emph{Econometric Society Monographs}, 97–141.

\bibitem[\protect\citeauthoryear{Bergemann and Ottaviani}{Bergemann and
  Ottaviani}{2021}]{bergemann2021information}
\textsc{Bergemann, D. and M.~Ottaviani} (2021): \enquote{Information markets
  and nonmarkets,} in \emph{Handbook of Industrial Organization}, Elsevier,
  vol.~4, 593--672.

\bibitem[\protect\citeauthoryear{Calzolari and Pavan}{Calzolari and
  Pavan}{2006{\natexlab{a}}}]{calzolari2006monopoly}
\textsc{Calzolari, G. and A.~Pavan} (2006{\natexlab{a}}): \enquote{Monopoly
  with resale,} \emph{The RAND Journal of Economics}, 37, 362--375.

\bibitem[\protect\citeauthoryear{Calzolari and Pavan}{Calzolari and
  Pavan}{2006{\natexlab{b}}}]{calzolari2006optimality}
---\hspace{-.1pt}---\hspace{-.1pt}--- (2006{\natexlab{b}}): \enquote{On the
  optimality of privacy in sequential contracting,} \emph{Journal of Economic
  theory}, 130, 168--204.

\bibitem[\protect\citeauthoryear{Carrillo and Mariotti}{Carrillo and
  Mariotti}{2000}]{carrillo2000strategic}
\textsc{Carrillo, J.~D. and T.~Mariotti} (2000): \enquote{Strategic ignorance
  as a self-disciplining device,} \emph{The Review of Economic Studies}, 67,
  529--544.

\bibitem[\protect\citeauthoryear{Crauel}{Crauel}{2002}]{crauel2002random}
\textsc{Crauel, H.} (2002): \emph{Random probability measures on Polish
  spaces}, vol.~11, CRC press.

\bibitem[\protect\citeauthoryear{Cummings, Ligett, Pai, and Roth}{Cummings
  et~al.}{2015}]{cummings2015strange}
\textsc{Cummings, R., K.~Ligett, M.~M. Pai, and A.~Roth} (2015): \enquote{The
  strange case of privacy in equilibrium models,} \emph{arXiv preprint
  arXiv:1508.03080}.

\bibitem[\protect\citeauthoryear{Deb and Said}{Deb and
  Said}{2015}]{deb2015dynamic}
\textsc{Deb, R. and M.~Said} (2015): \enquote{Dynamic screening with limited
  commitment,} \emph{Journal of Economic Theory}, 159, 891--928.

\bibitem[\protect\citeauthoryear{Decarolis, Spagnolo, and Pacini}{Decarolis
  et~al.}{2016}]{decarolis2016past}
\textsc{Decarolis, F., G.~Spagnolo, and R.~Pacini} (2016): \enquote{Past
  performance and procurement outcomes,} Tech. rep., National Bureau of
  Economic Research.

\bibitem[\protect\citeauthoryear{Doval and Skreta}{Doval and
  Skreta}{2022{\natexlab{a}}}]{doval2022mechanism}
\textsc{Doval, L. and V.~Skreta} (2022{\natexlab{a}}): \enquote{Mechanism
  design with limited commitment,} \emph{Econometrica}, 90, 1463--1500.

\bibitem[\protect\citeauthoryear{Doval and Skreta}{Doval and
  Skreta}{2022{\natexlab{b}}}]{doval2022markov}
---\hspace{-.1pt}---\hspace{-.1pt}--- (2022{\natexlab{b}}): \enquote{Mechanism
  Design with Limited Commitment: Markov Environments,}
  \href{https://www.dropbox.com/s/tqlqc2cqhelh1gs/markov-public.pdf?dl=0}{Click
  here}.

\bibitem[\protect\citeauthoryear{Doval and Skreta}{Doval and
  Skreta}{Forthcoming}]{doval2022constrained}
---\hspace{-.1pt}---\hspace{-.1pt}--- (Forthcoming): \enquote{Constrained
  information design,} \emph{Mathematics of Operations research}.

\bibitem[\protect\citeauthoryear{Dworczak and Martini}{Dworczak and
  Martini}{2019}]{dworczak2019simple}
\textsc{Dworczak, P. and G.~Martini} (2019): \enquote{The simple economics of
  optimal persuasion,} \emph{Journal of Political Economy}, 127, 1993--2048.

\bibitem[\protect\citeauthoryear{Eilat, Eliaz, and Mu}{Eilat
  et~al.}{2021}]{eilat2021bayesian}
\textsc{Eilat, R., K.~Eliaz, and X.~Mu} (2021): \enquote{Bayesian privacy,}
  \emph{Theoretical Economics}, 16, 1557--1603.

\bibitem[\protect\citeauthoryear{Ellison}{Ellison}{2005}]{ellison2005model}
\textsc{Ellison, G.} (2005): \enquote{A model of add-on pricing,} \emph{The
  Quarterly Journal of Economics}, 120, 585--637.

\bibitem[\protect\citeauthoryear{Es{\H{o}} and Szentes}{Es{\H{o}} and
  Szentes}{2007}]{eso2007optimal}
\textsc{Es{\H{o}}, P. and B.~Szentes} (2007): \enquote{Optimal information
  disclosure in auctions and the handicap auction,} \emph{The Review of
  Economic Studies}, 74, 705--731.

\bibitem[\protect\citeauthoryear{{Executive Office of the President of the
  United States}}{{Executive Office of the President of the United
  States}}{2015}]{obama2015bigdata}
\textsc{{Executive Office of the President of the United States}} (2015):
  \enquote{Big Data and Differential Pricing,}
  \href{https://obamawhitehouse.archives.gov/sites/default/files/whitehouse_files/docs/Big_Data_Report_Nonembargo_v2.pdf}{Click
  here}.

\bibitem[\protect\citeauthoryear{Fudenberg and Villas-Boas}{Fudenberg and
  Villas-Boas}{2006}]{fudenberg2006behavior}
\textsc{Fudenberg, D. and J.~M. Villas-Boas} (2006): \enquote{Behavior-based
  price discrimination and customer recognition,} \emph{Handbook on economics
  and information systems}, 1, 377--436.

\bibitem[\protect\citeauthoryear{Gentzkow and Kamenica}{Gentzkow and
  Kamenica}{2016}]{gentzkow2016rothschild}
\textsc{Gentzkow, M. and E.~Kamenica} (2016): \enquote{A Rothschild-Stiglitz
  approach to Bayesian persuasion,} \emph{American Economic Review}, 106,
  597--601.

\bibitem[\protect\citeauthoryear{Hart and Tirole}{Hart and
  Tirole}{1988}]{hart1988contract}
\textsc{Hart, O.~D. and J.~Tirole} (1988): \enquote{Contract renegotiation and
  Coasian dynamics,} \emph{The Review of Economic Studies}, 55, 509--540.

\bibitem[\protect\citeauthoryear{Hidir and Vellodi}{Hidir and
  Vellodi}{2021}]{hidir2020privacy}
\textsc{Hidir, S. and N.~Vellodi} (2021): \enquote{Privacy, personalization,
  and price discrimination,} \emph{Journal of the European Economic
  Association}, 19, 1342--1363.

\bibitem[\protect\citeauthoryear{Ichihashi}{Ichihashi}{2020}]{ichihashi2020online}
\textsc{Ichihashi, S.} (2020): \enquote{Online privacy and information
  disclosure by consumers,} \emph{American Economic Review}, 110, 569--95.

\bibitem[\protect\citeauthoryear{Itoh}{Itoh}{1983}]{itoh1983monopoly}
\textsc{Itoh, M.} (1983): \enquote{Monopoly, product differentiation and
  economic welfare,} \emph{Journal of Economic Theory}, 31, 88--104.

\bibitem[\protect\citeauthoryear{Ivaldi and Martimort}{Ivaldi and
  Martimort}{1994}]{ivaldi1994competition}
\textsc{Ivaldi, M. and D.~Martimort} (1994): \enquote{Competition under
  nonlinear pricing,} \emph{Annales d'Economie et de Statistique}, 71--114.

\bibitem[\protect\citeauthoryear{Johnson and Myatt}{Johnson and
  Myatt}{2003}]{johnson2003multiproduct}
\textsc{Johnson, J.~P. and D.~P. Myatt} (2003): \enquote{Multiproduct quality
  competition: Fighting brands and product line pruning,} \emph{American
  Economic Review}, 93, 748--774.

\bibitem[\protect\citeauthoryear{Johnson and Myatt}{Johnson and
  Myatt}{2015}]{johnson2015properties}
---\hspace{-.1pt}---\hspace{-.1pt}--- (2015): \enquote{The properties of
  product line prices,} \emph{International Journal of Industrial
  Organization}, 43, 182--188.

\bibitem[\protect\citeauthoryear{Johnson and Myatt}{Johnson and
  Myatt}{2018}]{johnson2018determinants}
---\hspace{-.1pt}---\hspace{-.1pt}--- (2018): \enquote{The determinants of
  product lines,} \emph{The RAND Journal of Economics}, 49, 541--573.

\bibitem[\protect\citeauthoryear{Kamenica}{Kamenica}{2008}]{kamenica2008contextual}
\textsc{Kamenica, E.} (2008): \enquote{Contextual inference in markets: On the
  informational content of product lines,} \emph{American Economic Review}, 98,
  2127--49.

\bibitem[\protect\citeauthoryear{Kolotilin}{Kolotilin}{2018}]{kolotilin2018optimal}
\textsc{Kolotilin, A.} (2018): \enquote{Optimal information disclosure: A
  linear programming approach,} \emph{Theoretical Economics}, 13, 607--635.

\bibitem[\protect\citeauthoryear{Laffont and Tirole}{Laffont and
  Tirole}{1988}]{laffont1988dynamics}
\textsc{Laffont, J.-J. and J.~Tirole} (1988): \enquote{The dynamics of
  incentive contracts,} \emph{Econometrica}, 1153--1175.

\bibitem[\protect\citeauthoryear{Li and Shi}{Li and
  Shi}{2017}]{li2017discriminatory}
\textsc{Li, H. and X.~Shi} (2017): \enquote{Discriminatory information
  disclosure,} \emph{American Economic Review}, 107, 3363--85.

\bibitem[\protect\citeauthoryear{Luo, Perrigne, and Vuong}{Luo
  et~al.}{2018}]{luo2018structural}
\textsc{Luo, Y., I.~Perrigne, and Q.~Vuong} (2018): \enquote{Structural
  analysis of nonlinear pricing,} \emph{Journal of Political Economy}, 126,
  2523--2568.

\bibitem[\protect\citeauthoryear{Milgrom and Segal}{Milgrom and
  Segal}{2002}]{milgrom2002envelope}
\textsc{Milgrom, P. and I.~Segal} (2002): \enquote{Envelope theorems for
  arbitrary choice sets,} \emph{Econometrica}, 70, 583--601.

\bibitem[\protect\citeauthoryear{Miravete}{Miravete}{2002}]{miravete2002estimating}
\textsc{Miravete, E.~J.} (2002): \enquote{Estimating demand for local telephone
  service with asymmetric information and optional calling plans,} \emph{The
  Review of Economic Studies}, 69, 943--971.

\bibitem[\protect\citeauthoryear{Mussa and Rosen}{Mussa and
  Rosen}{1978}]{mussa1978monopoly}
\textsc{Mussa, M. and S.~Rosen} (1978): \enquote{Monopoly and product quality,}
  \emph{Journal of Economic theory}, 18, 301--317.

\bibitem[\protect\citeauthoryear{Myerson}{Myerson}{1982}]{myerson1982optimal}
\textsc{Myerson, R.~B.} (1982): \enquote{Optimal coordination mechanisms in
  generalized principal--agent problems,} \emph{Journal of Mathematical
  Economics}, 10, 67--81.

\bibitem[\protect\citeauthoryear{Myerson}{Myerson}{1986}]{myerson1986multistage}
---\hspace{-.1pt}---\hspace{-.1pt}--- (1986): \enquote{Multistage games with
  communication,} \emph{Econometrica: Journal of the Econometric Society},
  323--358.

\bibitem[\protect\citeauthoryear{Pavan, Segal, and Toikka}{Pavan
  et~al.}{2014}]{pavan2014dynamic}
\textsc{Pavan, A., I.~Segal, and J.~Toikka} (2014): \enquote{Dynamic mechanism
  design: A myersonian approach,} \emph{Econometrica}, 82, 601--653.

\bibitem[\protect\citeauthoryear{Skreta}{Skreta}{2006}]{skreta2006sequentially}
\textsc{Skreta, V.} (2006): \enquote{Sequentially optimal mechanisms,}
  \emph{The Review of Economic Studies}, 73, 1085--1111.

\bibitem[\protect\citeauthoryear{Skreta}{Skreta}{2015}]{skreta2015optimal}
---\hspace{-.1pt}---\hspace{-.1pt}--- (2015): \enquote{Optimal auction design
  under non-commitment,} \emph{Journal of Economic Theory}, 159, 854--890.

\bibitem[\protect\citeauthoryear{Stantcheva}{Stantcheva}{2020}]{stantcheva2020dynamic}
\textsc{Stantcheva, S.} (2020): \enquote{Dynamic Taxation,} \emph{Annual Review
  of Economics}, 12, 801--831.

\bibitem[\protect\citeauthoryear{Stokey}{Stokey}{1979}]{stokey1979intertemporal}
\textsc{Stokey, N.~L.} (1979): \enquote{Intertemporal price discrimination,}
  \emph{The Quarterly Journal of Economics}, 355--371.

\bibitem[\protect\citeauthoryear{Sun}{Sun}{2014}]{sun2014dynamic}
\textsc{Sun, C.-J.} (2014): \enquote{Dynamic price discrimination with customer
  recognition,} \emph{The BE Journal of Theoretical Economics}, 14, 217--250.

\bibitem[\protect\citeauthoryear{Taylor}{Taylor}{2004}]{taylor2004consumer}
\textsc{Taylor, C.~R.} (2004): \enquote{Consumer privacy and the market for
  customer information,} \emph{RAND Journal of Economics}, 631--650.

\bibitem[\protect\citeauthoryear{Villas-Boas}{Villas-Boas}{2004}]{villas2004communication}
\textsc{Villas-Boas, J.~M.} (2004): \enquote{Communication strategies and
  product line design,} \emph{Marketing Science}, 23, 304--316.

\bibitem[\protect\citeauthoryear{Xu and Dukes}{Xu and
  Dukes}{2019}]{xu2019product}
\textsc{Xu, Z. and A.~Dukes} (2019): \enquote{Product line design under
  preference uncertainty using aggregate consumer data,} \emph{Marketing
  Science}, 38, 669--689.

\bibitem[\protect\citeauthoryear{Zhang}{Zhang}{2011}]{zhang2011perils}
\textsc{Zhang, J.} (2011): \enquote{The perils of behavior-based
  personalization,} \emph{Marketing Science}, 30, 170--186.

\end{thebibliography}

\end{document}